%% file: V1_long.tex
\documentclass[a4,12pt,draftcls,onecolumn]{IEEEtran}
\usepackage{bbm,dsfont,mathrsfs,fixmath}
\usepackage{amsmath,amssymb,graphicx}
\usepackage{cite}
\usepackage{amsmath,amstext,amsfonts,amssymb}
\usepackage{epsfig}
\usepackage{exscale}
\usepackage{enumerate}
\usepackage{cite}
\usepackage{amsmath}
\usepackage{graphicx}
\usepackage{amssymb}
\usepackage{array}
\usepackage{stfloats}
\usepackage{balance}
\usepackage{dsfont}
\usepackage{bbm}
\usepackage{ifpdf}
\usepackage{stfloats,color}
\usepackage{dsfont}

\include{packages}

\include{commandes}

\usepackage[T1]{fontenc}
\graphicspath{{./Figuras/}}

\begin{document}
\title{The Three-Terminal Interactive \\ Lossy Source Coding Problem}
\author{Leonardo Rey Vega,  Pablo Piantanida and Alfred O. Hero III
\thanks{The material in this paper was partially published in the IEEE International Symposium on Information Theory, Honolulu, Hawaii, USA, June 29 - July 4, 2014 and in the ncIEEE International Symposium on Information Theory, Hong-Kong, China, June 14 - June 19, 2015. The work of L. Rey Vega was partially supported by project UBACyT 2002013100751BA. The work of P. Piantanida  was partially supported by the FP7 Network of Excellence in Wireless COMmunications NEWCOM\#. The work of A. Hero was partially supported by a DIGITEO Chair from 2008 to 2013  and by US ARO grant W911NF-15-1-0479.}
\thanks{L. Rey Vega is with the Departments of Electronics (FIUBA) and CSC-CONICET, Buenos Aires, Argentina (e-mail: lrey@fi.uba.ar, cgalar@fi.uba.ar).}
\thanks{P. Piantanida is with the Laboratoire des Signaux et Syst\`emes (L2S), CentraleSupelec, 91192 Gif-sur-Yvette, France (e-mail: pablo.piantanida@supelec.fr).}
\thanks{Alfred O. Hero III  is with the  Department of Electrical Eng. \& CompSci University of Michigan, Ann Arbor, MI, USA (e-mail: hero@umich.edu).}}

\maketitle

\begin{abstract}
The three-node multiterminal lossy source coding problem is investigated. We derive an inner bound to the general rate-distortion region of this problem which is a natural extension of the seminal work by Kaspi'85 on the interactive two-terminal source coding problem. It is shown that this (rather involved) inner bound contains several rate-distortion regions of some relevant source coding settings. In this way, besides the non-trivial extension of the interactive two terminal problem, our results can be seen as a generalization and hence unification of several previous works in the field. Specializing to particular cases we obtain novel rate-distortion regions for several lossy source coding problems. We finish by describing some of the open problems and challenges. However, the general three-node multiterminal lossy source coding problem seems to offer a formidable mathematical complexity.
\end{abstract}


\IEEEpeerreviewmaketitle

\newpage

\section{Introduction}
\label{sec:intro}

\subsection{Motivation and related works}
\label{subsec:motivation}

Distributed source coding is an important branch of study in information theory with enormous relevance for the present and future technology. Efficient distributed data compression  may be the only way to guarantee acceptable levels of performance when energy and link bandwidth are severely limited as in many real world sensor networks. The distributed data collected by different nodes in a network can be highly correlated and this correlation can be exploited at the application layer, e.g., for target localization and tracking or anomaly detection. In such cases cooperative joint data-compression  can achieve a better overall rate-distortion trade-off that independent compression at each node.   

Complete answers to the optimal trade-offs between rate and distortion for distributed source coding are scarce and the solution to many problems remain elusive. Two of the most important results in information theory, Slepian-Wolf solution to the distributed lossless source coding problem \cite{Slepian_1973} and  Wyner-Ziv \cite{1055508} single letter solution for the rate-distortion region when side information is available at the decoder provided the kick-off for the study of these important problems. Berger and Tung \cite{berger-1977,PhD-Tung} generalized the Slepian-Wolf problem when lossy reconstructions are required at the decoder. It was shown that the region obtained, although not tight in general, is the optimal one in several special cases \cite{32119,490552,669162,4494707} and strictly suboptimal in others \cite{5895101}. Heegard and Berger \cite{heegard_rate_1985} considered the Wyner-Ziv problem when the side information at the decoder may be absent or when there are two decoders with degraded side information. Timo \emph{et al} \cite{timo_rate_2011} correctly extended the achievable region for many ($>2$) decoders. In \cite{6303916} and the references therein, the complementary delivery problem (closely related to the Heegard-Berger problem) is also studied. 
The use of interaction in a multiterminal source coding setting has not been so extensively studied as the problems mentioned above. Through the use of multiple rounds of interactive exchanges of information explicit cooperation can take place using  distributed/successive refinement source coding. Transmitting ``reduced pieces'' of information, and constructing an explicit sequential cooperative exchange of information, can be more efficient that transmitting the ``total information'' in one-shot.

The value of interaction for source coding problems was first recognized by Kaspi in his seminal work~\cite{kaspi_two-way_1985}, where the interactive two-terminal lossy source coding problem was introduced and solved under the assumption of a finite number of communication rounds. In \cite{5513293} it is shown that interaction strictly outperforms (in term of sum rate) the Wyner-Ziv rate function. There are also several extensions to the original Kaspi problem. In \cite{permuter_two-way_2010} the interactive source coding problem with a helper is solved when the sources satisfy a certain Markov chain property. In \cite{ma_results_2011,ma_interactive_2012, 6284202} other interesting cases where interactive cooperation can be beneficial are studied. To the best of our knowledge, a proper generalization of this setting to interactive multiterminal ($>2$) lossy source coding has not yet been observed.

\subsection{Main contributions}
\label{subsec:contributions}

\begin{figure}[t]
\centering
\ifpdf\includegraphics[angle=0,width=0.7\columnwidth,keepaspectratio,trim= 0mm 0mm 0mm 0mm,clip]{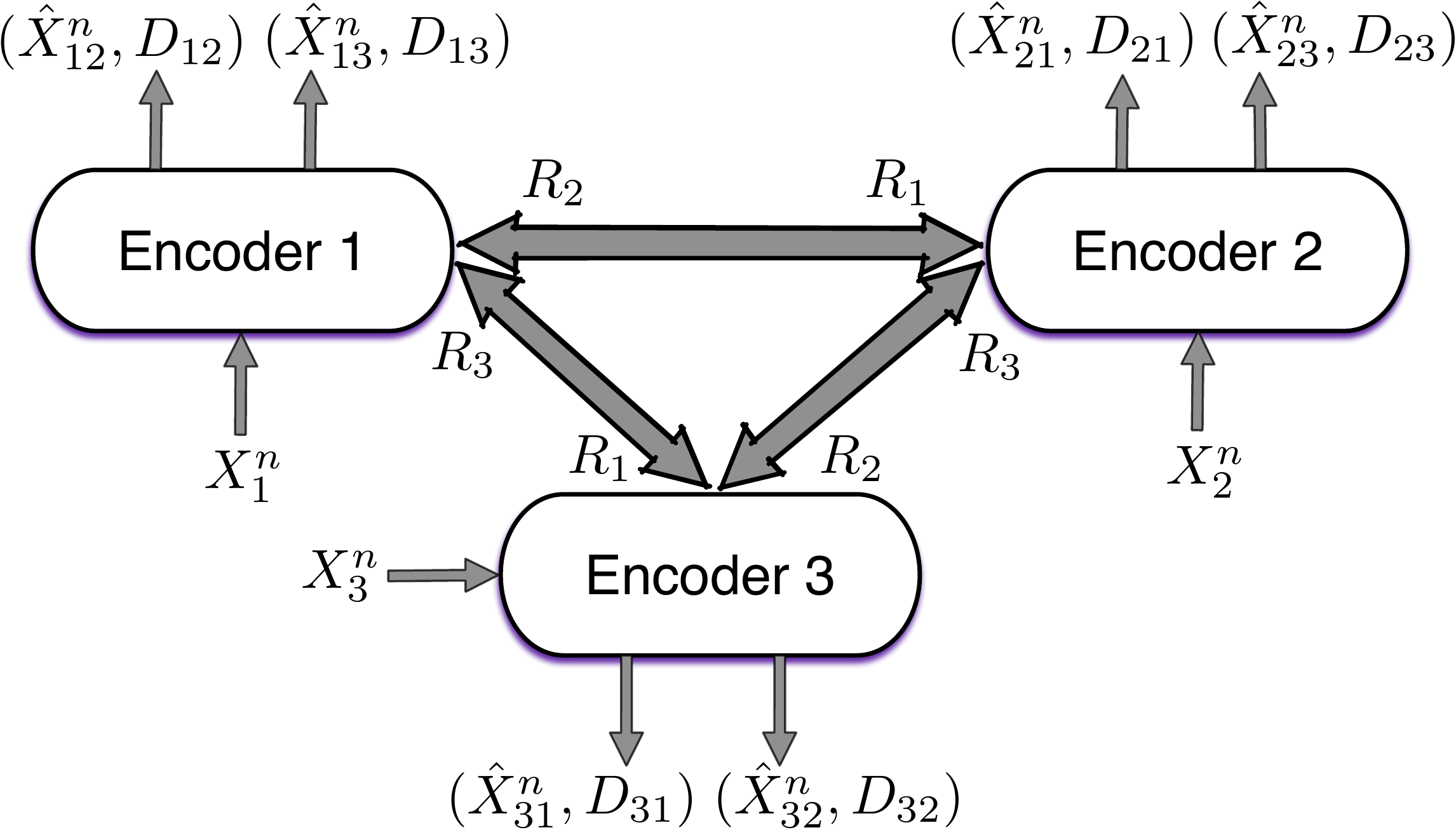} 
\else
	\includegraphics[angle=270,width=0.7\columnwidth,keepaspectratio,trim= 0mm 0mm 0mm 0mm,clip]{model.eps} 
\fi
\caption{Three-Terminal Interactive Source Coding. There is a single noiseless rate-limited broadcast
channel from each terminal to the other two terminals. $D_{ij}$ denotes the average per-letter distortion between the source $X_j^n$ and $\hat{X}_{ij}^n$ measured at the node $i$ for each pair $i\neq j$. } 
\label{fig:model}
\end{figure}

In this paper, we consider the three-terminal interactive lossy source coding problem presented in Fig.~\ref{fig:model}. We have a network composed of $3$ nodes which can interact through a broadcast rate-limited --error free-- channel. Each node measures the realization of a discrete memoryless source (DMS) and is required to reconstruct the sources from the other terminals with a fidelity criterion. Nodes are allowed to interact by interchanging descriptions of their observed sources realizations over a finite number of communication rounds. After the information exchange phase is over, the nodes try to reconstruct the realization of the sources at the other nodes using the recovered descriptions.

The general rate-distortion region seems to pose a formidable mathematical problem which encompass several known open problems. However, several properties of this problem are established in this paper.  

\subsection*{General achievable region}
We derive a general achievable region by assuming a finite number of rounds. This region is not a trivial extension of Kaspi's region~\cite{kaspi_two-way_1985}  and the main ideas behind its derivation are the exchange of common and private descriptions between the nodes in the network in order to exploit optimality the different side informations at the different nodes. As in the original Kaspi's formulation, the key to obtaining the achievable region is the natural cooperation between the nodes induced by the generation of new descriptions based on the past exchanged description. However, in comparison to Kaspi's  $2$ node case, the $3$ nodes interactions make significant differences in the optimal action of each node at the encoding and decoding procedure in a given round. At each encoding stage, each node need to communicate to two nodes with different side information. This is reminiscent of the Heegard-Berger problem \cite{heegard_rate_1985,timo_rate_2011}, whose complete solution is not known, when the side information at the decoders is not degraded. Moreover, the situation is a bit more complex because of the presence of $3$-way interaction.  This similarity between the Heegard-Berger problem leads us to consider the generation of two sets of messages at each node: common messages destined to all nodes and private messages destined to some restricted sets of nodes. On the other hand, when each node is acting as a decoder the nodes need to recover a set of common and private messages  generated at different nodes (i.e. at round $l$ node 3, needs to recover the common descriptions generated at nodes 1 and 2 and the private ones generated also at nodes 1 and 2). This is reminiscent of the Berger-Tung problem  \cite{PhD-Tung,32119,490552,gastpar_wyner-ziv_2004} which is also an open problem. Again, the situation is more involved because of the cooperation induced by the multiple rounds of exchanged information. Particularly important is the fact that, in the case of the common descriptions, there is  cooperation based on the conditioning on the previous exchanged descriptions  in addition to cooperation naturally induced by the encoding-decoding ordering imposed by the network. This explicit cooperation for the exchange of common messages is accomplished through the use of a special binning technique to be explained in Appendix \ref{app:coop_berger}.

Besides the complexity of the achievable  region,  we give an inner bound to the rate-distortion region that allows us to recover the two node Kaspi's region. We also recover several previous inner bounds and rate-distortion regions of some well-known cooperative and interactive  --as well as non-interactive-- lossy source coding problems. 

\subsection*{Special cases}
As the full problem seems to offer a formidable mathematical complexity, including  several special cases which are known to be long-standing open problems, we cannot give a full converse proving the optimality of the general achievable region obtained. However, in Section \ref{sec:new} we provide a complete answer to the rate-distortion regions of several specific  cooperative and interactive source coding problems: 

\begin{enumerate}[(1)]
\item Two encoders and one decoder subject to lossy/lossless reconstruction constraints without side information (see Fig.~\ref{fig:model-Coop-BT}).

\item Two encoders and three decoders subject to lossless/lossy reconstruction constraints with side information (see Fig.~\ref{fig:model-Coop-BT2}).

\item Two encoders and three decoders subject to lossless/lossy reconstruction constraints, reversal delivery and side information (see Fig.~\ref{fig:model-Coop-BT3}).

\item Two encoders and three decoders subject to lossy reconstruction constraints with degraded side information 
(see Fig.~\ref{fig:model-Coop-BT5}).

\item Three encoders and three decoders subject to  lossless/lossy reconstruction constraints with degraded side information (see Fig.~\ref{fig:model-Coop-BT4}).
\end{enumerate}

Interestingly enough, we show that for the two last problems, interaction through multiple rounds could be helpful. Whereas for the other three cases, it is shown that a single round of cooperatively exchanged descriptions suffices to 
 achieve optimality. Table~\ref{table:cases} summarizes the characteristics of each of the above mentioned cases. 

\begin{table}[ht!]
\begin{tabular}{| c | c | c  | c | p{3.5cm}| p{3.5cm} | p{3.5cm}|}
\hline
Cases  & $R_1$ & $R_2$ & $R_3$ & Constraints at Node 1 & Constraints at Node 2 & Constraints at Node 3 \\
 \hline
 (1)   &  $\neq 0$ & $\neq 0$ & $=0$ & \centering $\varnothing$ (is not reconstructing any source)&  \centering $\varnothing$ (is not reconstructing any source) & 
\vspace{-10mm}$$\begin{array}{l}
\mbox{Pr}\left(\hat{X}_{31}^n\neq X_1^n\right)\leq \epsilon\\
  \mathds{E}\left[d(\hat{X}_{32}^n, X_2^n)\right]\leq D_{32}
\end{array}\vspace{-4mm}
$$\\
  \hline
(2)   & $\neq 0$ & $\neq 0$ & $=0$ &  \centering $\mathds{E}\left[d(\hat{X}_{12}^n, X_2^n)\right]\leq D_{12}$ & $\mbox{Pr}\left(\hat{X}_{21}^n\neq X_1^n\right)\leq \epsilon$ & 
\vspace{-10mm}$$\begin{array}{l}
\mbox{Pr}\left(\hat{X}_{31}^n\neq X_1^n\right)\leq \epsilon\\
   \mathds{E}\left[d(\hat{X}_{32}^n, X_2^n)\right]\leq D_{32}
\end{array}\vspace{-4mm}
$$\\
   \hline 
(3)   & $\neq 0$ & $\neq 0$ & $=0$ &  \centering $\mathds{E}\left[d(\hat{X}_{12}^n, X_2^n)\right]\leq D_{12}$ & $\mbox{Pr}\left(\hat{X}_{21}^n\neq X_1^n\right)\leq \epsilon$ & 
\vspace{-10mm}$$\begin{array}{l}
\mbox{Pr}\left(\hat{X}_{32}^n\neq X_2^n\right)\leq \epsilon\\
      \mathds{E}\left[d(\hat{X}_{31}^n, X_1^n)\right]\leq D_{31}
  \end{array}\vspace{-4mm}
$$\\    
      \hline
 (4)   & $\neq 0$ & $\neq 0$ & $=0$ &  \centering $\mathds{E}\left[d(\hat{X}_{12}^n, X_2^n)\right]\leq D_{12}$ & $\mathds{E}\left[d(\hat{X}_{21}^n, X_1^n)\right]\leq D_{21}$ & 
 \vspace{-10mm}$$\begin{array}{l}
 \mathds{E}\left[d(\hat{X}_{31}^n, X_1^n)\right]\leq D_{31}\\
 \mathds{E}\left[d(\hat{X}_{32}^n, X_2^n)\right]\leq D_{32}
  \end{array}\vspace{-4mm}
$$\\  
 \hline
 (5)  & $\neq 0$ & $\neq 0$ & $\neq 0$ & \centering $\varnothing$ (is not decoding)& 
   \vspace{-10mm}$$\begin{array}{l}
  \mbox{Pr}\left(\hat{X}_{21}^n\neq X_1^n\right)\leq \epsilon\\
  \mathds{E}\left[d(\hat{X}_{23}^n, X_3^n)\right]\leq D_{23}
    \end{array}\vspace{-4mm}$$
   & 
 \vspace{-10mm}$$\begin{array}{l}  
  \mbox{Pr}\left(\hat{X}_{31}^n\neq X_1^n\right)\leq \epsilon\\ 
  \mathds{E}\left[d(\hat{X}_{32}^n, X_2^n)\right]\leq D_{32}
    \end{array}\vspace{-4mm}
$$\\  
  \hline
\end{tabular}
\caption{Special cases fully characterized in Section~\ref{sec:new}.}
\label{table:cases}
\end{table}

Next we summarize the contents of the paper. In Section \ref{sec:problem} we formulate the general problem. In Section \ref{sec:inner_bound} we present and discuss the inner bound of the general problem. In Section \ref{sec:known} we show how our inner bound contains several results previously obtained in the past. In Section \ref{sec:new} we present the converse results and their tightness with respect to the inner bound for the special cases mentioned above providing the optimal characterization for them. In Section \ref{sec:discuss} we present a discussion of the obtained results and their limitations and some numerical results concerning the new optimal cases from the previous Section. Finally in Section \ref{sec:summary} we provide some conclusions. The major mathematical details are relegated to the appendixes. 

\subsubsection*{Notation}
We summarize the notation. With ${x^n}$ and upper-case letters $X^n$ we denote vectors  and random vectors of $n$ components, respectively. The $i$-th component of vector $x^n$ is denoted as $x_i$. All alphabets  are assumed to be finite. 

Entropy is denoted by $H(\cdot)$ and mutual information by $I(\cdot;\cdot)$. $H_2(p)$ denotes the entropy associated with a Bernoulli random variable with parameter $p$. With $h(\cdot)$ we denote differential entropy. Let $X$, $Y$ and $V$ be three random variables on some alphabets with probability distribution~$p_{XYV}$. When clear from context we will simple denote $p_{X}(x)$ with $p(x)$. If the probability distribution of random variables $X,Y,V$ satisfies $p(x|yv)=p(x|y)$ for each $x,y,v$, then they form a Markov chain, which is denoted by $X\mkv Y\mkv V$. 

The probability of an event $\mathcal{A}$ is denoted by $\mbox{Pr}\left\{\mathcal{A}\right\}$, where the measure used to compute it will be understood from the context. Conditional probability of a set $\mathcal{A}$ with respect to a set $\mathcal{B}$ is denoted as $\mbox{Pr}\left\{\mathcal{A}\big|\mathcal{B}\right\}$. The set of strong typical sequences associated with random variable $X$ (see appendix \ref{app:strongly}) is denoted by  $\mathcal{T}^n_{[X]\epsilon}$, where $\epsilon>0$. We simply denote these sets as $\mathcal{T}_{\epsilon}^n$ when clear from the context. The cardinal of set $\mathcal{A}$ is denoted by $\|\mathcal{A}\|$. The complement of a set is denoted by $\bar{\mathcal{A}}$. With $\mathbb{Z}_{\geq \alpha}$ and $\mathbb{R}_{\geq \beta}$ we denote the integers and reals numbers greater than $\alpha$ and $\beta$ respectively.  
$\mbox{co}\left\{\mathcal{A}\right\}$ denotes the convex hull of a set $\mathcal{A}\in\mathbb{R}^N$, where $N\in\mathbb{N}$.

\section{Problem formulation}
\label{sec:problem}

Assume three discrete memoryless sources (DMS's) with alphabets and pmfs given by 
$
\big(\mathcal{X}_1\times\mathcal{X}_2\times\mathcal{X}_3, p_{X_1X_2X_3}\big)
$
and arbitrary  bounded distortion measures: $d_j:\mathcal{X}_j\times\mathcal{\hat{X}}_j\rightarrow\mathbb{R}_{\geq 0},\ \ j\in\mathcal{M}\triangleq \{1,2,3\}$ where $\{\hat{\mathcal{X}}_j\}_{j\in\mathcal{M}}$ are finite reconstruction alphabets\footnote{The problem can be easily generalized to the case in which there are different reconstruction alphabets at the terminals. It can also be shown that all the results are valid if we employ arbitrary bounded joint distortion functions, e.g. at node 1 we use $d(X_2,X_3;\hat{X}_2,\hat{X}_3)$.}. We consider the problem of characterizing the rate-distortion region of the interactive source coding scenario described in Fig.~\ref{fig:model}. In this setting, through $K$ rounds of information exchange between the nodes each one of them will attempt to recover a lossy description  of the sources that the others nodes observe, e.g., node 1 must reconstruct --while satisfying distortion constraints-- the realization of the sources $X_2^n$ and $X_3^n$ observed by nodes 2 and 3. Indeed, this setting can be seen as a generalization of the well-known Kaspi's problem \cite{kaspi_two-way_1985}. \vspace{1mm}
 
 \begin{definition}[$K$-step interactive source code]
 A $K$-step interactive $n$-length source code, denoted for the network model in Fig.~\ref{fig:model}, is defined by a sequence of encoder mappings: 
\begin{IEEEeqnarray}{lcl}
f_{1}^l &: &\, \mathcal{X}_1^n\times\left(\mathcal{J}_{2}^1\times\mathcal{J}_{3}^1\times\cdots\times\mathcal{J}_{2}^{l-1}\times\mathcal{J}_{3}^{l-1}\right) \longrightarrow \mathcal{J}_{1}^l \ , \label{eq-def-mappings1}\\
f_{2}^l &: &\,\mathcal{X}_2^n\times\left(\mathcal{J}_{1}^1\times\mathcal{J}_{3}^1\times\cdots\times\mathcal{J}_{3}^{l-1}\times\mathcal{J}_{1}^{l}\right) \longrightarrow \mathcal{J}_{2}^l  \ ,\label{eq-def-mappings2} \\
f_{3}^l &: &\,\mathcal{X}_3^n\times\left(\mathcal{J}_{1}^1\times\mathcal{J}_{2}^1\times\cdots\times\mathcal{J}_{1}^{l}\times\mathcal{J}_{2}^{l}\right) \longrightarrow \mathcal{J}_{3}^l \ , \label{eq-def-mappings3}
\end{IEEEeqnarray}
with $l\in[1:K]$ and message sets: $\mathcal{J}_{i}^l \triangleq \left\{1,2,\dots,\mathcal{I}_{i}^l\right\},\ \mathcal{I}_{i}^l\in\mathbb{Z}_{\geq 0},\ i\in\mathcal{M}$, and reconstruction mappings:
 \begin{equation}
 g_{ij}:\mathcal{X}_i^n\times\bigotimes_{m\in\mathcal{M},\ m\neq i}\left(\mathcal{J}_{m}^1\times\cdots\times\mathcal{J}_{m}^K\right)\longrightarrow\mathcal{\hat{X}}_{ij}^n,\ i\neq j.
 \end{equation} 
The average per-letter distortion and the corresponding distortion levels achieved at the node $i$ with respect to source $j$ are:
 \begin{equation}
 \mathds{E}\left[d_j\left(X_j^n,\hat{X}_{ij}^n\right)\right]\leq D_{ij}\ \ i,j\in\mathcal{M},\ \ i\neq j 
 \end{equation} 
with
 \begin{equation}
 d\left(x^n,y^n\right)\equiv\frac{1}{n}\sum_{m=1}^n d(x_m,y_m)\ .
  \end{equation} 
In compact form we denote a $K$-step interactive source coding by $(n,K,\mathcal{F},\mathcal{G})$ where $\mathcal{F}$ and $\mathcal{G}$ denote the sets of encoders and decoders mappings.
\end{definition}\vspace{1mm}
\begin{remark}
The code definition depends on the node ordering in the encoding procedure. Above we defined the encoding functions $\left\{f_{1}^l,f_{2}^l,f_{3}^l\right\}_{l=1}^K$ assuming that in each round node 1 acts first, followed by node 2, and finally by node 3, and the process beginning again at node 1.
\end{remark}
\begin{definition}[Achievability and rate-distortion region]
Consider $\mathbf{R}\triangleq\left(R_1,R_2,R_3\right)$ and $\mathbf{D}\triangleq (D_{12},D_{13},D_{21},D_{23},D_{31},D_{32})$. The rate vector $\mathbf{R}$ is  $(\mathbf{D},K)$-achievable if $\forall\varepsilon>0$ there is $n_0(\varepsilon,K)\in\mathbb{N}$ such that $\forall n>n_0(\varepsilon,K)$  there exists a $K$-step interactive source code $(n,K,\mathcal{F},\mathcal{G})$ with rates satisfying: 
 \begin{equation}
\frac{1}{n}\sum_{l=1}^K\log{\|\mathcal{J}_{i}^l\|}\leq R_i+\epsilon,\ i\in\mathcal{M}
 \end{equation} 
and with average per-letter distortions at node $i$ with respect to source $j$: 
 \begin{equation}
\mathds{E} \left[d_j(X_j^n,\hat{X}^n_{ij})\right]\leq D_{ij}+\epsilon,\ i,j\in\mathcal{M},\ i\neq j \ ,
 \end{equation} 
where $\hat{X}_{ij}^n\equiv g_{ij}\left(X_{i}^n,\bigotimes_{m\in\mathcal{M},\ m\neq i}\left(\mathcal{J}_{m}^1\times\cdots\times\mathcal{J}_{m}^K\right)\right),\ i\neq j\in\mathcal{M}$.
The rate-distortion region $\mathcal{R}_3(\mathbf{D},K)$ is defined by:
\begin{equation}
\mathcal{R}_3(\mathbf{D},K)=\Big\{\mathbf{R}: \mathbf{R} \ \mbox{is $(\mathbf{D},K)$-achievable}\Big\}
\end{equation}
Similarly, the $\mathbf{D}$-achievable region $\mathcal{R}_3(\mathbf{D})$ is given by $\mathcal{R}_3(\mathbf{D})=\bigcup_{K=1}^\infty\mathcal{R}_3(\mathbf{D},K)$\footnote{Notice that this limit exists because it is the union of a monotone increasing sequence of sets.}, that is:
\begin{equation}
\mathcal{R}_3(\mathbf{D})=\Big\{\mathbf{R}: \mathbf{R} \ \mbox{is $(\mathbf{D},K)$-achievable for some $K\in\mathbb{Z}_{\geq 1}$}\Big\}\ .
\end{equation}
\end{definition}
\vspace{1mm} 
\begin{remark}
By definition $\mathcal{R}_3(\mathbf{D},K)$ is closed and using a time-sharing argument it is easy to show that it is also convex $\forall K\in\mathbb{Z}_{\geq 1}$.
\end{remark}
\vspace{1mm}
\begin{remark}
$\mathcal{R}_3(\mathbf{D},K)$ depends on the node ordering in the encoding procedure. Above we defined the encoding functions $\left\{f_{1}^l,f_{2}^l,f_{3}^l\right\}_{l=1}^K$ assuming that in each round node 1 acts first, followed by node 2, and finally by node 3, and the process beginning again at node 1.  In this paper we restrict the analysis to the \emph{canonical ordering}  ($1\rightarrow 2\rightarrow 3$). However, there are $3!=6$ different orderings that generally  lead to different  regions and the $(\mathbf{D},K)$-achievable region defined above is more explicitly denoted $\mathcal{R}_3(\mathbf{D},K,\sigma_c)$, where $\sigma_c$ is the trivial permutation for $\mathcal{M}$. The correct  $(\mathbf{D},K)$-achievable region is:
\begin{equation}
\label{eq:real_R}
\mathcal{R}_3(\mathbf{D},K)=\bigcup_{\sigma\in\Sigma(\mathcal{M})}\mathcal{R}_3(\mathbf{D},K,\sigma)
\end{equation}
where $\Sigma(\mathcal{M})$ contains all the permutations of set $\mathcal{M}$. The theory presented in this paper for determining $\mathcal R_3(\mathbf D, K, \sigma_c)$ can be used on the other permutations $\sigma \neq \sigma_c$ to compute  (\ref{eq:real_R})\footnote{It should be mentioned that this is not the most general setting of the problem. The most general encoding procedure will follow from the definition of the transmission order by a sequence
$t_1 , t_2 , t_3 , \dots,t_{\|\mathcal{M}\|\times K}$ with $t_i\in\mathcal{M}$. This will cover even the situation in which the order can be changed in each round. To keep the mathematical presentation simpler we will not consider this more general setting. }.
\end{remark}
\vspace{1mm} 

\section{Inner Bound on the Rate-Distortion Region}
\label{sec:inner_bound}

In this Section, we provide a general achievable rate-region on the rate-distortion region.  

\subsection{Inner bound}

We first present a general achievable rate-region where each node at a given round $l$ will generate descriptions destined to the other nodes based on the realization of its own source, the past descriptions generated by a particular node and the descriptions generated at the other nodes and recovered by the node up to the present round.  In order to precisely describe the complex rate-region, we need to introduce some definitions. For a set $\mathcal{A}$, let $\mathcal{C}\left(\mathcal{A}\right)=2^{\mathcal{A}}\setminus\left\{\mathcal{A},\emptyset\right\}$ be the set of all subsets of $\mathcal{A}$ minus $\mathcal{A}$ and the empty set. Denote the auxiliary random variables:\vspace{-1mm}
\begin{equation}
U_{i\rightarrow S, l}, \ \ S\in\mathcal{C}\left(\mathcal{M}\right),\  i\notin S, \ l=1,\dots, K.\vspace{-1mm}
\label{eq:auxiliary_1}
\end{equation}
Auxiliary random variables $\{U_{i\rightarrow S, l}\}$ will be used to denote the descriptions generated at node $i$ and at round $l$ and destined to a set of nodes $S\in\mathcal{C}\left(\mathcal{M}\right)$ with  $i\notin S$. For example, $U_{1\rightarrow 23,l}$ denote the description generated at node 1 and at round $l$ and destined to nodes 2 and 3. Similarly, $\{U_{1\rightarrow 2,l}\}$ will be used to denote the descriptions generated at node 1 at round $l$ and destined only to node 2. We define  variables:
\begin{description}
\item{\hspace{-13mm}$\mathcal{W}_{[i,l]}$} $\equiv$ Common information\footnote{Not to be confused with the Wyner's definition of \emph{common information} \cite{common_wyner}.} shared by the three nodes available at node $i$ at round $l$ before encoding 
\item{\hspace{-13mm}$\mathcal{V}_{[S,l,i]}$} $\equiv$ Private information shared by nodes in $S  \in  \mathcal{C} \left( \mathcal{M} \right)$ available at node $i \in  S$, at round $l$, before encoding
\end{description}
In precise terms, the quantities introduced above for our problem are defined by: 
\begin{IEEEeqnarray*}{rcl}
\mathcal{W}_{[1,l]} & =& \left\{U_{1\rightarrow 23,k},U_{2\rightarrow 13,k},U_{3\rightarrow 12,k}\right\}_{k=1}^{l-1} \ , \\
\mathcal{W}_{[2,l]}&=& \mathcal{W}_{[1,l]}\cup U_{1\rightarrow 23,l}\ ,  \\
\mathcal{W}_{[3,l]}&=&\mathcal{W}_{[2,l]}\cup U_{2\rightarrow 13,l}\ ,\\
\mathcal{V}_{[12,l,1]} &=& \left\{U_{1\rightarrow 2,k},U_{2\rightarrow 1,k}\right\}_{k=1}^{l-1},\ \mathcal{V}_{[12,l,2]}=\mathcal{V}_{[12,l,1]}\cup U_{1\rightarrow 2,l}\ ,\\
\mathcal{V}_{[13,l,1]}& =& \left\{U_{1\rightarrow 3,k},U_{3\rightarrow 1,k}\right\}_{k=1}^{l-1},\ \mathcal{V}_{[13,l,3]}=\mathcal{V}_{[13,l,1]}\cup U_{1\rightarrow 3,l}\ ,\\
\mathcal{V}_{[23,l,2]}& = &\left\{U_{2\rightarrow 3,k},U_{3\rightarrow 2,k}\right\}_{k=1}^{l-1}, \ \mathcal{V}_{[23,l,3]}=\mathcal{V}_{[23,l,2]}\cup U_{2\rightarrow 3,l}\ .
\end{IEEEeqnarray*}

Before presenting the general inner bound, we provide the basic idea of the \emph{random coding} scheme that achieves the rate-region in Theorem~\ref{theo-main-theorem} for the case of $K$ communication rounds. 

Assume that all codebooks are randomly generated and known to all the nodes before the information exchange begins and consider the encoding ordering given by $1\rightarrow 2\rightarrow 3$ so that we begin at round $l=1$ in node $1$. Also, and in order to maintain the explanation simple and to help the reader to grasp the essentials of the coding scheme employed, we will consider that all terminal are able to recover the descriptions generated at other nodes (which will be the case under the conditions in our Theorem \ref{theo-main-theorem}).  From the observation of the source $X_1^n$, node $1$ generates a set of descriptions for each of the other nodes connected to it.  In particular it generates a common description to be recovered at nodes  2 and 3 in addition to  two private descriptions for node 2 and 3, respectively, generated from a conditional codebook given the common description. Then, node $2$ tries to recover the descriptions destined to it (the common description generated at 1 and its corresponding private description), using $X_2^n$ as side information, and generates its own descriptions, based on source $X_2^n$ and the recovered descriptions from node $1$. Again, it generates a common description for nodes 1 and 3, a private description for node 3 and another one for node 1. The same process goes on until node $3$, which tries to recover jointly the common descriptions generated by node $1$ and node $2$, and then the private descriptions destined to him by node 1 and 2.  Then generates its own descriptions (common and private ones) destined to nodes $1$ and $2$. Finally, node $1$ tries to recover all the descriptions destined to it generated by nodes $2$ and $3$ in the same way as previously done by node 3. After this, round $l=1$ is over, and round $l=2$ begins with node $1$ generating new descriptions using $X_1^n$, its encoding history (from previous round) and the recovered descriptions from the other nodes. 

The process continues in a similar manner until we reach round $l=K$ where node $3$ recovers the descriptions from the other nodes and generates its own ones. Node $1$ recovers the last descriptions destined to it from nodes $2$ and $3$ but does not generate new ones. The same holds for node $2$ who only recovers the descriptions generated by node $3$ and thus terminating the information exchange procedure. Notice that at the end of round $K$ the decoding in node 1 and node 2 can be done simultaneously. This is due to the fact that node 1 is not generating a new description destined to node 2. However, in order to simplify the analysis and notation in the appendix we will consider that the last decoding of node 2 occurs in round $K+1$\footnote{This is clearly a fictitious round, in the sense that there is not descriptions generation on it. In this way, there is not modification of the final rates achieved by the procedure described if we consider this additional round.}. After all the exchanges are done, each node recovers an estimate of the other nodes, source realizations by using all the available recovered descriptions from the $K$ previous rounds.

\begin{theorem}[Inner bound]\label{theo-main-theorem}
Let $\bar{\mathcal{R}}_3(\mathbf{D},K)$ be the closure of set of all rate tuples satisfying:
\begin{IEEEeqnarray}{rCl}
R_1 &=& \sum_{l=1}^K \big(R_{1\rightarrow 23}^{(l)}+ R_{1\rightarrow 2}^{(l)}+ R_{1\rightarrow 3}^{(l)}\big)\\
R_2 &=& \sum_{l=1}^K \big(R_{2\rightarrow 13}^{(l)}+ R_{2\rightarrow 1}^{(l)}+  R_{2\rightarrow 3}^{(l)}\big)\\
R_3 &=& \sum_{l=1}^K \big(R_{3\rightarrow 12}^{(l)}+ R_{3\rightarrow 1}^{(l)}+ R_{3\rightarrow 2}^{(l)}\big)\\
R_1+R_2 &=& \sum_{l=1}^K \big(R_{1\rightarrow 23}^{(l)}+R_{2\rightarrow 13}^{(l)}+ R_{1\rightarrow 3}^{(l)}+ R_{2\rightarrow 3}^{(l)}+ R_{1\rightarrow 2}^{(l)}+R_{2\rightarrow 1}^{(l)}\big) \\
R_1+R_3 &=&  \sum_{l=1}^{K+1} \big(R_{1\rightarrow 23}^{(l)}+R_{3\rightarrow 12}^{(l-1)} + R_{1\rightarrow 2}^{(l)}+R_{3\rightarrow 2}^{(l-1)}+ R_{1\rightarrow 3}^{(l)} + R_{3\rightarrow 1}^{(l)} \big) \\
R_2+R_3& =&  \sum_{l=1}^K \big(R_{2\rightarrow 13}^{(l)}+R_{3\rightarrow 12}^{(l)}+ R_{2\rightarrow 1}^{(l)}+R_{3\rightarrow 1}^{(l)}+  R_{2\rightarrow 3}^{(l)}+ R_{3\rightarrow 2}^{(l)} \big)
\end{IEEEeqnarray}\hspace{-2mm}
where\footnote{Notice that these definitions are motivated by the fact that at round 1, node 2 only recovers the descriptions generated by node 1 and at round $K+1$ only recovers what node 3 already generated at round $K$.} for each $l\in[1:K]$:
\begin{IEEEeqnarray}{rCl}
R_{1\rightarrow 23}^{(l)} &> & I\left(X_1;U_{1\rightarrow 23,l}\Big| X_2\mathcal{W}_{[1,l]}\mathcal{V}_{[12,l,1]}\mathcal{V}_{[23,l-1,3]}\right)\vspace{1mm} \\
R_{2\rightarrow 13}^{(l)} &> & I\left(X_2;U_{2\rightarrow 13,l}\Big| X_3\mathcal{W}_{[2,l]}\mathcal{V}_{[13,l,1]}\mathcal{V}_{[23,l,2]}\right)\vspace{1mm} \\
R_{3\rightarrow 12}^{(l)} &> &  I\left(X_3;U_{3\rightarrow 12,l}\Big| X_1\mathcal{W}_{[3,l]}\mathcal{V}_{[12,l,2]}\mathcal{V}_{[13,l,3]}\right)\vspace{1mm} \\
R_{1\rightarrow 23}^{(l)} +R_{2\rightarrow 13}^{(l)} &>& I\left(X_1X_2;U_{1\rightarrow 23,l}U_{2\rightarrow 13,l}\Big|X_3\mathcal{W}_{[1,l]}\mathcal{V}_{[13,l,1]}\mathcal{V}_{[23,l,2]}\right) \vspace{1mm}\\
R_{2\rightarrow 13}^{(l)} +R_{3\rightarrow 12}^{(l)} &> &I\left(X_2X_3;U_{2\rightarrow 13,l}U_{3\rightarrow 12,l}\Big|X_1\mathcal{W}_{[2,l]}\mathcal{V}_{[12,l,2]}\mathcal{V}_{[13,l,3]}\right) \vspace{1mm}\\
R_{1\rightarrow 23}^{(l)} +R_{3\rightarrow 12}^{(l-1)} &>& I\left(X_1X_3;U_{1\rightarrow 23,l}U_{3\rightarrow 12,l-1}\Big|X_2\mathcal{W}_{[3,l-1]}\mathcal{V}_{[12,l,1]}\mathcal{V}_{[23,l-1,3]}\right) \vspace{1mm}\\
  R_{3\rightarrow 2}^{(l-1)}&>& I\left(X_3;U_{3\rightarrow 2,l-1}\Big|X_2\mathcal{W}_{[2,l]}\mathcal{V}_{[23,l-1,3]}\mathcal{V}_{[12,l,2]}\right)\vspace{1mm} \\%
R_{1\rightarrow 2}^{(l)} &>& I\left(X_1;U_{1\rightarrow2,l}\Big|X_2\mathcal{W}_{[2,l]}\mathcal{V}_{[23,l,2]}\mathcal{V}_{[12,l,1]}\right)\vspace{1mm}  \\
 R_{1\rightarrow 2}^{(l)}+R_{3\rightarrow 2}^{(l-1)} &>& I\left(X_1X_3;U_{1\rightarrow 2,l}U_{3\rightarrow 12,l-1}\Big|X_2\mathcal{W}_{[2,l]}\mathcal{V}_{[23,l-1,3]}\mathcal{V}_{[12,l,1]}\right)\vspace{1mm} \\
    R_{1\rightarrow 3}^{(l)} &>& I\left(X_1;U_{1\rightarrow 3,l}\Big|X_3\mathcal{W}_{[3,l]}\mathcal{V}_{[23,l,3]}\mathcal{V}_{[13,l,1]}\right) \vspace{1mm} \\
   R_{2\rightarrow 3}^{(l)}&>& I\left(X_2;U_{2\rightarrow 3,l}\Big|X_3\mathcal{W}_{[3,l]}\mathcal{V}_{[23,l,2]}\mathcal{V}_{[13,l,3]}\right) \vspace{1mm} \\
   R_{1\rightarrow 3}^{(l)} + R_{2\rightarrow 3}^{(l)} &>& I\left(X_1X_2;U_{1\rightarrow 3,l}U_{2\rightarrow 3,l}\Big|X_3\mathcal{W}_{[3,l]}\mathcal{V}_{[23,l,2]}\mathcal{V}_{[13,l,1]}\right) \vspace{1mm} \\
   R_{2\rightarrow 1}^{(l)} &>& I\left(X_2;U_{2\rightarrow 1,l}\Big|X_1\mathcal{W}_{[1,l+1]}\mathcal{V}_{[12,l,2]}\mathcal{V}_{[13,l+1,1]}\right)\vspace{1mm} \\
     R_{3\rightarrow 1}^{(l)} &>& I\left(X_3;U_{3\rightarrow 1,l}\Big|X_1\mathcal{W}_{[1,l+1]}\mathcal{V}_{[12,l+1,1]}\mathcal{V}_{[13,l,3]}\right)\vspace{1mm} \\
     R_{2\rightarrow 1}^{(l)} + R_{3\rightarrow 1}^{(l)} &>&  I\left(X_2X_3;U_{2\rightarrow 1,l}U_{3\rightarrow 1,l}\Big|X_1\mathcal{W}_{[1,l+1]}\mathcal{V}_{[12,l,2]}\mathcal{V}_{[13,l,3]}\right) 
     \label{eq:rates_theorem}
    \end{IEEEeqnarray}
with $R_{i\rightarrow S}^{(0)}=R_{i\rightarrow S}^{(K+1)}=0$ and $U_{i\rightarrow S,0}=U_{i\rightarrow S,K+1}=\varnothing$ for $S\in\mathcal{C}\left(\mathcal{M}\right)$ and $i\notin S$. 

With these definitions the rate-distortion region satisfies\footnote{It is straightforward to show that the LHS of equation (\ref{eq-inner-bound}) is convex, which implies that the convex hull operation is not needed. }:
\begin{equation}
 \bigcup_{p\in\mathcal{P}\left(\mathbf{D},K\right)}\bar{\mathcal{R}}_3(\mathbf{D},K)\subseteq \mathcal{R}_3(\mathbf{D},K)\ ,\label{eq-inner-bound}
\end{equation}
where $\mathcal{P}(\mathbf{D},K)$ denotes the set of all joint probability measures associated with the following Markov chains for every $l\in[1:K]$: \vspace{1mm} 

\begin{enumerate} 
\item $U_{1\rightarrow 23,l}\mkv (X_1,\mathcal{W}_{[1,l]})\mkv (X_2,X_3,\mathcal{V}_{[12,l,1]},\mathcal{V}_{[13,l,1]},\mathcal{V}_{[23,l,2]})$\ ,\vspace{1mm} 
\item $U_{1\rightarrow 2,l}\mkv (X_1,\mathcal{W}_{[2,l]},\mathcal{V}_{[12,l,1]})\mkv (X_2,X_3,\mathcal{V}_{[13,l,1]},\mathcal{V}_{[23,l,2]})$\ ,\vspace{1mm} 
\item $U_{1\rightarrow 3,l}\mkv (X_1,\mathcal{W}_{[2,l]},\mathcal{V}_{[13,l,1]})\mkv (X_2,X_3,\mathcal{V}_{[12,l,2]},\mathcal{V}_{[23,l,2]})$\ ,\vspace{1mm} 
\item $U_{2\rightarrow 13,l}\mkv (X_2,\mathcal{W}_{[2,l]})\mkv (X_1,X_3,\mathcal{V}_{[12,l,2]},\mathcal{V}_{[13,l,3]},\mathcal{V}_{[23,l,2]})$\ ,\vspace{1mm} 
\item $U_{2\rightarrow 1,l}\mkv (X_2,\mathcal{W}_{[3,l]},\mathcal{V}_{[12,l,2]})\mkv (X_1,X_3,\mathcal{V}_{[13,l,3]},\mathcal{V}_{[23,l,2]})$\ ,\vspace{1mm} 
\item $U_{2\rightarrow 3,l}\mkv (X_2,\mathcal{W}_{[3,l]},\mathcal{V}_{[23,l,2]})\mkv (X_1,X_3,\mathcal{V}_{[12,l+1,1]},\mathcal{V}_{[13,l,3]})$\ ,\vspace{1mm} 
\item $U_{3\rightarrow 12,l} \mkv (X_3,\mathcal{W}_{[3,l]})\mkv(X_1,X_2,\mathcal{V}_{[12,l+1,1]},\mathcal{V}_{[13,l,3]},\mathcal{V}_{[23,l,3]})$\ ,\vspace{1mm} 
\item $U_{3\rightarrow 1,l} \mkv (X_3,\mathcal{W}_{[1,l+1]},\mathcal{V}_{[13,l,3]}) \mkv (X_1,X_2,\mathcal{V}_{[12,l+1,1]},\mathcal{V}_{[23,l,3]})$\ ,\vspace{1mm} 
\item $U_{3\rightarrow 2,l} \mkv (X_3,\mathcal{W}_{[1,l+1]},\mathcal{V}_{[23,l,3]}) \mkv 1 (X_1,X_2,\mathcal{V}_{[12,l+1,1]},\mathcal{V}_{[13,l+1,1]})$\ ,\vspace{1mm} 
\end{enumerate}
and such that there exist reconstruction mappings:
\begin{equation}
g_{ij} \left( X_i,\mathcal{V}_{[ij,K+1,1]},\!\mathcal{W}_{[1,K+1]} \right)=\hat{X}_{ij}\vspace{-2mm}
\end{equation}
with $\mathds{E} \left[d_j(X_j,\hat{X}_{ij}) \right] \leq  D_{ij}$ for each $i,j\in\mathcal{M}$ and $i\neq j$. 
\end{theorem}

The proof of this theorem is relegated to Appendix~\ref{app:proof_theo_1} and relies on the auxiliary results presented in Appendix~\ref{app:strongly} and the theorem on the cooperative Berger-Tung problem with side information presented in Appendix~\ref{app:coop_berger}. 
\vspace{1mm}
\begin{remark}
It is worth mentioning here that  our coding scheme is constrained to use \emph{successive decoding}, i.e., by recovering first the coding layer of  common descriptions and then coding layer of private descriptions (at each coding layer each node employ \emph{joint-decoding}). Obviously, this is a sub-optimum procedure since the best scheme would be to use \emph{joint decoding} where both common and private informations can be jointly recovered. However, the analysis of this scheme is much more involved. The associated achievable rate region involves a large number of equations that combine rates belonging to private and common messages from different nodes. Also, several mutual information terms in each of these rate equations cannot be combined, leading to a proliferation of many equations that offer little insight to the problem.
\end{remark}
\vspace{1mm}
\begin{remark}
 The idea behind our derivation of the achievable region can be extended to any number $M$ ($>3$) of nodes in the network.  This can be accomplished by generating a greater number of superimposed coding layers. First a layer of codes that generates descriptions destined to be decoded by all nodes. The next layer corresponding to all subsets of size $M-1$, etc, until we reach the final layer composed by codes that generate private descriptions for each of nodes. Again, \emph{successive decoding} is used at the nodes to recover the descriptions in these layers destined to them. Of course, the number of required descriptions will increase with the number of nodes as well as the obtained  rate-distortion region. 
 \end{remark}
 \vspace{1mm}
\begin{remark}
It is interesting to compare the main ideas of our scheme with those of Kaspi \cite{kaspi_two-way_1985}. The main idea in \cite{kaspi_two-way_1985} is to have a single coding \emph{tree} shared by the two nodes. Each \emph{leaf} in the coding tree is codeword generated either at node 1 or 2. At a given round each node knows (assuming no errors at the encoding and decoding procedures) the path followed in the tree. For example, at round $l$, node 1, using the knowledge of the path until round $l$ and its source realization generate a leaf (from a set of possible ones) using joint typicality encoding and binning. Node 2, using the same path known at node 1 and its source realization, uses joint typicality decoding to estimate the leaf generate at node 1. If there is no error at these encoding and decoding steps, the previous path is updated with the new leaf and both -node 1 and 2- know the updated path. Node 2 repeats the procedure. This is done until round $K$ where the final path is known at both nodes and used to reconstruct the desired sources.

In the case of three nodes the situation is more involved. At a given round, the encoder at an arbitrary node is seeing two decoders with different side information\footnote{Because at each node the source realizations are different, and the recovered previous descriptions can also be different.}. In order to simplify the explanation consider that we are at round $l$ in the encoder 1, and that the listening nodes are nodes 2 and 3. This situation forces node 1 to encode two sets of descriptions: one common for the other two nodes and a set of private ones  associated with each of the listening nodes 2 and 3. Following the ideas of Kaspi, it is then natural to consider three different coding trees followed by node 1. One coding tree has leaves that are the common descriptions generated and shared by all the nodes in the network. The second tree is composed by leaves that are the private descriptions generated and shared with node 2. The third tree is composed by leaves that are the private descriptions generated and shared with node 3. As the private descriptions \emph{refine} the common ones, depending on the quality of the side information of the node that is the intended recipient, it is clear that descriptions are correlated. For example, the private description destined to node 2, should depend not only on the past private descriptions generated and shared by nodes 1 and 2, but also on the common descriptions generated at all previous rounds in all the nodes and on the common description generated at the present round in node 1. Something similar happens for the private description destined to node 3.  It is clear that as the common descriptions are to be recovered by all the nodes in the network, they can only be conditioned with respect to the past common descriptions generated at previous rounds and with respect to the common descriptions generated at the present round by a node who acted before (i.e. at round $l$ node 1 acts before node 2). The private descriptions, as they are only required to be recovered at some set of nodes, can be generated conditioned on the past exchanged common descriptions and the past private descriptions generated and recovered in the corresponding set of nodes (i.e., the private descriptions exchanged between nodes 1 and 2 at round $l$, can only be generated conditioned on the past common descriptions generated at nodes 1, 2 and 3 and on the past private descriptions exchanged only between 1 and 2). 

We can see clearly that there are basically four paths to be cooperatively followed in the network:
\begin{itemize}
\item One path of common descriptions shared by nodes 1, 2 and 3.
\item One path of private descriptions shared by nodes 1 and 2.
\item One path of private descriptions shared by nodes 1 and 3.
\item One path of private descriptions shared by nodes 2 and 3.
\end{itemize} 
It is also clear that each node only follows three of these paths simultaneously. The exchange of common descriptions deserves special mention. Consider the case at round $l$ in node 3. This node needs to recover the common descriptions generated at nodes 1 and 2. But at the moment node 2 generated its own common description, it also recovered the common one generated at node 1. This allows for a natural explicit cooperation between nodes 1 and 2 in order to help node 3 to recover both descriptions. Clearly, this is not the case for private descriptions from nodes 1 and 2 to be recovered at node 3. Node 2 does not recover the private description from node 1 to 3 and cannot generate an explicit collaboration to help node 3 to recover both private descriptions. Note, however, that as both private descriptions will be dependent on previous common descriptions an implicit collaboration (intrinsic to the code generation) is also in force. In appendix \ref{app:coop_berger} we consider the problem (not in the interactive setting) of generating the explicit cooperation for the common descriptions through the use of what we call a \emph{super-binning} procedure, in order to use the results for our interactive three-node problem. 

\end{remark}
 
%
%
\section{Known Cases and Related Work}
\label{sec:known}

Several inner bounds and rate-distortion regions on multiterminal source coding problems can be derived by specializing the inner bound~\eqref{eq-inner-bound}. Below we summarize only a few of them.
 
 \vspace{1mm}

\subsubsection{Distributed source coding with side information~\cite{PhD-Tung,gastpar_wyner-ziv_2004}}

Consider the distributed source coding problem where two nodes encode separately  sources $X_1^n$ and $X_2^n$ to rates $(R_1,R_2)$ and  a decoder by using side information $X_3^n$ must reconstruct both sources with average distortion less than $D_1$ and $D_2$, respectively. By considering only one-round/one-way information exchange from nodes $1$ and $2$ (the encoders) to node $3$ (the decoder), the results in~\cite{PhD-Tung, gastpar_wyner-ziv_2004} can be recovered as a special case of the inner bound~\eqref{eq-inner-bound}. Specifically, we set: 
\begin{IEEEeqnarray*}{rcl}
U_{1\rightarrow 23,l}&=&U_{2\rightarrow 13,l}=U_{3\rightarrow 12,l}=U_{1\rightarrow 2,l}=U_{2\rightarrow 1,l}=U_{3\rightarrow 1,l}=U_{3\rightarrow 2,l}=\varnothing, \,\, \forall l \\
U_{1\rightarrow 3,l}&=&U_{2\rightarrow 3,l}=\varnothing, \,\,\forall l>1 \ .
\end{IEEEeqnarray*}
In this case, the Markov chains of Theorem \ref{theo-main-theorem} reduce to: 
\begin{IEEEeqnarray}{rcl}
 U_{1\rightarrow 3,1} &\mkv&  X_1\mkv (X_2,X_3,U_{2\rightarrow 3,1}) \ ,\\
U_{2\rightarrow 3,1} & \mkv & X_2\mkv (X_1,X_3,U_{1\rightarrow 3,1}) \ ,
\end{IEEEeqnarray}
and thus the inner bound from Theorem \ref{theo-main-theorem}  recovers the results in~\cite{gastpar_wyner-ziv_2004}
\begin{IEEEeqnarray}{rcl}
R_1 &>& I(X_1;U_{1\rightarrow 3,1}|X_3U_{2\rightarrow 3,1})\ ,\\
R_2 &>& I(X_2;U_{2\rightarrow 3,1}|X_3U_{1\rightarrow 3,1})\ , \\
R_1+R_2 &>& I(X_1X_2;U_{1\rightarrow 3,1}U_{2\rightarrow 3,1}|X_3)\ .
\end{IEEEeqnarray}

\vspace{1mm}

\subsubsection{Source coding with side information at $2$-decoders~\cite{heegard_rate_1985,timo_rate_2011}} 

Consider the setting where one encoder $X_1^n$ transmits descriptions to two decoders with different side informations $(X_2^n,X_3^n)$ and distortion requirements $D_2$ and $D_3$. Again we consider only one way/round information exchange from node $1$ (the encoder) to nodes $2$ and $3$ (the decoders). 

In this case, we set: 
\begin{IEEEeqnarray*}{rcl}
U_{2\rightarrow 13,l}&=&U_{3\rightarrow 12,l}=U_{2\rightarrow 1,l}=U_{3\rightarrow 1,l}=U_{3\rightarrow 2,l}=U_{2\rightarrow 3,l}=\varnothing, \,\,\forall l \\
U_{1\rightarrow 23,l}&=&U_{1\rightarrow 23,l}=U_{1\rightarrow 2,l}=U_{1\rightarrow 3,l}=\varnothing,  \,\,\forall l>1\ .
\end{IEEEeqnarray*}
The above Markov chains imply 
\begin{IEEEeqnarray}{rcl}
(U_{1\rightarrow 23,1},U_{1\rightarrow 2,1},U_{1\rightarrow 3,1})\mkv X_1\mkv (X_2,X_3)
\end{IEEEeqnarray}
and thus the inner bound from Theorem \ref{theo-main-theorem} reduces to the results in~\cite{heegard_rate_1985,timo_rate_2011}
\begin{IEEEeqnarray}{rcl}
R_1 &>&\max{\big\{I(X_1;U_{1\rightarrow 23,1}|X_2)\,,\,I(X_1;U_{1\rightarrow 23,1}|X_3)\big\}}\nonumber\\
&+& I(X_1;U_{1\rightarrow 2,1}|X_2U_{1\rightarrow 23,1})+ I(X_1;U_{1\rightarrow 3,1}|X_3U_{1\rightarrow 23,1})\ .
\end{IEEEeqnarray}

\vspace{1mm}

\subsubsection{Two terminal interactive source coding~\cite{kaspi_two-way_1985}} 

Our inner bound~\eqref{eq-inner-bound} is basically the generalization of the two terminal problem to the three-terminal setting. Assume only two encoders-decoders $X_1^n$ and $X_2^n$ which must reconstruct the other terminal source $3$ with distortion constraints $D_1$ and $D_2$, and after $K$ rounds of information exchange. Let us set:  
\begin{IEEEeqnarray*}{rcl}
U_{1\rightarrow 23,l} &=& U_{2\rightarrow 13,l}=U_{3\rightarrow 12,l}=U_{1\rightarrow 3,l}=U_{3\rightarrow 1,l}=U_{2\rightarrow 3,l}=U_{3\rightarrow 2,l}=\varnothing,\,\, \forall l\\
X_3&=&\varnothing\ . 
\end{IEEEeqnarray*}
The Markov chains become 
\begin{IEEEeqnarray}{rcl}
U_{1\rightarrow 2,l}\mkv (X_1,\mathcal{V}_{[12,l,1]})\mkv X_2 \, ,\\
U_{2\rightarrow 1,l}\mkv (X_2,\mathcal{V}_{[12,l,2]})\mkv X_2\, ,
\end{IEEEeqnarray}
for $l\in[1:K]$ and thus the inner bound from Theorem \ref{theo-main-theorem} permit us to obtain  the results in~\cite{kaspi_two-way_1985}
\begin{IEEEeqnarray}{rcl}
R_1 &>& I(X_1;\mathcal{V}_{[12,K+1,1]}|X_2)\ ,\\ 
R_2 &>& I(X_2;\mathcal{V}_{[12,K+1,2]}|X_1)\ .
\end{IEEEeqnarray}

\vspace{1mm}

\subsubsection{Two terminal interactive source coding with a helper~\cite{permuter_two-way_2010}}

Consider now two encoders/decoders, namely $X_2^n$ and $X_3^n$, that must reconstruct the other terminal source with distortion constraints $D_2$ and $D_3$, respectively, using $K$ communication rounds. Assume also that  another encoder $X_1^n$ provides both nodes $(2,3)$ with a common description before beginning the information exchange and then remains silent. Such common description can be exploited as coded side information. Let us set:
\begin{IEEEeqnarray*}{rcl}
U_{2\rightarrow 13,l} &=& U_{3\rightarrow 12,l}=U_{1\rightarrow 3,l}
=U_{1\rightarrow 2,l}=U_{1\rightarrow 3,l}=U_{2\rightarrow 1,l}=U_{3\rightarrow 1,l}=\varnothing,\,\, \forall l \\
U_{1\rightarrow 23,l}&=&\varnothing,\,\,  \forall l>1\ .
\end{IEEEeqnarray*}
The Markov chains reduce to:
\begin{IEEEeqnarray}{lcl}
U_{1\rightarrow 23,1} &\mkv& X_1\mkv (X_2,X_3)\ , \\
U_{2\rightarrow 3,l} &\mkv& (X_2,U_{1\rightarrow 23,1},\mathcal{V}_{[23,l,2]})\mkv (X_1,X_3)\ ,\\
U_{3\rightarrow 2,l} &\mkv& (X_3,U_{1\rightarrow 23,1},\mathcal{V}_{[23,l,3]})\mkv (X_1,X_2)\ .
\end{IEEEeqnarray}
An inner bound to the rate-distortion region for this problem reduces to (using the rate equations in our Theorem \ref{theo-main-theorem})
\begin{IEEEeqnarray}{lcl}
R_1&>& \max{\big\{I(X_1;U_{1\rightarrow 23,1}|X_2),I(X_1;U_{1\rightarrow 23,1}|X_3)\big\}}\ ,\\
R_2&>& I(X_2;\mathcal{V}_{[23,K+1,2]}|X_3U_{1\rightarrow 23,1})\ ,\\
R_3&>& I(X_3;\mathcal{V}_{[23,K+1,2]}|X_2U_{1\rightarrow 23,1})\ .
\end{IEEEeqnarray}
This region contains as a special case the region  in~\cite{permuter_two-way_2010}. In that paper it is further assumed (in order to have a converse result) that $X_1\mkv X_3\mkv X_2$. Then, the value of $R_1$ satisfies $R_1>I(X_1;U_{1\rightarrow 23,1}|X_2)$. Obviously, with the same extra Markov chain we obtain the same limiting value for $R_1$ and the above region is the rate-distortion region.


\section{New Results on Interactive and Cooperative Source Coding}
\label{sec:new}

\subsection{Two encoders and one decoder subject to lossy/lossless reconstruction constraints without side information} 
 
\begin{figure}[th]
\centering
\ifpdf\includegraphics[angle=0,width=0.75\columnwidth,keepaspectratio,trim= 0mm 0mm 0mm 0mm,clip]{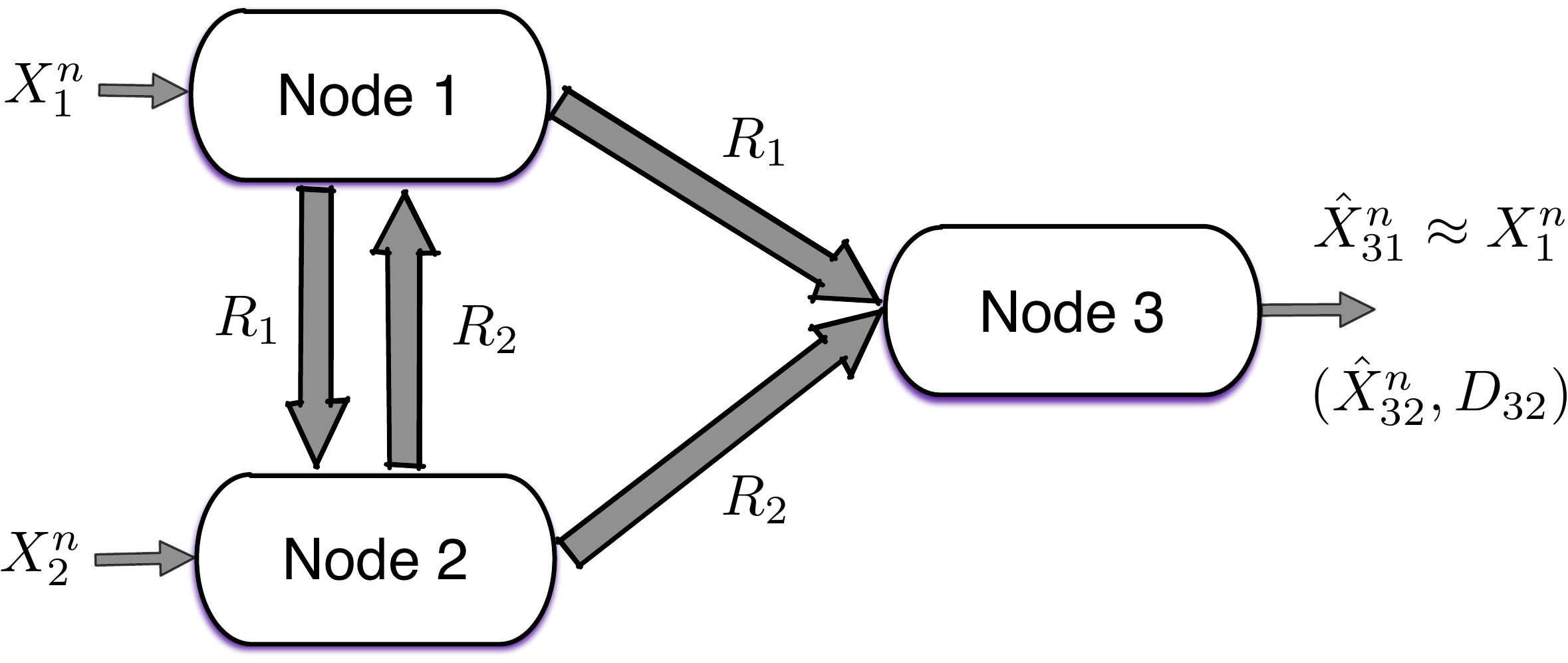} 
\else
\includegraphics[angle=270,width=0.75\columnwidth,keepaspectratio,trim= 0mm 0mm 0mm 0mm,clip]{model.eps} 
\fi
\caption{Two encoders and one decoder subject to lossy/lossless reconstruction constraints without side information.}

\label{fig:model-Coop-BT}
\end{figure}

Consider now the problem described in Fig.~\ref{fig:model-Coop-BT} where encoder $1$ wishes to 
 communicate the source $X_1^n$ to node $3$ in a \emph{lossless} manner while encoder $2$ wishes to send a \emph{lossy} description of the source $X_2^n$ to node $3$ with distortion constraint $D_{31}$. To achieve this, the encoders use $K$ communication rounds.  This problem can be seen as the cooperating encoders version of  the well-known  Berger-Yeung~\cite{32119} problem. 

\begin{theorem}\label{theo-Coop-BT}
The rate-distortion region of the setting described in Fig.~\ref{fig:model-Coop-HB} is given by the union over all joint probability measures $p_{X_1X_2U_{2\rightarrow 13}}$ such that there exists a reconstruction mapping:
\begin{equation}
g_{32} \left( X_1, U_{2\rightarrow 13} \right)=\hat{X}_{32}\,\,\,\, \textrm{ with }\,\,\,\,\mathds{E} \left[d(X_2,\hat{X}_{32}) \right] \leq  D_{32}\ ,
\end{equation}
of the set of all tuples satisfying: 
\begin{IEEEeqnarray}{rCl}
R_1 &\geq& H(X_1|X_2) \label{eq-BT-rate-1}\ ,\\
R_2 &\geq& I(X_2; U_{2\rightarrow 13} |X_1)\ ,\\
R_1+R_2 &\geq& H(X_1) + I(X_2; U_{2\rightarrow 13} |X_1)\ .
\end{IEEEeqnarray}
The auxiliary random variable $ U_{2\rightarrow 13}$ has a cardinality bound of $\| \mathcal{U}_{2\rightarrow 13}\|\leq \|\mathcal{X}_1\|\|\mathcal{X}_2\|+1$.
\end{theorem}
\vspace{1mm}

\begin{remark}
It is worth emphasizing that the rate-distortion region in Theorem~\ref{theo-Coop-BT} outperforms the non-cooperative rate-distortion region first derived in~\cite{32119}. This is due to two facts: the conditional entropy given in the rate constraint~\eqref{eq-BT-rate-1} which is strictly smaller than the entropy $H(X_1)$ present in the rate-region in~\cite{32119}, and the fact that the random description $U_{2\rightarrow 13}$ may be arbitrarily dependent on both sources $(X_1,X_2)$ which is not the case without cooperation~\cite{32119}. Therefore, cooperation between encoders $1$ and $2$ reduces the rate needed to communicate the source $X_1$ while increasing the optimization set of all admissible source descriptions. 
\end{remark}
\vspace{1mm}
\begin{remark}
Notice that the rate-distortion region in Theorem~\ref{theo-Coop-BT} is achievable with a single round  of interactions $K=1$, which implies that multiple rounds do not improve the rate-distortion region in this case. This holds because of the fact that node 3 reconstruct $X_1$ in a lossless fashion.
\end{remark}
\vspace{1mm}
\begin{remark}
Although in the considered setting of Fig.~\ref{fig:model-Coop-HB} node $1$ is not supposed to decode neither a lossy description nor the complete  source $X_2^n$, if nodes $1$ and $3$ wish to  recover the same descriptions the optimal rate-region remains  the same as given in Theorem~\ref{theo-Coop-BT}.  The only difference relies on the fact that node $1$ is now able to find a function $g_{12} \left( X_1, U_{2\rightarrow 13} \right)=\hat{X}_{12}$ which  must satisfy an additional  distortion constraint $\mathds{E} \left[d(X_2,\hat{X}_{12}) \right] \leq  D_{12}$. In order to show this, it is enough to check that in the converse proof given below the specific choice of the auxiliary random variable already allows node $1$ to recover a general function $\hat{X}_{12[t]}=g_{12} \left( X_{1[t]}, U_{2\rightarrow 13[t]} \right)$ for each time $t\in\{1,\dots,n\}$.
\end{remark}
\vspace{1mm}

\begin{IEEEproof}
The direct part of the proof simply follows by choosing: 
\begin{IEEEeqnarray*}{rcl}
U_{3\rightarrow 12,l}&=&U_{1\rightarrow 3,l}=U_{1\rightarrow 2,l}=U_{
  2\rightarrow 1,l}=U_{2\rightarrow 3,l}=U_{3\rightarrow 1,l}=U_{3\rightarrow 2,l}=\varnothing,\,\, \forall l\\
U_{1\rightarrow 23,1} &=& X_1 \ , \,\,\,\,\, U_{1\rightarrow 23,l} = U_{2\rightarrow 13,l} =\varnothing\ \forall\,\,\, l>1 \ ,
\end{IEEEeqnarray*}
and thus the rate-distortion region~\eqref{eq-inner-bound} reduces to the desired region in Theorem~\ref{theo-Coop-BT} where for simplicity we dropped the round index.  We now proceed to the proof of the converse. 

If a pair of rates $(R_1,R_2)$  and distortion $D_{32}$ are admissible for the $K$-steps interactive cooperative distributed source coding setting described in Fig.~\ref{fig:model-Coop-HB}, then  for all $\varepsilon >0$ there exists $n_0(\varepsilon,K)$, such that $\forall\,n>n_0(\varepsilon,K)$  there exists a $K$-steps interactive source code $(n,K,\mathcal{F},\mathcal{G})$ with intermediate rates satisfying:  
\begin{equation}
\frac{1}{n}\sum_{l=1}^K\log{\|\mathcal{J}_{i}^l\|}\leq R_i+\varepsilon\ ,\ i\in\{1,2\}
 \end{equation} 
and with average per-letter distortions  with respect to the source $2$ and perfect reconstruction with respect to the source $1$ at node $3$: 
\begin{IEEEeqnarray}{rcl}
&&\mathds{E} \left[d(X_2^n,\hat{X}^n_{32})\right]\leq D_{32}+\varepsilon\ ,\label{eq-converse-BT-distortion}\\
&&\mbox{Pr}\left(  X_1^n\neq \hat{X}_{31}^n\right)\leq \varepsilon\ ,\label{eq-converse-BT-error}
\end{IEEEeqnarray}
where 
\begin{IEEEeqnarray}{rcl}
&& \hat{X}_{32}^n \equiv g_{32}\left(\mathcal{J}_{1}^{[1:K]}, \mathcal{J}_{2}^{[1:K]}\right)\ ,\,\,\,\, \hat{X}_{31}^n \equiv g_{31}\left(\mathcal{J}_{1}^{[1:K]}, \mathcal{J}_{2}^{[1:K]}\right)\ .
\end{IEEEeqnarray}

For each $t\in\{1,\dots,n\}$, define random variables $U_{2\rightarrow 13[t]}$ as follows:
\begin{IEEEeqnarray}{rCl}
U_{2\rightarrow 13[t]}&\triangleq & \left(\mathcal{J}_{1}^{[1:K]}, \mathcal{J}_{2}^{[1:K]}, X_{1[1:t-1]},X_{1[t+1:n]}\right) \label{eq:defW} \ .
\end{IEEEeqnarray}
By the condition~\eqref{eq-converse-BT-error} which says that $\mbox{Pr}\left(  X_1^n\neq \hat{X}_{31}^n\right)\leq \varepsilon$ and Fano's inequality~\cite{cover2006elements}, we have 
\begin{IEEEeqnarray}{rCl}
H(X_1^n| \hat{X}_{31}^n)  &\leq &\mbox{Pr}\left(  X_1^n\neq \hat{X}_{31}^n\right) \log_2 (\|\mathcal{X}_1^n \|-1) + H_2\left( \mbox{Pr}\big(  X_1^n\neq \hat{X}_{31}^n\big)  \right)  \triangleq    n \epsilon_n   \ ,
\label{eq-converse-BT-fano}
\end{IEEEeqnarray}
where $\epsilon_n(\varepsilon) \rightarrow 0$ provided that $\varepsilon\rightarrow 0$ and $n \rightarrow \infty$.

\subsubsection{Rate at node 1}
For the first rate, we have
\begin{IEEEeqnarray}{rCl}
n(R_1+\varepsilon)
	&\geq& 	H\left(\mathcal{J}_{1}^{[1:K]}\right) \\
	&\geq&	I\left(\mathcal{J}_{1}^{[1:K]}; X_1^n| X_2^n  \right)  \\
	&=& n H(X_1| X_2) -  H\left(X_1^n | X_2^n \mathcal{J}_{1}^{[1:K]}  \right) \\
	&\stackrel{(a)}{=}& n H(X_1| X_2) -  H\left(X_1^n | X_2^n \mathcal{J}_{1}^{[1:K]} \mathcal{J}_{2}^{[1:K]} \right)\\
	&\stackrel{(b)}{\geq }&  n H(X_1| X_2) -  H(X_1^n| \hat{X}_{31}^n) \\
	&\stackrel{(b)}{\geq }& n H(X_1| X_2) - n \epsilon_n\ ,
\end{IEEEeqnarray}
where
\begin{itemize}
\item step~$(a)$ follows from the fact that by definition of the code the sequence $\mathcal{J}_{2}^{[1:K]}$ is a function of the source $X_2^n$ and the vector of messages $\mathcal{J}_{1}^{[1:K]}$,
\item step~$(b)$ follows from the code assumption that guarantees the existence of a reconstruction function $\hat{X}_{31}^n \equiv g_{31}\left(\mathcal{J}_{1}^{[1:K]}, \mathcal{J}_{2}^{[1:K]}\right)$, 
\item step~$(c)$ follows from Fano's inequality in~\eqref{eq-converse-BT-fano}.
\end{itemize}

\subsubsection{Rate at node 2}
For the second rate, we have
\begin{IEEEeqnarray}{rCl}
n(R_2+\varepsilon)
	&\geq& 				H\left(\mathcal{J}_{2}^{[1:K]}\right) \\
	&\geq&		I\left(\mathcal{J}_{2}^{[1:K]}; X_1^nX_2^n\right)\\
	&=&  I\left(\mathcal{J}_{2}^{[1:K]}; X_1^n\right)+ I\left(\mathcal{J}_{2}^{[1:K]}; X_2^n| X_1^n\right)  \\
	&\stackrel{(a)}{\geq}& I\left(\mathcal{J}_{2}^{[1:K]}; X_1^n\right)+  \sum_{t=1}^n I\left(\mathcal{J}_{2}^{[1:K]}; X_{2[t]}| X_{1[t]}X_{1[t+1:n]} X_{1[1:t-1]}  X_{2[1:t-1]} \right) \\
	&\stackrel{(b)}{=}&  I\left(\mathcal{J}_{2}^{[1:K]}; X_1^n\right)+ \sum_{t=1}^n I\left(\mathcal{J}_{2}^{[1:K]}X_{1[t+1:n]} X_{1[1:t-1]}  X_{2[1:t-1]}; X_{2[t]}| X_{1[t]} \right) \\
	&\stackrel{(c)}{\geq }& I\left(\mathcal{J}_{2}^{[1:K]}; X_1^n\right)+  \sum_{t=1}^n I\left(U_{2\rightarrow 13[t]} ; X_{2[t]}| X_{1[t]} \right) \\
	&\stackrel{(d)}{= }& I\left(\mathcal{J}_{2}^{[1:K]}; X_1^n\right)+  \sum_{t=1}^n I\left(U_{2\rightarrow 13[Q]} ; X_{2[Q]}| X_{1[Q]}, Q=t \right) \\ 
	&\stackrel{(e)}{= }& I\left(\mathcal{J}_{2}^{[1:K]}; X_1^n\right)+  n I\left(U_{2\rightarrow 13[Q]} ; X_{2[Q]}| X_{1[Q]}, Q\right)  \\ 
		&\stackrel{(f)}{\geq  }&  I\left(\mathcal{J}_{2}^{[1:K]}; X_1^n\right)+n I\left(\widetilde{U}_{2\rightarrow 13} ; X_{2}| X_{1}\right)\label{eq-converse-BT-connection}
		\\ &\stackrel{(g)}{\geq  }&   n I\left(\widetilde{U}_{2\rightarrow 13} ; X_{2}| X_{1}\right)\ , 
\end{IEEEeqnarray}
where
\begin{itemize}
\item step~$(a)$ follows  from the chain rule for conditional mutual information and non-negativity of mutual information,
\item step~$(b)$ follows  from the memoryless property across time of the sources $(X_1^n,X_2^n)$, 
\item step~$(c)$ follows  from the non-negativity of mutual information and definitions~\eqref{eq:defW}, 
\item step~$(d)$ follows  from the use of a time sharing random variable $Q$ uniformly distributed over the set $\{1,\dots,n\}$, 
\item step~$(e)$  follows  from the definition of the conditional mutual information, 
\item step~$(f)$ follows by letting a new random variable $\widetilde{U}_{2\rightarrow 13}\triangleq (U_{2\rightarrow 13[Q]},Q)$,
\item step~$(g)$ follows from the non-negativity of mutual information.
\end{itemize}

\subsubsection{Sum-rate of nodes $1$ and $2$}
For the sum-rate, we have
\begin{IEEEeqnarray}{rCl}
n(R_1 + R_2+2\varepsilon)
	&\geq& 				H\left(\mathcal{J}_{1}^{[1:K]}\right)+ n (R_2 + \varepsilon)\\
	&\stackrel{(a)}{\geq}&		H\left(\mathcal{J}_{1}^{[1:K]}\right)+ I\left(\mathcal{J}_{2}^{[1:K]}; X_1^n\right) + n I\left(\widetilde{U}_{2\rightarrow 13} ; X_{2}| X_{1}\right)  \\
	& = & H\left(\mathcal{J}_{1}^{[1:K]} | \mathcal{J}_{2}^{[1:K]}  \right) +    I\left(\mathcal{J}_{1}^{[1:K]}; \mathcal{J}_{2}^{[1:K]} \right)\nonumber\\ 
	&+& I\left(\mathcal{J}_{2}^{[1:K]}; X_1^n\right) +n I\left(\widetilde{U}_{2\rightarrow 13} ; X_{2}| X_{1}\right) \\ 
        &\geq & I\left(\mathcal{J}_{1}^{[1:K]}; X_1^n | \mathcal{J}_{2}^{[1:K]} \right) +  I\left(\mathcal{J}_{1}^{[1:K]}; \mathcal{J}_{2}^{[1:K]} \right)\nonumber\\ 
        &+& I\left(\mathcal{J}_{2}^{[1:K]}; X_1^n\right) + n I\left(\widetilde{U}_{2\rightarrow 13} ; X_{2}| X_{1}\right)\\ 
        & = & I\left(\mathcal{J}_{1}^{[1:K]}; X_1^n\right)  +  I\left(X_1^n\mathcal{J}_{1}^{[1:K]}; \mathcal{J}_{2}^{[1:K]}  \right)+n I\left(\widetilde{U}_{2\rightarrow 13} ; X_{2}| X_{1}\right)  \\
        & \stackrel{(b)}{=} & n \left[H(X_1) + I\left(\widetilde{U}_{2\rightarrow 13} ; X_{2}| X_{1}\right)\right]\nonumber\\
 &-&    H\left(X_1^n|\mathcal{J}_{1}^{[1:K]}\right) +  I\left(X_1^n\mathcal{J}_{1}^{[1:K]}; \mathcal{J}_{2}^{[1:K]}  \right) \,\,\,\,           \\ 
               & \stackrel{(c)}{\geq } & n\left[H(X_1) + I\left(\widetilde{U}_{2\rightarrow 13} ; X_{2}| X_{1}\right) \right]
- H\left(X_1^n|\mathcal{J}_{1}^{[1:K]}\mathcal{J}_{2}^{[1:K]}\right)     \\
             & \stackrel{(d)}{\geq } & n  \left[ H(X_1) + I\left(\widetilde{U}_{2\rightarrow 13} ; X_{2}| X_{1}\right)\right] 
 - H(X_1^n| \hat{X}_{31}^n)    \\
         & \stackrel{(e)}{\geq } &  n\left[H(X_1) + I\left(\widetilde{U}_{2\rightarrow 13} ; X_{2}| X_{1}\right) -  \epsilon_n\right] \ ,
\end{IEEEeqnarray}
where
\begin{itemize}
\item step~$(a)$ follows from inequality~\eqref{eq-converse-BT-connection},
\item step~$(b)$ follows  from the memoryless property across time of the source $X_1^n$,
\item step~$(c)$ follows from non-negativity of mutual information, 
\item step~$(d)$ follows  from the code assumption that guarantees the existence of reconstruction function $\hat{X}_{31}^n \equiv g_{31}\left(\mathcal{J}_{1}^{[1:K]}, \mathcal{J}_{2}^{[1:K]}\right)$ and from the fact that unconditioning increases entropy, 
\item step~$(e)$ from Fano's inequality in~\eqref{eq-converse-BT-fano}. 
\end{itemize}

\subsubsection{Distortion at node 3}

Node $3$ reconstructs lossless $\hat{X}_{31}^n \equiv g_{31}\left(\mathcal{J}_{1}^{[1:K]}, \mathcal{J}_{2}^{[1:K]}\right)$ and lossy $\hat{X}_{32}^n \equiv g_{32}\left(\mathcal{J}_{1}^{[1:K]} , \mathcal{J}_{2}^{[1:K]}\right)$. For each $t\in\{1,\dots,n\}$, define a function $\hat{X}_{32[t]}$ as beging the $t$-th coordinate of this estimate:
\begin{IEEEeqnarray}{rCl}
\hat{X}_{32[t]}\left(U_{2\rightarrow 13[t]}\right) \triangleq   g_{32[t]} \left(\mathcal{J}_{1}^{[1:K]} , \mathcal{J}_{2}^{[1:K]}\right) \ .
\end{IEEEeqnarray}
The component-wise mean distortion thus verifies
\begin{IEEEeqnarray}{rCl}
D_{32} + \varepsilon
	&\geq&	\bE\left[ d\left (X_2,g_{31}\big(\mathcal{J}_{1}^{[1:K]}, \mathcal{J}_{2}^{[1:K]}\big) \right) \right] \\
	&=& 	\frac1n \sum_{t=1}^n \bE\left[ d\left (X_{2[t]},\hat{X}_{32[t]}\left(U_{2\rightarrow 13[t]}\right) \right) \right] \\
	&=& 	\frac1n \sum_{t=1}^n \bE\left[ d\left (X_{2[Q]}, \hat{X}_{32[Q]}\left(U_{2\rightarrow 13[Q]}\right) \right)\ \middle|\ Q=t \right] \\
	&=& 	\bE\left[ d\left (X_{2[Q]}, \hat{X}_{32[Q]}\left(U_{2\rightarrow 13[Q]}\right) \right) \right] \\
	&=& 	\bE\left[ d\left (X_2, \widetilde{X}_{32}\left(\widetilde{U}_{2\rightarrow 13}\right) \right) \right] \ ,
\end{IEEEeqnarray}
where we defined function $\widetilde{X}_{32}$ by 
\begin{IEEEeqnarray}{rCl}
\widetilde{X}_{32}\left(\widetilde{U}_{2\rightarrow 13}\right) = \widetilde{X}_{32}\left(Q, U_{2\rightarrow 13[Q]}\right) \triangleq \hat{X}_{32[Q]}\left(U_{2\rightarrow 13[Q]}\right) \ .
\end{IEEEeqnarray}
This concludes the proof of the converse and thus that of the theorem. 
\end{IEEEproof}

\subsection{Two encoders and three decoders subject to lossless/lossy reconstruction constraints with side information} 

\begin{figure}[th!]
\centering
\ifpdf\includegraphics[angle=0,width=0.75\columnwidth,keepaspectratio,trim= 0mm 0mm 0mm 0mm,clip]{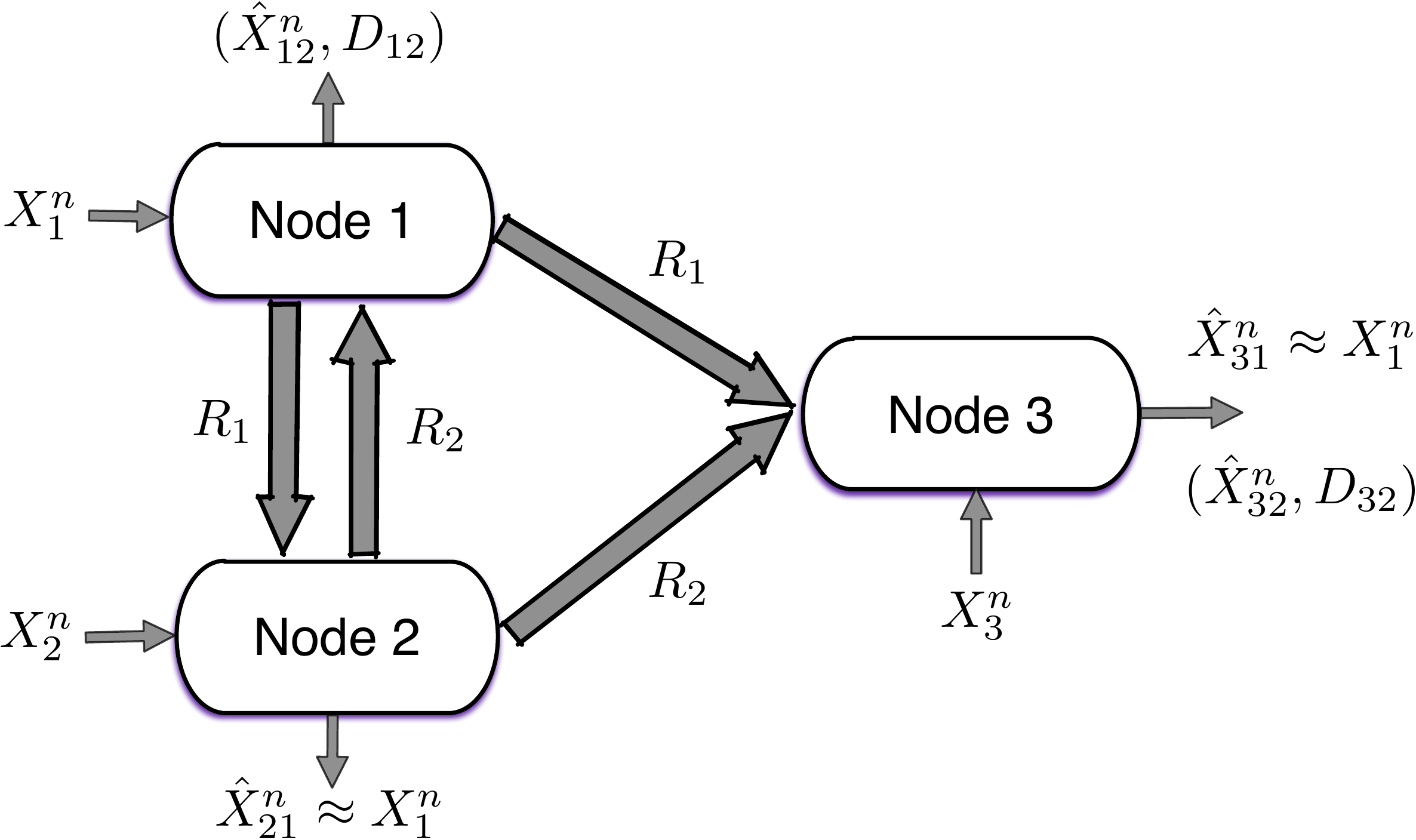} 
\else
\includegraphics[angle=270,width=0.75\columnwidth,keepaspectratio,trim= 0mm 0mm 0mm 0mm,clip]{model.eps} 
\fi
\caption{Two encoders and three decoders subject to lossless/lossy reconstruction constraints with side information.}

\label{fig:model-Coop-BT2}
\end{figure}

Consider now the problem described in Fig.~\ref{fig:model-Coop-BT2} where encoder $1$ wishes to communicate the \emph{lossless} the source $X_1^n$ to nodes $2$ and $3$ while encoder $2$ wishes to send a \emph{lossy} description of its source $X_2^n$ to nodes $1$ and $3$ with distortion constraints $D_{12}$ and $D_{32}$, respectively. In addition to this, the encoders overhead the communication using $K$ communication rounds.  This problem can be seen as a generalization of the settings previously investigated in~\cite{berger-1977,32119}.

\vspace{1mm}
 \begin{theorem}\label{theo-Coop-BT2}
The rate-distortion region of the setting described in Fig.~\ref{fig:model-Coop-BT2} is given by the union over all joint probability measures $p_{X_1X_2X_3U_{2\rightarrow 13}U_{2\rightarrow 3}}$ satisfying the Markov chain
\begin{equation}
(U_{2\rightarrow 13}, U_{2\rightarrow 3}) \mkv (X_1,X_2) \mkv  X_3
\label{eq:two_dist_crit_mark}
\end{equation}
and such that there exists reconstruction mappings:
\begin{IEEEeqnarray}{rcl}
g_{32} \left( X_1, X_3 , U_{2\rightarrow 13} ,U_{2\rightarrow 3}\right)&=&\hat{X}_{32}\,\,\,\, \textrm{ with }\,\,\,\,\mathds{E} \left[d(X_2,\hat{X}_{32}) \right] \leq  D_{32}\ ,\\
g_{12} \left( X_1,  U_{2\rightarrow 13} \right)&=&\hat{X}_{12}\,\,\,\, \textrm{ with }\,\,\,\,\mathds{E} \left[d(X_2,\hat{X}_{12}) \right] \leq  D_{12}\ ,
\end{IEEEeqnarray} 
of the set of all tuples satisfying: 
\begin{IEEEeqnarray}{rCl}
R_1 &\geq& H(X_1|X_2) \ ,\\
R_2 &\geq& I(U_{2\rightarrow 13};X_2 |X_1) + I(U_{2\rightarrow 3};X_2| U_{2\rightarrow 13}X_1 X_3) \ ,\\
R_1+R_2 &\geq& H(X_1|X_3) +I(U_{2\rightarrow 13}U_{2\rightarrow 3};X_2| X_1 X_3). 
\end{IEEEeqnarray}
The auxiliary random variables have cardinality bounds: $\| \mathcal{U}_{2\rightarrow 13}\|\leq \|\mathcal{X}_1\|\|\mathcal{X}_2\|+2$, $\| \mathcal{U}_{2\rightarrow 3}\|\leq \|\mathcal{X}_1\|\|\mathcal{X}_2\|\|\mathcal{U}_{2\rightarrow 13}\|+1$.
\end{theorem}
\vspace{1mm}
\begin{remark}
Notice that the rate-distortion region in Theorem~\ref{theo-Coop-BT2} is achievable with a single round  of interactions $K=1$, which implies that multiple rounds do not improve the rate-distortion region in this case. 
\end{remark}
\vspace{1mm}
\begin{remark}
It is worth mentioning that cooperation between encoders reduces the rate needed to communicate the source $X_2$ while increasing the optimization set of all admissible source descriptions. 
\end{remark}

\vspace{1mm}
\begin{IEEEproof}
The direct part of the proof follows by choosing:
\begin{IEEEeqnarray*}{rcl}
U_{3\rightarrow 12,l}&=&U_{1\rightarrow 3,l}=U_{1\rightarrow 2,l}=U_{
  2\rightarrow 1,l}=U_{3\rightarrow 1,l}=U_{3\rightarrow 2,l}=\varnothing,\,\, \forall l\\
U_{1\rightarrow 23} &\equiv&  U_{1\rightarrow 23,1} = X_1 \ , \,\,\,\,\, U_{1\rightarrow 23,l} = U_{2\rightarrow 13,l}=U_{2\rightarrow 3,l} =\varnothing\ \forall\,\,\, l>1 \ . 
\end{IEEEeqnarray*}
and $U_{2\rightarrow 13,1}\equiv U_{2\rightarrow 13}$ and $U_{2\rightarrow 3,1}\equiv U_{2\rightarrow 3}$ are auxiliary random variables that according to Theorem \ref{theo-main-theorem} should satisfy:
\begin{equation}
U_{2\rightarrow 13} \mkv (X_1,X_2) \mkv  X_3,\ \ U_{2\rightarrow 3}, \mkv (U_{2\rightarrow 13},X_1,X_2) \mkv  X_3.
\end{equation}
Notice, however that these Markov chains are equivalent to (\ref{eq:two_dist_crit_mark}). From the rate equations in Theorem \ref{theo-main-theorem}, and the above choices for the auxiliary random variables we obtain:
\begin{IEEEeqnarray}{rcl}
R_{1\rightarrow 23}&>& H(X_1|X_2)\ ,\\
R_{2\rightarrow 13}&>&\max{\left\{I(X_2;U_{2\rightarrow 13}|X_1),I(X_2;U_{2\rightarrow 13}|X_1X_3)\right\}}\\&=&I(X_2;U_{2\rightarrow 13}|X_1)\ ,\\
R_{1\rightarrow 23}+R_{2\rightarrow 13}&>&H(X_1|X_3)+I(X_2;U_{2\rightarrow 13}|X_1X_3)\ ,\\
R_{2\rightarrow 3}&>&I(X_2;U_{2\rightarrow 3}|U_{2\rightarrow 13}X_1X_3)\ .
\end{IEEEeqnarray}
Noticing that $R_1\equiv R_{1\rightarrow 23}$ and $R_2\equiv R_{2\rightarrow 13}+R_{2\rightarrow 3}$ the rate-distortion region~\eqref{eq-inner-bound} reduces to the desired region in Theorem~\ref{theo-Coop-BT2}, where for simplicity we dropped the round index. We now proceed to the proof of the converse. 

If a pair of rates $(R_1,R_2)$  and distortions $(D_{12},D_{32})$ are admissible for the $K$-steps interactive cooperative distributed source coding setting described in Fig.~\ref{fig:model-Coop-BT2}, then  for all $\varepsilon >0$ there exists $n_0(\varepsilon,K)$, such that $\forall\,n>n_0(\varepsilon,K)$  there is a $K$-steps interactive source code $(n,K,\mathcal{F},\mathcal{G})$ with intermediate rates satisfying:  
\begin{equation}
\frac{1}{n}\sum_{l=1}^K\log{\|\mathcal{J}_{i}^l\|}\leq R_i+\varepsilon\ ,\ i\in\{1,2\}
 \end{equation} 
and with average per-letter distortions  with respect to the source $2$ and perfect reconstruction with respect to the source $1$ at all nodes: 
\begin{IEEEeqnarray}{rcl}
&&\mathds{E} \left[d(X_2^n,\hat{X}^n_{32})\right]\leq D_{32}+\varepsilon\ ,\label{eq-converse-BT-distortion2}\\
&&\mbox{Pr}\left(  X_1^n\neq \hat{X}_{21}^n\right)\leq \varepsilon\ ,\label{eq-converse-BT-error3}\\
&&\mathds{E} \left[d(X_2^n,\hat{X}^n_{12})\right]\leq D_{12}+\varepsilon\ ,\label{eq-converse-BT-distortion3}\\
&&\mbox{Pr}\left(  X_1^n\neq \hat{X}_{31}^n\right)\leq \varepsilon\ ,\label{eq-converse-BT-error2}
\end{IEEEeqnarray}
where 
\begin{IEEEeqnarray}{rCl}
\hat{X}_{32}^n &\equiv & g_{32} \left(\mathcal{J}_{1}^{[1:K]}, \mathcal{J}_{2}^{[1:K]}, X_3^n\right)\ ,\,\,\,\,\,
\hat{X}_{12}^n  \equiv  g_{12}\left( \mathcal{J}_{2}^{[1:K]}, X_1^n\right)\ ,\label{eq-BT2-12} \\
\hat{X}_{31}^n & \equiv & g_{31}\left(\mathcal{J}_{1}^{[1:K]}, \mathcal{J}_{2}^{[1:K]}, X_3^n\right)\ ,\,\,\,\,\,
\hat{X}_{21}^n  \equiv  g_{21}\left( \mathcal{J}_{1}^{[1:K]}, X_2^n\right)\ . \label{eq-BT2-21a}
\end{IEEEeqnarray}

For each $t\in\{1,\dots,n\}$, define random variables $U_{2\rightarrow 13[t]}$ and $U_{2\rightarrow 3[t]}$ as follows:
\begin{IEEEeqnarray}{rCl}
U_{2\rightarrow 13[t]}&\triangleq & \left(\mathcal{J}_{1}^{[1:K]}, \mathcal{J}_{2}^{[1:K]}, X_{1[1:t-1]},X_{1[t+1:n]}, X_{3[1:t-1]}\right) \label{eq:MarkovBT2-1a} \ , \\
U_{2\rightarrow 3[t]} &\triangleq & \left(U_{2\rightarrow 13[t]}, X_{3[t+1:n]}, X_{2[1:t-1]}\right) \label{eq:MarkovBT2-2} \ .
\end{IEEEeqnarray}
The fact that these choices of the auxiliary random variables satisfy the Markov chain (\ref{eq:two_dist_crit_mark}) can be obtained from point 6) in Lemma \ref{lemma:Markov_interactive_encoding_1}. By the conditions~\eqref{eq-converse-BT-error2} and~\eqref{eq-converse-BT-error3}, and Fano's inequality, we have 
\begin{IEEEeqnarray}{rCl}
H(X_1^n| \hat{X}_{31}^n)  &\leq &\mbox{Pr}\left(  X_1^n\neq \hat{X}_{31}^n\right) \log_2 (\|\mathcal{X}_1^n \|-1) + H_2\left( \mbox{Pr}\big(  X_1^n\neq \hat{X}_{31}^n\big)  \right)  \triangleq    n \epsilon_n   \ , \label{eq-converse-BT-fano2-1}\\
H(X_1^n| \hat{X}_{21}^n)  &\leq &\mbox{Pr}\left(  X_1^n\neq \hat{X}_{21}^n\right) \log_2 (\|\mathcal{X}_1^n \|-1) + H_2\left( \mbox{Pr}\big(  X_1^n\neq \hat{X}_{21}^n\big)  \right)  \triangleq    n \epsilon_n   \ ,\label{eq-converse-BT-fano2-2}
\end{IEEEeqnarray}
where $\epsilon_n(\varepsilon) \rightarrow 0$ provided that $\varepsilon\rightarrow 0$ and $n \rightarrow \infty$.

\subsubsection{Rate at node 1}

For the first rate, we have
\begin{IEEEeqnarray}{rCl}
n(R_1+\varepsilon)
	&\geq& 					H\left(\mathcal{J}_{1}^{[1:K]}\right) \\
	&\geq&	H\left(\mathcal{J}_{1}^{[1:K]}| X_2^n \right)   \\
	&\stackrel{(a)}{=}&	I\left(\mathcal{J}_{1}^{[1:K]}; X_1^n| X_2^n  \right)  \\
	&=& n H(X_1| X_2) -  H\left(X_1^n | X_2^n \mathcal{J}_{1}^{[1:K]}  \right) \\
	&\stackrel{(b)}{\geq }&  n H(X_1| X_2) -  H(X_1^n| \hat{X}_{21}^n) \\
	&\stackrel{(c)}{\geq }& n\left[ H(X_1| X_2) -  \epsilon_n\right]\ ,
\end{IEEEeqnarray}
where
\begin{itemize}
\item step~$(a)$ follows from the fact that by definition of the code the sequence $\mathcal{J}_{1}^{[1:K]}$ is a function of the both sources $(X_1^n,X_2^n)$, 
\item step~$(b)$ follows from the code assumption in~\eqref{eq-BT2-21a} that guarantees the existence of a reconstruction function $\hat{X}_{21}^n \equiv g_{21}\left(\mathcal{J}_{1}^{[1:K]},X_2^n\right)$, 
\item step~$(c)$ follows from Fano's inequality in~\eqref{eq-converse-BT-fano2-2}.
\end{itemize}

\subsubsection{Rate at node 2}
For the second rate, we have
\begin{IEEEeqnarray}{rCl}
n(R_2+\varepsilon)
	&\geq& 				H\left(\mathcal{J}_{2}^{[1:K]}\right) \\
	&\stackrel{(a)}{=}&		I\left(\mathcal{J}_{2}^{[1:K]}; X_1^nX_2^nX_3^n\right)\\
	&\stackrel{(b)}{\geq }&		I\left(\mathcal{J}_{2}^{[1:K]}; X_2^nX_3^n|X_1^n\right)\\	
 	&\stackrel{(c)}{= }&		I\left(\mathcal{J}_{1}^{[1:K]}\mathcal{J}_{2}^{[1:K]}; X_2^nX_3^n|X_1^n\right)\\	
	&= &		I\left(\mathcal{J}_{1}^{[1:K]}\mathcal{J}_{2}^{[1:K]};  X_3^n|X_1^n\right) + I\left(\mathcal{J}_{1}^{[1:K]}\mathcal{J}_{2}^{[1:K]};  X_2^n|X_1^nX_3^n\right) \\
	&\stackrel{(d)}{=}&  
	 \sum_{t=1}^n \Big[I\left(\mathcal{J}_{1}^{[1:K]}\mathcal{J}_{2}^{[1:K]} ; X_{3[t]} |  X_1^n, X_{3[1:t-1]} \right) \nonumber \\
	   \IEEEeqnarraymulticol{3}{c}{ + I\left(\mathcal{J}_{1}^{[1:K]}\mathcal{J}_{2}^{[1:K]}  ; X_{2[t]} | X_1^nX_3^nX_{2[1:t-1]} \right)\Big] }\\
	&\stackrel{(e)}{=}&   \sum_{t=1}^n \Big[ I\left(\mathcal{J}_{1}^{[1:K]}\mathcal{J}_{2}^{[1:K]} X_{1[1:t-1]}X_{1[t+1:n]}X_{3[1:t-1]}; X_{3[t]} |  X_{1[t]} \right) \nonumber \\
	 &+&  I\left(\mathcal{J}_{1}^{[1:K]}\mathcal{J}_{2}^{[1:K]}  X_{1[1:t-1]} X_{1[t+1:n]} X_{3[1:t-1]} X_{3[t+1:n]} X_{2[1:t-1]} ; X_{2[t]} | X_{1[t]}X_{3[t]} \right)	\Big]\,\,\,\,\,\,\,\, \\
		&\stackrel{(f)}{=}&   \sum_{t=1}^n \Big[ I\left(U_{2\rightarrow 13[t]}; X_{3[t]} |  X_{1[t]} \right) +  I\left(U_{2\rightarrow 13[t]} ; X_{2[t]} | X_{1[t]}X_{3[t]} \right)\nonumber	\\
	    \IEEEeqnarraymulticol{3}{c}{ +  I\left( U_{2\rightarrow 3[t]}; X_{2[t]} | X_{1[t]}X_{3[t]} U_{2\rightarrow 13[t]} \right)	 \Big]} \,\,\,\,\,\,\,\, \\
	  &=&   \sum_{t=1}^n \Big[ I\left(U_{2\rightarrow 13[t]} ; X_{2[t]} X_{3[t]} |  X_{1[t]} \right)	 +  I\left( U_{2\rightarrow 3[t]}; X_{2[t]} | X_{1[t]}X_{3[t]}U_{2\rightarrow 13[t]} \right)	 \Big] \,\,\,\,\,\,\,\, \\
	  	  &\stackrel{(g)}{=}&   \sum_{t=1}^n \Big[ I\left(U_{2\rightarrow 13[t]}; X_{2[t]}  |  X_{1[t]} \right) +  I\left( U_{2\rightarrow 3[t]} ; X_{2[t]} | X_{1[t]}X_{3[t]} U_{2\rightarrow 13[t]} \right)	 \Big]\\
			  &\stackrel{(h)}{=}&   \sum_{t=1}^n \Big[ I\left(U_{2\rightarrow 13[Q]}; X_{2[Q]}  |  X_{1[Q]} ,Q=t \right)\nonumber\\
			  \IEEEeqnarraymulticol{3}{c}{  +   I\left( U_{2\rightarrow 3[Q]} ; X_{2[Q]} | X_{1[Q]}X_{3[Q]} U_{2\rightarrow 13[Q]} ,Q=t \right)	 \Big]} \\
&\stackrel{(i)}{\geq }&  n\Big[ I\left(\widetilde{U}_{2\rightarrow 13}; X_{2}  |  X_{1} \right) +  I\left( \widetilde{U}_{2\rightarrow 3} ; X_{2} | X_{1}X_{3} \widetilde{U}_{2\rightarrow 13}\right)	 \Big]\ ,
\end{IEEEeqnarray}
where
\begin{itemize}
\item step~$(a)$ follows from the fact that $\mathcal{J}_{2}^{[1:K]}$ is a function of the sources $(X_1^n,X_2^n)$, 
\item step~$(b)$ follows  from the non-negativity of mutual information, 
\item step~$(c)$ follows  from the fact that $\mathcal{J}_{1}^{[2:K]}$ is a function of $\mathcal{J}_{2}^{[1:K]}$ and the source $X_1^n$, 
\item step~$(d)$ follows  from the chain rule for conditional mutual information,
\item step~$(e)$ follows  from the memoryless property across time of the sources $(X_1^n,X_2^n,X_3^n)$, 
\item step~$(f)$  follows  from the chain rule for conditional mutual information and the definitions~\eqref{eq:MarkovBT2-1a} and ~\eqref{eq:MarkovBT2-2}, 
\item step~$(g)$ follows from the Markov chain $U_{2\rightarrow 13[t]} \mkv (X_{1[t]},X_{2[t]}) \mkv X_{3[t]} $, for all $t\in\{1,\dots,n\}$,
\item step~$(h)$ follows from the use of a time sharing random variable $Q$ uniformly distributed over the set $\{1,\dots,n\}$, 
\item step~$(i)$ follows by letting new random variables $\widetilde{U}_{2\rightarrow 13} \triangleq (U_{2\rightarrow 13[Q]},Q)$ and $\widetilde{U}_{2\rightarrow 3}\triangleq (U_{2\rightarrow 3[Q]},Q)$.
\end{itemize}

\subsubsection{Sum-rate of nodes $1$ and $2$}
For the sum-rate, we have
\begin{IEEEeqnarray}{rCl}
n(R_1 + R_2+2\varepsilon)
	&\geq& 				H\left(\mathcal{J}_{1}^{[1:K]}\right)+ H\left(\mathcal{J}_{2}^{[1:K]}\right)\\
&=& H\left(\mathcal{J}_{1}^{[1:K]}\mathcal{J}_{2}^{[1:K]} \right) +  I\left(\mathcal{J}_{1}^{[1:K]} ; \mathcal{J}_{2}^{[1:K]}\right)\\
	&\stackrel{(a)}{=}&	I\left(\mathcal{J}_{1}^{[1:K]}\mathcal{J}_{2}^{[1:K]}; X_1^nX_3^nX_2^n \right) +  I\left(\mathcal{J}_{1}^{[1:K]} ; \mathcal{J}_{2}^{[1:K]}\right)\\
& \stackrel{(b)}{\geq } & I\left(\mathcal{J}_{1}^{[1:K]}\mathcal{J}_{2}^{[1:K]}; X_1^nX_2^n | X_3^n \right)\\
&=& I\left(\mathcal{J}_{1}^{[1:K]}\mathcal{J}_{2}^{[1:K]}; X_1^n | X_3^n \right)+I\left(\mathcal{J}_{1}^{[1:K]}\mathcal{J}_{2}^{[1:K]}; X_2^n | X_1^nX_3^n \right)\\
&=& H\left(X_1^n | X_3^n \right) - H\left(X_1^n | \mathcal{J}_{1}^{[1:K]}\mathcal{J}_{2}^{[1:K]}X_3^n \right)\nonumber\\
  \IEEEeqnarraymulticol{3}{c}{   +I\left(\mathcal{J}_{1}^{[1:K]}\mathcal{J}_{2}^{[1:K]}; X_2^n | X_1^nX_3^n \right)}\\
  & \stackrel{(c)}{\geq } & H\left(X_1^n | X_3^n \right) - H(X_1^n | \hat{X}_{31}^n ) +I\left(\mathcal{J}_{1}^{[1:K]}\mathcal{J}_{2}^{[1:K]}; X_2^n | X_1^nX_3^n \right)\\
    & \stackrel{(d)}{\geq } & n\left[H\left(X_1 | X_3 \right) - \epsilon_n \right]+I\left(\mathcal{J}_{1}^{[1:K]}\mathcal{J}_{2}^{[1:K]}; X_2^n | X_1^nX_3^n \right)\\
    & \stackrel{(e)}{= } &   \sum_{t=1}^n  I\left(\mathcal{J}_{1}^{[1:K]}\mathcal{J}_{2}^{[1:K]} X_{1[1:t-1]}X_{1[t+1:n]}X_{3[1:t-1]}X_{3[t+1:n]}X_{2[1:t-1]}; X_{2[t]} |  X_{1[t]}X_{3[t]} \right) \nonumber\\
    &+& n\left[H\left(X_1 | X_3 \right) - \epsilon_n \right] \\
        & \stackrel{(f)}{= } & n\left[H\left(X_1 | X_3 \right) - \epsilon_n \right] +  \sum_{t=1}^n  I\left( U_{2\rightarrow 13[t]}U_{2\rightarrow 3[t]}; X_{2[t]} |  X_{1[t]}X_{3[t]} \right) \\
                & \stackrel{(g)}{= } & n\left[H\left(X_1 | X_3 \right) - \epsilon_n+ I\left( U_{2\rightarrow 13[Q]}U_{2\rightarrow 3[Q]}; X_{2[Q]} |  X_{1[Q]}X_{3[Q]},Q \right)  \right] \\
                & \stackrel{(h)}{= } & n\left[H\left(X_1 | X_3 \right) - \epsilon_n  + I\left( \widetilde{U}_{2\rightarrow 13}\widetilde{U}_{2\rightarrow 3}; X_{2} |  X_{1}X_{3} \right) \right]   \ ,              
\end{IEEEeqnarray}
where
\begin{itemize}
\item step~$(a)$ follows from the fact that $\mathcal{J}_{1}^{[1:K]}$ and $\mathcal{J}_{2}^{[1:K]}$ are functions of the sources $(X_1^n,X_2^n,X_3^n)$, to emphasize 
\item step~$(b)$ follows non-negativity of mutual information, 
\item step~$(c)$ follows  from the code assumption in~\eqref{eq-BT2-21a} that guarantees the existence of reconstruction function $\hat{X}_{31}^n \equiv g_{31}\left(\mathcal{J}_{1}^{[1:K]},\mathcal{J}_{2}^{[1:K]},X^n_3\right)$,  
\item step~$(d)$ follows from Fano's inequality in~\eqref{eq-converse-BT-error2},  
\item step~$(e)$ follows  from the chain rule of conditional mutual information and the memoryless property across time of the source $(X_1^n,X_2^n,X_3^n)$,
\item step~$(f)$ from follows from the definitions~\eqref{eq:MarkovBT2-1a} and ~\eqref{eq:MarkovBT2-2}, 
\item step~$(g)$ follows from the use of a time sharing random variable $Q$ uniformly distributed over the set $\{1,\dots,n\}$,
\item step~$(h)$ follows by letting new random variables $\widetilde{U}_{2\rightarrow 13} \triangleq (U_{2\rightarrow 13[Q]},Q)$ and $\widetilde{U}_{2\rightarrow 3}\triangleq (U_{2\rightarrow 3[Q]},Q)$.
\end{itemize}

\subsubsection{Distortion at node 1}

Node $1$ reconstructs a lossy  $\hat{X}_{12}^n \equiv g_{12}\left( \mathcal{J}_{2}^{[1:K]},X_1^n\right)$. It is clear that we write without loss of generality $\hat{X}_{12}^n \equiv g_{12}\left(\mathcal{J}_{1}^{[1:K]}, \mathcal{J}_{2}^{[1:K]},X_1^n\right)$. For each $t\in\{1,\dots,n\}$, define a function $\hat{X}_{12[t]}$ as beging the $t$-th coordinate of this estimate:
\begin{IEEEeqnarray}{rCl}
\hat{X}_{12[t]}\left(U_{2\rightarrow 13[t]}, X_{1[t]} \right) \triangleq   g_{12[t]} \left(\mathcal{J}_{1}^{[1:K]} , \mathcal{J}_{2}^{[1:K]},X_1^n \right) \ .
\end{IEEEeqnarray}
The component-wise mean distortion thus verifies
\begin{IEEEeqnarray}{rCl}
D_{12} + \varepsilon
	&\geq&	\bE\left[ d\left (X_2,g_{12}\big(\mathcal{J}_{1}^{[1:K]}, \mathcal{J}_{2}^{[1:K]},X_1^n \big) \right) \right] \\
	&=& 	\frac1n \sum_{t=1}^n \bE\left[ d\left (X_{2[t]},\hat{X}_{12[t]}\left(U_{2\rightarrow 13[t]},X_{1[t]}\right) \right) \right] \\
	&=& 	\frac1n \sum_{t=1}^n \bE\left[ d\left (X_{2[Q]}, \hat{X}_{12[Q]}\left(U_{2\rightarrow 13[Q]},X_{1[Q]}\right) \right)\ \middle|\ Q=t \right] \\
	&=& 	\bE\left[ d\left (X_{2[Q]}, \hat{X}_{12[Q]}\left(U_{2\rightarrow 13[Q]},X_{1[Q]}\right) \right) \right] \\
	&=& 	\bE\left[ d\left (X_2, \widetilde{X}_{12}\left(\widetilde{U}_{2\rightarrow 13},X_{1}\right) \right) \right] \ ,
\end{IEEEeqnarray}
where we defined function $\widetilde{X}_{12}$ by 
\begin{IEEEeqnarray}{rCl}
\widetilde{X}_{12}\left(\widetilde{U}_{2\rightarrow 13},X_{1}\right) = \widetilde{X}_{12}\left(Q, U_{2\rightarrow 13[Q]},X_{1[Q]}\right) \triangleq \hat{X}_{12[Q]}\left(U_{2\rightarrow 13[Q]},X_{1[Q]}\right) \ .
\end{IEEEeqnarray}

\subsubsection{Distortion at node 3}
Node $3$ reconstructs a  lossy description $\hat{X}_{32}^n \equiv g_{32}\left(\mathcal{J}_{1}^{[1:K]}, \mathcal{J}_{2}^{[1:K]},X^n_3\right)$. For each $t\in\{1,\dots,n\}$, define a function $\hat{X}_{32[t]}$ as beging the $t$-th coordinate of this estimate:
\begin{IEEEeqnarray}{rCl}
\hat{X}_{32[t]}\left(U_{2\rightarrow 13[t]},U_{2\rightarrow 3[t]},X_{3[t]}\right) \triangleq   g_{32[t]} \left(\mathcal{J}_{1}^{[1:K]} , \mathcal{J}_{2}^{[1:K]},X^n_3\right) \ .
\end{IEEEeqnarray}
The component-wise mean distortion thus verifies
\begin{IEEEeqnarray}{rCl}
D_{32} + \varepsilon
	&\geq&	\bE\left[ d\left (X_2,g_{32}\big(\mathcal{J}_{1}^{[1:K]}, \mathcal{J}_{2}^{[1:K]},X^n_3\big) \right) \right] \\
	&=& 	\frac1n \sum_{t=1}^n \bE\left[ d\left (X_{2[t]},\hat{X}_{32[t]}\left(U_{2\rightarrow 13[t]},U_{2\rightarrow 3[t]},X_{3[t]}\right) \right) \right] \\
	&=& 	\frac1n \sum_{t=1}^n \bE\left[ d\left (X_{2[Q]}, \hat{X}_{32[Q]}\left(U_{2\rightarrow 13[Q]},U_{2\rightarrow 3[Q]},X_{3[Q]}\right) \right)\ \middle|\ Q=t \right] \\
	&=& 	\bE\left[ d\left (X_{2[Q]}, \hat{X}_{32[Q]}\left(U_{2\rightarrow 13[Q]},U_{2\rightarrow 3[Q]},X_{3[Q]}\right) \right) \right] \\
	&=& 	\bE\left[ d\left (X_2, \widetilde{X}_{32}\left(\widetilde{U}_{2\rightarrow 13},\widetilde{U}_{2\rightarrow 3},X_3\right) \right) \right] \ ,
\end{IEEEeqnarray}
where we defined function $\widetilde{X}_{32}$ by 
\begin{IEEEeqnarray}{rCl}
\widetilde{X}_{32}\left(\widetilde{U}_{2\rightarrow 13},\widetilde{U}_{2\rightarrow 3},X_3\right)& =& \widetilde{X}_{32}\left(Q, U_{2\rightarrow 13[Q]},U_{2\rightarrow 3[Q]}, X_{3[Q]}\right)\nonumber  \\
&\triangleq & \hat{X}_{32[Q]}\left(U_{2\rightarrow 13[Q]},U_{2\rightarrow 3[Q]},X_{3[Q]} \right) \ .
\end{IEEEeqnarray}
This concludes the proof of the converse and thus that of the theorem. 
\end{IEEEproof}

\subsection{Two encoders and three decoders subject to lossless/lossy reconstruction constraints, reversal delivery and side information}

\begin{figure}[th!]to emphasize 
\centering
\includegraphics[angle=0,width=0.7\columnwidth,keepaspectratio,trim= 0mm 0mm 0mm 0mm,clip]{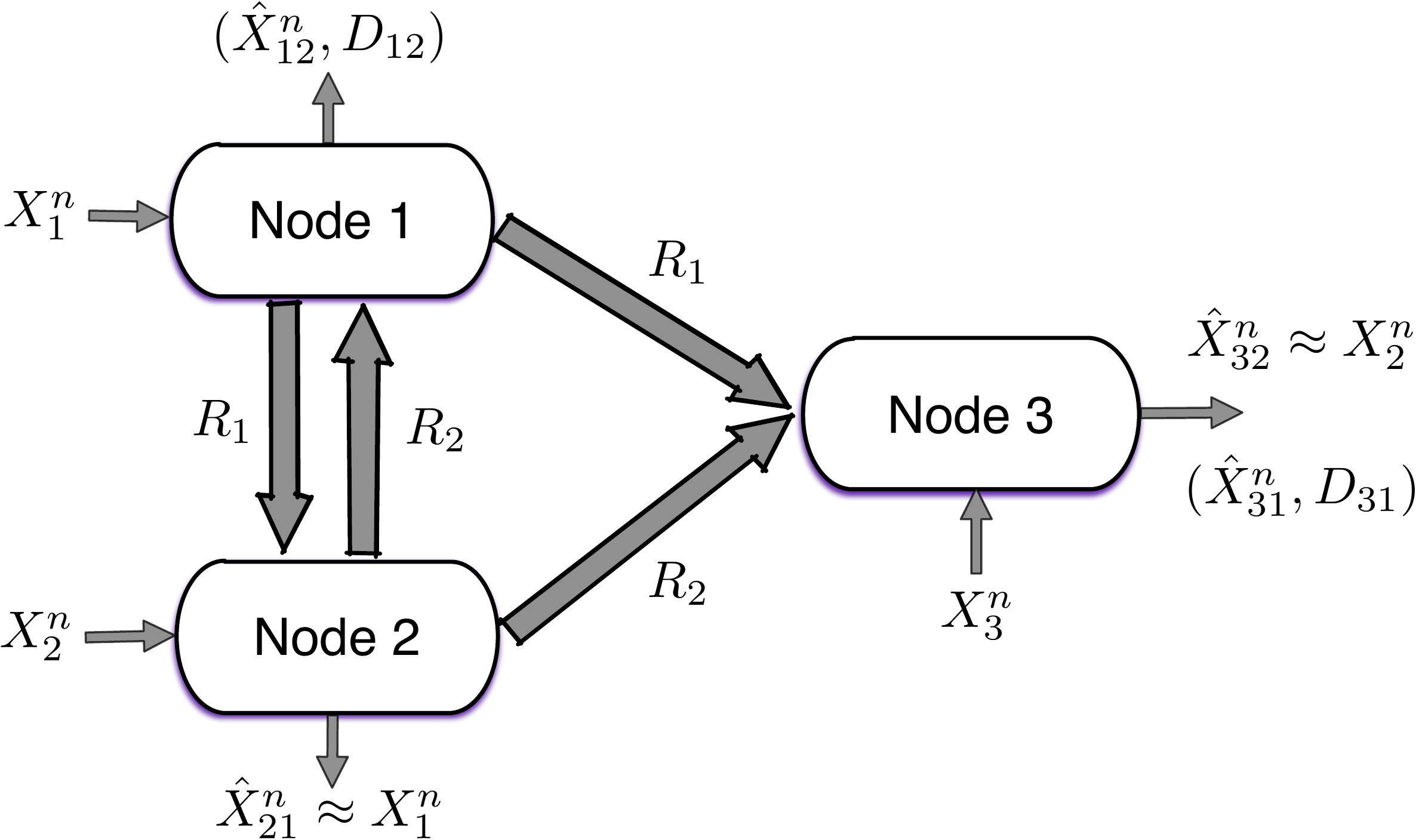} 
\caption{Two encoders and three decoders subject to lossless/lossy reconstruction constraints, reversal delivery and side information.}
\label{fig:model-Coop-BT3}
\end{figure}

Consider now the problem described in Fig.~\ref{fig:model-Coop-BT3} where encoder $1$ wishes to communicate the \emph{lossless} the source $X_1^n$ to node $2$ and a lossy description to node $3$. Encoder $2$ wishes to send a \emph{lossy} description of its source $X_2^n$ to node $1$ and a lossless one to node $3$. The corresponding distortion at node $1$ and $3$  are $D_{12}$ and $D_{31}$, respectively. In addition to this, the encoders accomplish the communication using $K$ communication rounds.  This problem is very similar to the problem described in Fig.~\ref{fig:model-Coop-BT2}, with the difference that the decoding at node $3$ is inverted. 

\vspace{1mm}
 \begin{theorem}\label{theo-Coop-BT3}
The rate-distortion region of the setting described in Fig.~\ref{fig:model-Coop-BT3} is given by the union over all joint probability measures $p_{X_1X_2X_3U_{2\rightarrow 13}}$ satisfying the Markov chain
\begin{equation}
U_{2\rightarrow 13}\mkv (X_1,X_2) \mkv  X_3
\label{eq:two_dist_crit_mark2}
\end{equation}
and such that there exists reconstruction mappings:
\begin{IEEEeqnarray}{rcl}
g_{31} \left(X_2,X_3, U_{2\rightarrow 13}\right)&=&\hat{X}_{31}\,\,\,\, \textrm{ with }\,\,\,\,\mathds{E} \left[d(X_1,\hat{X}_{31}) \right] \leq  D_{31}\ ,\\
g_{12} \left( X_1,  U_{2\rightarrow 13} \right)&=&\hat{X}_{12}\,\,\,\, \textrm{ with }\,\,\,\,\mathds{E} \left[d(X_2,\hat{X}_{12}) \right] \leq  D_{12}\ ,
\end{IEEEeqnarray} 
of the set of all tuples satisfying: 
\begin{IEEEeqnarray}{rcl}
R_1 &\geq& H(X_1|X_2) \ ,\\
R_2 &\geq& I(U_{2\rightarrow 13};X_2 |X_1) + H(X_2| U_{2\rightarrow 13}X_1 X_3) \ ,\\
R_1+R_2 &\geq& H(X_1X_2|X_3)\ .
\end{IEEEeqnarray}
The auxiliary random variable has cardinality bounds: $\| \mathcal{U}_{2\rightarrow 13}\|\leq \|\mathcal{X}_1\|\|\mathcal{X}_2\|+3$.
\end{theorem}
\vspace{1mm}
\begin{remark}
Notice that the rate-distortion region in Theorem~\ref{theo-Coop-BT3} is achievable with a single round  of interactions $K=1$, which implies that multiple rounds do not improve the rate-distortion region in this case. 
\end{remark}
\vspace{1mm}
\begin{remark}
Notice that, although node $3$ requires only the lossy recovery of $X_1$, it can in fact recover $X_1$ perfectly. That is, as node 3 requires the perfect recovery of $X_2$, it has the same information that node 2 who recover $X_1$ perfectly. This explains the sum-rate term. We also see, that the cooperation helps in the Wyner-Ziv problem that exists between node 2 and 1, with an increasing of the optimization region thanks to the Markov chain (\ref{eq:two_dist_crit_mark2}).
\end{remark}

\vspace{1mm}
\begin{IEEEproof}
The direct part of the proof follows by choosing:
\begin{IEEEeqnarray*}{rcl}
U_{3\rightarrow 12,l}&=&U_{1\rightarrow 3,l}=U_{1\rightarrow 2,l}=U_{
  2\rightarrow 1,l}=U_{3\rightarrow 1,l}=U_{3\rightarrow 2,l}=\varnothing,\,\, \forall l\\
U_{1\rightarrow 23} &\equiv&  U_{1\rightarrow 23,1} = X_1 \ , U_{2\rightarrow 3} \equiv  U_{2\rightarrow 3,1} = X_2 \,\,\,\,\, U_{1\rightarrow 23,l} = U_{2\rightarrow 13,l}=U_{2\rightarrow 3,l} =\varnothing\ \forall\,\,\, l>1 
\end{IEEEeqnarray*}
and with $U_{2\rightarrow 13,1}\equiv U_{2\rightarrow 13}$ auxiliary random variable that according to Theorem~\ref{theo-main-theorem} should satisfy:
\begin{equation}
U_{2\rightarrow 13} \mkv (X_1,X_2) \mkv  X_3\ .
\end{equation}
 From the rate equations in Theorem~\ref{theo-main-theorem}, and the above choices for the auxiliary random variables we obtain:
\begin{IEEEeqnarray}{rll}
R_{1\rightarrow 23}&>& H(X_1|X_2)\ ,\\
R_{2\rightarrow 13}&>&\max{\left\{I(X_2;U_{2\rightarrow 13}|X_1),I(X_2;U_{2\rightarrow 13}|X_1X_3)\right\}}\\&=&I(X_2;U_{2\rightarrow 13}|X_1)\ ,\\
R_{1\rightarrow 23}+R_{2\rightarrow 13}&>&H(X_1|X_3)+I(X_2;U_{2\rightarrow 13}|X_1X_3)\ ,\\
R_{2\rightarrow 3}&>&H(X_2|U_{2\rightarrow 13}X_1X_3)\ .
\end{IEEEeqnarray}
Noticing that $R_1\equiv R_{1\rightarrow 23}$ and $R_2\equiv R_{2\rightarrow 13}+R_{2\rightarrow 3}$ the rate-distortion region~\eqref{eq-inner-bound} reduces to the desired region in Theorem~\ref{theo-Coop-BT3}, where for simplicity we dropped the round index. We now proceed to the proof of the converse. 

If a pair of rates $(R_1,R_2)$  and distortions $(D_{12},D_{31})$ are admissible for the $K$-steps interactive cooperative distributed source coding setting described in Fig.~\ref{fig:model-Coop-BT3}, then  for all $\varepsilon >0$ there exists $n_0(\varepsilon,K)$, such that $\forall\,n>n_0(\varepsilon,K)$  there exists a $K$-steps interactive source code $(n,K,\mathcal{F},\mathcal{G})$ with intermediate rates satisfying:  
\begin{equation}
\frac{1}{n}\sum_{l=1}^K\log{\|\mathcal{J}_{i}^l\|}\leq R_i+\varepsilon\ ,\ i\in\{1,2\}
 \end{equation} 
and with reconstruction constraints: 
\begin{IEEEeqnarray}{rcl}
&&\mathds{E} \left[d(X_1^n,\hat{X}^n_{31})\right]\leq D_{31}+\varepsilon\ ,\label{eq-converse-BT3-distortion2}\\
&&\mbox{Pr}\left(  X_1^n\neq \hat{X}_{21}^n\right)\leq \varepsilon\ ,\label{eq-converse-BT3-error3}\\
&&\mathds{E} \left[d(X_2^n,\hat{X}^n_{12})\right]\leq D_{12}+\varepsilon\ ,\label{eq-converse-BT3distortion3}\\
&&\mbox{Pr}\left(  X_2^n\neq \hat{X}_{32}^n\right)\leq \varepsilon\ ,\label{eq-converse-BT3-error2}
\end{IEEEeqnarray}
where 
\begin{IEEEeqnarray}{rCl}
\hat{X}_{32}^n &\equiv & g_{32} \left(\mathcal{J}_{1}^{[1:K]}, \mathcal{J}_{2}^{[1:K]}, X_3^n\right)\ ,\,\,\,\,\,
\hat{X}_{12}^n  \equiv  g_{12}\left( \mathcal{J}_{2}^{[1:K]}, X_1^n\right)\ ,\label{eq-BT3-12} \\
\hat{X}_{31}^n & \equiv & g_{31}\left(\mathcal{J}_{1}^{[1:K]}, \mathcal{J}_{2}^{[1:K]}, X_3^n\right)\ ,\,\,\,\,\,
\hat{X}_{21}^n  \equiv  g_{21}\left( \mathcal{J}_{1}^{[1:K]}, X_2^n\right)\ . \label{eq-BT3-21}
\end{IEEEeqnarray}

For each $t\in\{1,\dots,n\}$, define random variables $U_{2\rightarrow 13[t]}$ as follows:
\begin{IEEEeqnarray}{rCl}
U_{2\rightarrow 13[t]}&\triangleq & \left(\mathcal{J}_{1}^{[1:K]}, \mathcal{J}_{2}^{[1:K]}, X_{1[1:t-1]},X_{1[t+1:n]}, X_{3[1:t-1]}\right) \label{eq:MarkovBT2-1} \ .
\end{IEEEeqnarray}
Using point 6) in Lemma \ref{lemma:Markov_interactive_encoding_1}  we can see that this choice satisfies (\ref{eq:two_dist_crit_mark2}). By the conditions~\eqref{eq-converse-BT3-error3} and~\eqref{eq-converse-BT3-error2}, and Fano's inequality, we have 
\begin{IEEEeqnarray}{rCl}
H(X_1^n| \hat{X}_{21}^n)  &\leq &\mbox{Pr}\left(  X_1^n\neq \hat{X}_{21}^n\right) \log_2 (\|\mathcal{X}_1^n \|-1) + H_2\left( \mbox{Pr}\big(  X_1^n\neq \hat{X}_{21}^n\big)  \right)  \triangleq    n \epsilon_n   \ ,\label{eq-converse-BT3-fano2-2}\\
H(X_2^n| \hat{X}_{32}^n)  &\leq &\mbox{Pr}\left(  X_2^n\neq \hat{X}_{32}^n\right) \log_2 (\|\mathcal{X}_2^n \|-1) + H_2\left( \mbox{Pr}\big(  X_2^n\neq \hat{X}_{32}^n\big)  \right)  \triangleq    n \epsilon_n   \ , \label{eq-converse-BT3-fano2-1}
\end{IEEEeqnarray}
where $\epsilon_n(\varepsilon) \rightarrow 0$ provided that $\varepsilon\rightarrow 0$ and $n \rightarrow \infty$. 

\subsubsection{Rate at node 1}

For the first rate, from cut-set arguments similar to the ones used in Theorem \ref{theo-Coop-BT2} and Fano inequality, we can easily obtain:
\begin{equation}
n(R_1+\varepsilon)\geq n\left[ H(X_1| X_2) -  \epsilon_n\right] 
\end{equation}

\subsubsection{Rate at node 2}
For the second rate, we have
\begin{IEEEeqnarray}{rCl}
n(R_2+\varepsilon)
	&\geq& 				H\left(\mathcal{J}_{2}^{[1:K]}\right) \\
	&\stackrel{(a)}{=}&		I\left(\mathcal{J}_{2}^{[1:K]}; X_1^nX_2^nX_3^n\right)\\
	&\stackrel{(b)}{\geq }&		I\left(\mathcal{J}_{2}^{[1:K]}; X_2^nX_3^n|X_1^n\right)\\	
 	&\stackrel{(c)}{= }&		I\left(\mathcal{J}_{1}^{[1:K]}\mathcal{J}_{2}^{[1:K]}; X_2^nX_3^n|X_1^n\right)\\	
	&= &		I\left(\mathcal{J}_{1}^{[1:K]}\mathcal{J}_{2}^{[1:K]};  X_3^n|X_1^n\right) + I\left(\mathcal{J}_{1}^{[1:K]}\mathcal{J}_{2}^{[1:K]};  X_2^n|X_1^nX_3^n\right) \\
	&\stackrel{(d)}{=}&  
	 \sum_{t=1}^n \Big[I\left(\mathcal{J}_{1}^{[1:K]}\mathcal{J}_{2}^{[1:K]} ; X_{3[t]} |  X_1^n, X_{3[1:t-1]} \right) \nonumber \\
	   \IEEEeqnarraymulticol{3}{c}{ + I\left(\mathcal{J}_{1}^{[1:K]}\mathcal{J}_{2}^{[1:K]}  ; X_{2[t]} | X_1^nX_3^nX_{2[1:t-1]} \right)\Big] }\\
	&\stackrel{(e)}{=}&   \sum_{t=1}^n \Big[ I\left(\mathcal{J}_{1}^{[1:K]}\mathcal{J}_{2}^{[1:K]} X_{1[1:t-1]}X_{1[t+1:n]}X_{3[1:t-1]}; X_{3[t]} |  X_{1[t]} \right) \nonumber \\
	 &+&  I\left(\mathcal{J}_{1}^{[1:K]}\mathcal{J}_{2}^{[1:K]}  X_{1[1:t-1]} X_{1[t+1:n]} X_{3[1:t-1]} X_{3[t+1:n]} X_{2[1:t-1]} ; X_{2[t]} | X_{1[t]}X_{3[t]} \right)	\Big]\,\,\,\,\,\,\,\, \\
		&\stackrel{(f)}{=}&   \sum_{t=1}^n \Big[ I\left(U_{2\rightarrow 13[t]}; X_{3[t]} |  X_{1[t]} \right)+
		 I\left(U_{2\rightarrow 13[t]} X_{3[t+1:n]} X_{2[1:t-1]}; X_{2[t]} |  X_{1[t]} X_{3[t]} \right)\nonumber	\\
	  &=&   \sum_{t=1}^n \Big[ I\left(U_{2\rightarrow 13[t]} ; X_{2[t]} X_{3[t]} |  X_{1[t]} \right)\nonumber\\
	    \IEEEeqnarraymulticol{3}{r}{   +  I\left( X_{3[t+1:n]} X_{2[1:t-1]}; X_{2[t]} | X_{1[t]}X_{3[t]}U_{2\rightarrow 13[t]} \right)	 \Big]}\,\,\,\,\,\,\,\, \\
	  	  &\stackrel{(g)}{=}&   \sum_{t=1}^n \Big[ I\left(U_{2\rightarrow 13[t]}; X_{2[t]}  |  X_{1[t]} \right) + H\left(X_{2[t]} | X_{1[t]}X_{3[t]} U_{2\rightarrow 13[t]}\right)\nonumber\\
	  	 \IEEEeqnarraymulticol{3}{c}{  	-  H\left( X_{2[t]} | X_{1[t]}X_{3[t:n]} U_{2\rightarrow 13[t]}X_{2[1:t-1]}\right)\Big]}\\
	  	 &\stackrel{(h)}{\geq}&   \sum_{t=1}^n \Big[ I\left(U_{2\rightarrow 13[t]}; X_{2[t]}  |  X_{1[t]} \right)
	  	 			 +   H\left(X_{2[t]} | X_{1[t]}X_{3[t]} U_{2\rightarrow 13[t]}\right)-\epsilon_n	 \Big] \\
			  &\stackrel{(i)}{=}&   \sum_{t=1}^n \Big[ I\left(U_{2\rightarrow 13[Q]}; X_{2[Q]}  |  X_{1[Q]} ,Q=t \right)\nonumber\\
			   \IEEEeqnarraymulticol{3}{c}{  +   H\left(X_{2[Q]} | X_{1[Q]}X_{3[Q]} U_{2\rightarrow 13[Q]} ,Q=t \right)-\epsilon_n\Big] }\\
&\stackrel{(j)}{\geq }&  n\Big[ I\left(\widetilde{U}_{2\rightarrow 13}; X_{2}  |  X_{1} \right) +  H\left( X_{2} | X_{1}X_{3} \widetilde{U}_{2\rightarrow 13}\right)	-\epsilon_n \Big]\ ,
\end{IEEEeqnarray}
where
\begin{itemize}
\item step~$(a)$ follows from the fact that $\mathcal{J}_{2}^{[1:K]}$ is a function of the sources $(X_1^n,X_2^n)$, 
\item step~$(b)$ follows  from the non-negativity of mutual information, 
\item step~$(c)$ follows  from the fact that $\mathcal{J}_{1}^{[2:K]}$ is a function of $\mathcal{J}_{2}^{[1:K]}$ and the source $X_1^n$, 
\item step~$(d)$ follows  from the chain rule for conditional mutual information,
\item step~$(e)$ follows  from the memoryless property across time of the sources $(X_1^n,X_2^n,X_3^n)$, 
\item step~$(f)$  follows  from the definition~\eqref{eq:MarkovBT2-1} 
\item step~$(g)$ follows from the Markov chain $U_{2\rightarrow 13[t]} \mkv (X_{1[t]},X_{2[t]}) \mkv X_{3[t]} $, for all $t\in\{1,\dots,n\}$ and the usual decomposition of mutual information,
\item step~$(h)$ follows from the fact that $\mbox{Pr}\left(  X_{2[t]}\neq \hat{X}_{32[t]}\right)\leq \epsilon\ \forall  t\in\{1,\dots,n\}$ , $\hat{X}_{32}[t]\equiv g_{32[t]}(U_{2\rightarrow 13[t]},X_{3[t:n]})$ and Fano inequality.
\item step~$(i)$ follows from the use of a time sharing random variable $Q$ uniformly distributed over the set $\{1,\dots,n\}$, 
\item step~$(j)$ follows by defining new random variables $\widetilde{U}_{2\rightarrow 13} \triangleq (U_{2\rightarrow 13[Q]},Q)$.
\end{itemize}

\subsubsection{Sum-rate of nodes $1$ and $2$}
From cut-set arguments and Fano inequality, we can easily obtain:
\begin{equation}
n(R_1+R_2+2\varepsilon)\geq n\left[ H(X_1| X_2X_3) -  \epsilon_n\right] 
\end{equation}

\subsubsection{Distortion at node 1}

Node $1$ reconstructs a lossy  $\hat{X}_{12}^n \equiv g_{12}\left( \mathcal{J}_{2}^{[1:K]},X_1^n\right)$. For each $t\in\{1,\dots,n\}$, define a function $\hat{X}_{12[t]}$ as the $t$-th coordinate of this estimate:
\begin{IEEEeqnarray}{rCl}
\hat{X}_{12[t]}\left(U_{2\rightarrow 13[t]}, X_{1[t]} \right) \triangleq   g_{12[t]} \left( \mathcal{J}_{2}^{[1:K]},X_1^n \right) \ .
\end{IEEEeqnarray}
The component-wise mean distortion thus verifies
\begin{IEEEeqnarray}{rCl}
D_{12} + \varepsilon
	&\geq&	\bE\left[ d\left (X_2^n,g_{12}\big( \mathcal{J}_{2}^{[1:K]},X_1^n \big) \right) \right] \\
	&=& 	\frac1n \sum_{t=1}^n \bE\left[ d\left (X_{2[t]},\hat{X}_{12[t]}\left(U_{2\rightarrow 13[t]},X_{1[t]}\right) \right) \right] \\
	&=& 	\frac1n \sum_{t=1}^n \bE\left[ d\left (X_{2[Q]}, \hat{X}_{12[Q]}\left(U_{2\rightarrow 13[Q]},X_{1[Q]}\right) \right)\ \middle|\ Q=t \right] \\
	&=& 	\bE\left[ d\left (X_{2[Q]}, \hat{X}_{12[Q]}\left(U_{2\rightarrow 13[Q]},X_{1[Q]}\right) \right) \right] \\
	&=& 	\bE\left[ d\left (X_2, \widetilde{X}_{12}\left(\widetilde{U}_{2\rightarrow 13},X_{1}\right) \right) \right] \ ,
\end{IEEEeqnarray}
where we defined function $\widetilde{X}_{12}$ by 
\begin{IEEEeqnarray}{rCl}
\widetilde{X}_{12}\left(\widetilde{U}_{2\rightarrow 13},X_{1}\right) = \widetilde{X}_{12}\left(Q, U_{2\rightarrow 13[Q]},X_{1[Q]}\right) \triangleq \hat{X}_{12[Q]}\left(U_{2\rightarrow 13[Q]},X_{1[Q]}\right) \ .
\end{IEEEeqnarray}

\subsubsection{Distortion at node 3}
Node $3$ reconstructs a  lossy description $\hat{X}_{31}^n \equiv g_{31}\left(\mathcal{J}_{1}^{[1:K]}, \mathcal{J}_{2}^{[1:K]},X^n_3\right)$. For each $t\in\{1,\dots,n\}$, define a function $\hat{X}_{31[t]}$ as:
\begin{IEEEeqnarray}{rCl}
\bar{X}_{31[t]}\left(U_{2\rightarrow 13[t]},\hat{X}_{32[t]},X_{3[t]},X_{3[t+1:n]}\right) \triangleq g_{31[t]} \left(\mathcal{J}_{1}^{[1:K]} , \mathcal{J}_{2}^{[1:K]},X^n_3\right)\ t\in\{1,\dots,n\}.
\end{IEEEeqnarray}
This can be done because $\hat{X}_{32[t]}$ is also a function of $ \left(\mathcal{J}_{1}^{[1:K]} , \mathcal{J}_{2}^{[1:K]},X^n_3\right)$.
The component-wise mean distortion thus verifies
\begin{IEEEeqnarray}{rCl}
D_{32} + \varepsilon
	&\geq&	\bE\left[ d\left (X_1^n,g_{31}\big(\mathcal{J}_{1}^{[1:K]}, \mathcal{J}_{2}^{[1:K]},X^n_3\big) \right) \right] \\
	&=& \frac{1}{n} \sum_{t=1}^n \bE\left[ d\left (X_{1[t]},\bar{X}_{31[t]}\left(U_{2\rightarrow 13[t]},\hat{X}_{32[t]},X_{3[t]},X_{3[t+1:n]}\right) \right) \right] \\
	&\stackrel{(a)}{=}& \frac{1}{n} \sum_{t=1}^n \bE\left[ d\left (X_{1[t]},\bar{X}_{31[t]}\left(U_{2\rightarrow 13[t]},X_{2[t]},X_{3[t]},X_{3[t+1:n]}\right) \right) \right] \\
& \stackrel{(b)}{\geq}&	\frac{1}{n} \sum_{t=1}^n\bE\left[ d\left (X_{1[t]},\hat{X}_{31[t]}\left(U_{2\rightarrow 13[t]},X_{2[t]},X_{3[t]}\right)\right)\right] \\
		&=& 	\frac{1}{n} \sum_{t=1}^n \bE\left[ d\left (X_{1[Q]}, \hat{X}_{31[Q]}\left(U_{2\rightarrow 13[Q]},X_{2[Q]},X_{3[Q]}\right) \right)\ \middle|\ Q=t \right] \\
	&=& 	\bE\left[ d\left (X_{1[Q]}, \hat{X}_{32[Q]}\left(U_{2\rightarrow 13[Q]},X_{2[Q]},X_{3[Q]}\right) \right) \right] \\
	&=& 	\bE\left[ d\left (X_1, \widetilde{X}_{32}\left(\widetilde{U}_{2\rightarrow 13},X_{2},X_3\right) \right) \right] \ ,
\end{IEEEeqnarray}
where
\begin{itemize}
\item step~$(a)$ follows from the fact that $\bar{X}_{31[t]}\left(U_{2\rightarrow 13[t]},\hat{X}_{32[t]},X_{3[t]},X_{3[t+1:n]}\right)$ can be trivially expressed as a function of  $\left(U_{2\rightarrow 13[t]},X_{2[t]},X_{3[t]},X_{3[t+1:n]}\right)$ as follows:
\[\bar{X}_{31[t]}\left(U_{2\rightarrow 13[t]}X_{2[t]}X_{3[t]}X_{3[t+1:n]}\right)=\left\{\begin{array}{cc}
\bar{X}_{31[t]}\left(U_{2\rightarrow 13[t]}X_{2[t]}X_{3[t]}X_{3[t+1:n]}\right) & \mbox{if}\ X_{2[t]}=\hat{X}_{32[t]}\\
\bar{X}_{31[t]}\left(U_{2\rightarrow 13[t]}\hat{X}_{32[t]}X_{3[t]}X_{3[t+1:n]}\right) & \mbox{if}\ X_{2[t]}\neq\hat{X}_{32[t]}
\end{array}\right.\]
\item step~$(b)$ follows from the fact that  $X_{1[t]}\mkv (U_{2\rightarrow 13[t]}X_{2[t]}X_{3[t]})\mkv X_{3[t+1:n]}$, which implies that for all $t\in\{1,\dots,n\}$ exists $\hat{X}_{31[t]}\left(U_{2\rightarrow 13[t]},X_{2[t]},X_{3[t]}\right)$ such that 
\[\bE\left[ d\left (X_{1[t]},\hat{X}_{31[t]}\left(U_{2\rightarrow 13[t]}X_{2[t]}X_{3[t]}\right)\right)\right] \leq \bE\left[ d\left (X_{1[t]},\bar{X}_{31[t]}\left(U_{2\rightarrow 13[t]}X_{2[t]}X_{3[t]}X_{3[t+1:n]}\right) \right) \right]\]
We also defined function $\widetilde{X}_{32}$ by 
\begin{IEEEeqnarray}{rCl}
\widetilde{X}_{32}\left(\widetilde{U}_{2\rightarrow 13},X_{2},X_3\right)& =& \widetilde{X}_{32}\left(Q, U_{2\rightarrow 13[Q]},X_{2[Q]}, X_{3[Q]}\right)\nonumber  \\
&\triangleq & \hat{X}_{32[Q]}\left(U_{2\rightarrow 13[Q]},X_{2[Q]},X_{3[Q]} \right) \ .
\end{IEEEeqnarray}
\end{itemize}
This concludes the proof of the converse and thus that of the theorem. 
\end{IEEEproof}

\subsection{Two encoders and three decoders subject to lossy reconstruction constraints with degraded side information} 

\begin{figure}[th!]
\centering
\includegraphics[angle=0,width=0.7\columnwidth,keepaspectratio,trim= 0mm 0mm 0mm 0mm,clip]{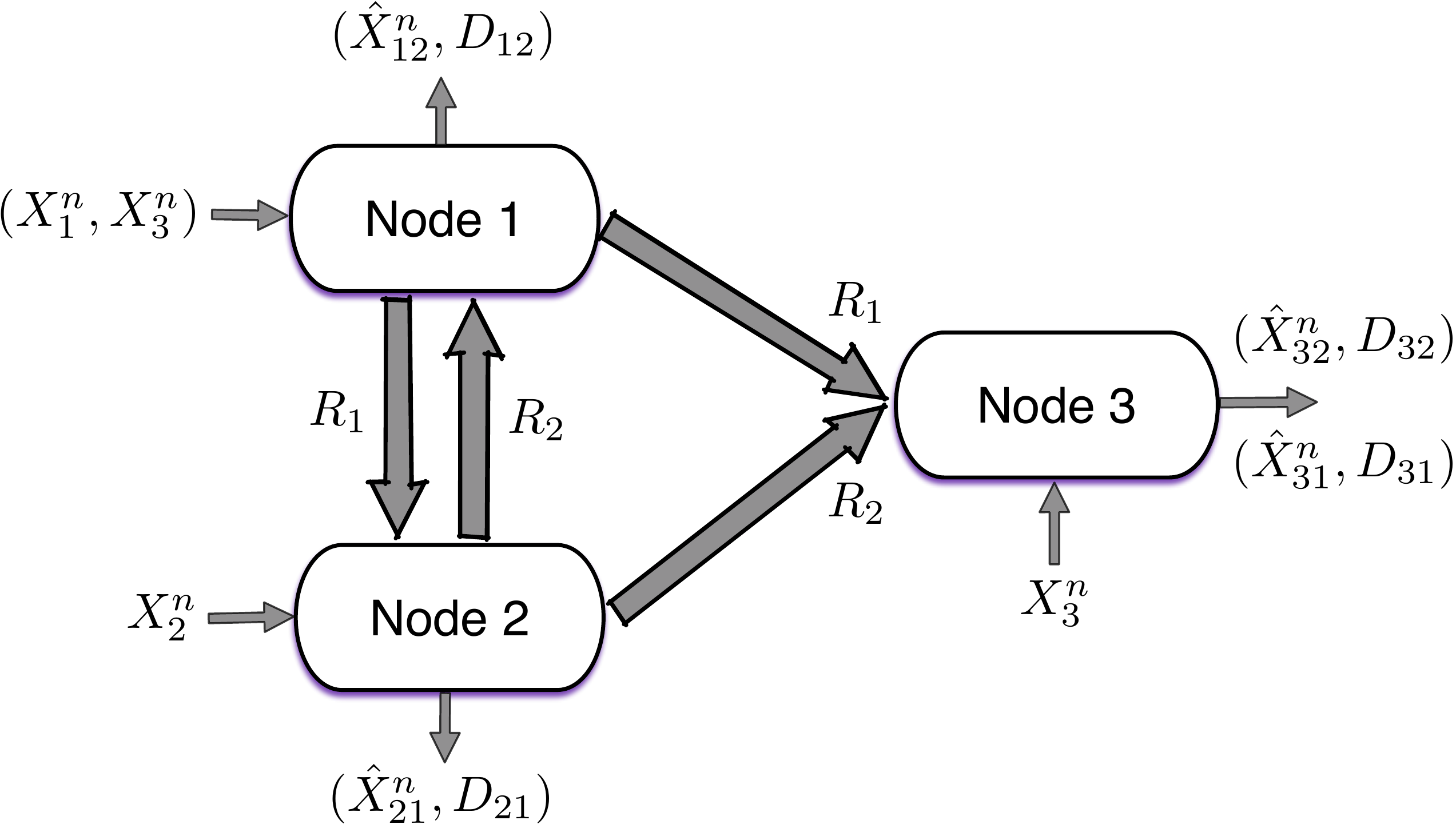} 
\caption{Two encoders and three decoders subject to lossy reconstruction constraints with degraded side information.}
\label{fig:model-Coop-BT5}
\end{figure}

Consider now the problem described in Fig.~\ref{fig:model-Coop-BT5} where encoder $1$ has access to $X_1$ and $X_3$ and wishes to communicate a \emph{lossy} description of $X_1$ to nodes $2$ and $3$  with distortion constraints $D_{21}$ and $D_{31}$, while encoder $2$ wishes to send a \emph{lossy} description of its source $X_2^n$ to nodes $1$ and $3$ with distortion constraints $D_{12}$ and $D_{32}$. In addition to this, the encoders overhead the communication using $K$ communication rounds. This problem can be seen as a generalization of the settings previously investigated in~\cite{gastpar_wyner-ziv_2004}.  This setup is motivated by the following application. Consider that node 1 transmits  a probing signal $X_3$ which is used to explore a spatial  region (i.e. a radar transmitter). After transmission of this probing signal, node 1 measures the response ($X_1$) at its location. Similarly, in a different location node 2 measures the response $X_2$. Responses $X_1$ and $X_2$ have to be sent to node 3 (e.g. the fusion center) which has knowledge of the probing signal $X_3$ and wants to reconstruct a lossy  estimate of them. Nodes 1 and 2 cooperate through multiple rounds to accomplish this task. 

\vspace{1mm}
 \begin{theorem}\label{theo-Coop-BT5}
The rate-distortion region of the setting described in Fig.~\ref{fig:model-Coop-BT5} where $X_1 \mkv  X_3 \mkv  X_2$ form a Markov chain  is given by the union over all joint probability measures $p_{X_1X_2X_3\mathcal{W}_{[1,K+1]} U_{1\rightarrow 3,K}}$ satisfying the following Markov chains:
\begin{IEEEeqnarray}{rCl}
U_{1\rightarrow 23,l}  \mkv (X_1,X_3,\mathcal{W}_{[1,l]} ) \mkv  X_2 \ , \label{eq:two_dist_crit_mark5-1}\\
U_{2\rightarrow 13,l}  \mkv (X_2,\mathcal{W}_{[2,l]} ) \mkv  (X_1,X_3) \ , \label{eq:two_dist_crit_mark5-2}\\
U_{1\rightarrow 3,K}  \mkv (X_1,X_3,\mathcal{W}_{[2,K]} ) \mkv  X_2 \ , \label{eq:two_dist_crit_mark5-3}
\end{IEEEeqnarray}
for all $l=[1:K]$,  and such that there exist reconstruction mappings:
\begin{IEEEeqnarray*}{rCl}
g_{12} \left( X_1, X_3,U_{1\rightarrow 3,K} ,\mathcal{W}_{[1,K+1]}\right)&=&\hat{X}_{12}\,\,\,\, \textrm{ with }\,\,\,\,\mathds{E} \left[d(X_2,\hat{X}_{12}) \right] \leq  D_{12}\ ,\\
g_{21} \left( X_2, \mathcal{W}_{[1,K+1]} \right)&=&\hat{X}_{21}\,\,\,\, \textrm{ with }\,\,\,\,\mathds{E} \left[d(X_1,\hat{X}_{21}) \right] \leq  D_{21}\ ,\\
g_{31} \left( X_3, \mathcal{W}_{[1,K+1]} ,U_{1\rightarrow 3,K}   \right)&=&\hat{X}_{31}\,\,\,\, \textrm{ with }\,\,\,\,\mathds{E} \left[d(X_1,\hat{X}_{31}) \right] \leq  D_{31}\ ,\\
g_{32} \left( X_3, \mathcal{W}_{[1,K+1]} ,U_{1\rightarrow 3,K}   \right)&=&\hat{X}_{32}\,\,\,\, \textrm{ with }\,\,\,\,\mathds{E} \left[d(X_2,\hat{X}_{32}) \right] \leq  D_{32}\ ,
\end{IEEEeqnarray*}
with $\mathcal{W}_{[1,l]}=\left\{U_{1\rightarrow 23,l},U_{2\rightarrow 13,l}\right\}_{k=1}^{l-1}$for all $l=[1:K]$,\footnote{Notice that $U_{3\rightarrow 12,l}=\varnothing$ for all $l$ because $R_3=0$.}
of the set of all tuples satisfying: 
\begin{IEEEeqnarray}{rCl}
R_1 &\geq& I(\mathcal{W}_{[1,K+1]}  ;X_1X_3 |X_2) + I(U_{1\rightarrow 3,K};X_1| \mathcal{W}_{[1,K+1]} X_3) \ ,\\
R_2 &\geq& I(\mathcal{W}_{[1,K+1]} ;X_2 |X_3) \ .
\end{IEEEeqnarray}
The auxiliary random variables have cardinality bounds: 
\begin{equation}
\| \mathcal{U}_{1\rightarrow 23,l}\|\leq \|\mathcal{X}_1\|\|\mathcal{X}_3\|\prod_{i=1}^{l-1}\| \mathcal{U}_{1\rightarrow 23,i}\|\| \mathcal{U}_{2\rightarrow 13,i}\|+1,\ \ l\in[1:K]
\end{equation}
\begin{equation}
\| \mathcal{U}_{2\rightarrow 13,l}\|\leq \|\mathcal{X}_2\|\| \mathcal{U}_{1\rightarrow 23,l}\|\prod_{i=1}^{l-1}\| \mathcal{U}_{1\rightarrow 23,i}\|\| \mathcal{U}_{2\rightarrow 13,i}\|+1,\ \ l\in[1:K]
\end{equation}
\begin{equation}
\| \mathcal{U}_{1\rightarrow 3,l}\|\leq \|\mathcal{X}_1\|\|\mathcal{X}_3\|\prod_{i=1}^{K}\| \mathcal{U}_{1\rightarrow 23,i}\|\| \mathcal{U}_{2\rightarrow 13,i}\|+3.
\end{equation}

\end{theorem}
\begin{remark}
Notice that multiple rounds are needed to achieve the rate-distortion region in Theorem~\ref{theo-Coop-BT5}. It is worth to mention that first encoders $1$ and $2$ cooperate over the $K$ rounds while on the last round only node 1 send a private description to node 3. Because of  the Markov chain assumed for the sources we observe the following:
\begin{itemize}
\item Only node 1 send a private description to node 3. This is due to the fact that node 3 has better side information than 2.
\item For the transmissions from node 2, both node 1 and 3 can  be thought as an unique node and there is not reason for node 2 to send a private description to node 1 or node 3.
\item Notice that the there is not sum-rate.  Node 3  recovers the descriptions generated at nodes 1 and 2 without resorting to \emph{joint-decoding}. That is, node 3 can recover the descriptions generated at nodes 1 and 2 separately and independently. 
\end{itemize}
\end{remark}

\begin{IEEEproof}
The direct part of the proof follows by choosing:
\begin{IEEEeqnarray*}{rcl}
U_{1\rightarrow 2,l}&=& U_{
  2\rightarrow 1,l}=U_{3\rightarrow 12,l}=U_{3\rightarrow 1,l}=U_{3\rightarrow 2,l}=U_{2\rightarrow 3,l} =\varnothing,\,\, \forall l\\
U_{1\rightarrow 3,l} &=& \varnothing\,\,\, l<K \ .
\end{IEEEeqnarray*}
and $U_{1\rightarrow 23,l}$ and $U_{2\rightarrow 13,l}$ and $U_{1\rightarrow 3,K}$ are auxiliary random variables that according to Theorem \ref{theo-main-theorem} should satisfy the Markov chains \eqref{eq:two_dist_crit_mark5-1}-\eqref{eq:two_dist_crit_mark5-3}. Cumbersome but straightforward calculations allows to obtain the desired results.
We now proceed to the proof of the converse. 

If a pair of rates $(R_1,R_2)$  and distortions $(D_{12},D_{21},D_{31},D_{32})$ are admissible for the $K$-steps interactive cooperative distributed source coding setting described in Fig.~\ref{fig:model-Coop-BT5}, then  for all $\varepsilon >0$ there exists $n_0(\varepsilon,K)$, such that $\forall\,n>n_0(\varepsilon,K)$  there exists a $K$-steps interactive source code $(n,K,\mathcal{F},\mathcal{G})$ with intermediate rates satisfying:  
\begin{equation}
\frac{1}{n}\sum_{l=1}^K\log{\|\mathcal{J}_{i}^l\|}\leq R_i+\varepsilon\ ,\ i\in\{1,2\}
 \end{equation} 
and with average per-letter distortions
\begin{IEEEeqnarray}{rcl}
&&\mathds{E} \left[d(X_1^n,\hat{X}^n_{21})\right] \leq D_{21}+\varepsilon\ ,\\
&&\mathds{E} \left[d(X_1^n,\hat{X}^n_{31})\right] \leq D_{31}+\varepsilon\ ,\\
&&\mathds{E} \left[d(X_2^n,\hat{X}^n_{12})\right] \leq D_{12}+\varepsilon\ ,\\
&&\mathds{E} \left[d(X_2^n,\hat{X}^n_{32})\right] \leq D_{32}+\varepsilon\ ,
\end{IEEEeqnarray}
where 
\begin{IEEEeqnarray}{rCl}
\hat{X}_{32}^n &\equiv & g_{32} \left(\mathcal{J}_{1}^{[1:K]}, \mathcal{J}_{2}^{[1:K]}, X_3^n\right)\ ,\,\,\,\,\,
\hat{X}_{12}^n  \equiv  g_{12}\left( \mathcal{J}_{1}^{[1:K]}, \mathcal{J}_{2}^{[1:K]}, X_1^n,X_3^n\right)\ ,\label{eq-BT2-12} \\
\hat{X}_{31}^n & \equiv & g_{31}\left(\mathcal{J}_{1}^{[1:K]}, \mathcal{J}_{2}^{[1:K]}, X_3^n\right)\ ,\,\,\,\,\,
\hat{X}_{21}^n  \equiv  g_{21}\left( \mathcal{J}_{1}^{[1:K]}, \mathcal{J}_{2}^{[1:K]}, X_2^n\right)\ . \label{eq-BT2-21}
\end{IEEEeqnarray}
For each $t\in\{1,\dots,n\}$, define random variables $(U_{1 \rightarrow 3,[t]},U_{2 \rightarrow 3,[t]})$ and the sequences of random variables $(U_{1\rightarrow 23,k,[t]},U_{2\rightarrow 13,k,[t]})_{k=[1:K]}$ as follows:
\begin{IEEEeqnarray}{rCl}
U_{1\rightarrow 23,1,[t]} &\triangleq & \left(\mathcal{J}_{1}^1, X_{3[1:t-1]},X_{2[t+1:n]} \right) \ , \label{eq-def-theo5-1} \\
U_{2\rightarrow 13,1,[t]} &\triangleq & \mathcal{J}_{2}^1\ , \label{eq-def-theo5-2} \\
U_{1\rightarrow 23,k,[t]} &\triangleq & \mathcal{J}_{1}^k \ , \textrm{ $\forall$ $k=[2:K]$} \ ,\label{eq-def-theo5-3} \\
U_{2\rightarrow 13,k,[t]} &\triangleq & \mathcal{J}_{2}^k   \ , \textrm{ $\forall$ $k=[2:K]$} \ , \label{eq-def-theo5-4}\\
U_{1 \rightarrow 3,K,[t]} &\triangleq & X_{3[t+1:n]} \ , \label{eq-def-theo5-5}
\end{IEEEeqnarray}
From Corollary~\ref{coro:Markov_interactive_encoding_1} in the Appendices we see that these choices satisfy equations (\ref{eq:two_dist_crit_mark5-1}), (\ref{eq:two_dist_crit_mark5-2}) and (\ref{eq:two_dist_crit_mark5-3}).  

\subsubsection{Rate at node 1}

For the first rate, we have
\begin{IEEEeqnarray}{rCl}
n(R_1+\varepsilon)
	&\geq& 				H\left(\mathcal{J}_{1}^{[1:K]}\right) \\
	&\stackrel{(a)}{=}&		I\left(\mathcal{J}_{1}^{[1:K]}; X_1^nX_2^nX_3^n\right)\\
	&\stackrel{(b)}{\geq }&		I\left(\mathcal{J}_{1}^{[1:K]}; X_1^nX_3^n|X_2^n\right)\\	
 	&\stackrel{(c)}{= }&		I\left(\mathcal{J}_{1}^{[1:K]}\mathcal{J}_{2}^{[1:K]}; X_1^nX_3^n|X_2^n\right)\\	
	&= &		 I\left(\mathcal{J}_{1}^{[1:K]}\mathcal{J}_{2}^{[1:K]};  X_1^n|X_2^nX_3^n\right) +I\left(\mathcal{J}_{1}^{[1:K]}\mathcal{J}_{2}^{[1:K]};  X_3^n|X_2^n\right)\\
	&\stackrel{(d)}{=}&  
	 \sum_{t=1}^n I\left(\mathcal{J}_{1}^{[1:K]} \mathcal{J}_{2}^{[1:K]} X_{2[1:t-1]}  X_{2[t+1:n]} X_{3[1:t-1]}  X_{3[t+1:n]}  X_{1[1:t-1]} ; X_{1[t]} | X_{2[t]} X_{3[t]} \right) \nonumber \\
&+&  \sum_{t=1}^n  I\left(\mathcal{J}_{1}^{[1:K]} \mathcal{J}_{2}^{[1:K]}   X_{3[1:t-1]} X_{2[1:t-1]}  X_{2[t+1:n]} ; X_{3[t]} |  X_{2[t]}  \right)  \\
 &\stackrel{(e)}{\geq}&    \sum_{t=1}^n I\left(\mathcal{J}_{1}^{[1:K]} \mathcal{J}_{2}^{[1:K]}  X_{2[t+1:n]} X_{3[1:t-1]}  X_{3[t+1:n]}  X_{1[1:t-1]} ; X_{1[t]} | X_{2[t]} X_{3[t]} \right) \nonumber \\
 &+&  \sum_{t=1}^n  I\left(\mathcal{J}_{1}^{[1:K]} \mathcal{J}_{2}^{[1:K]}   X_{3[1:t-1]}  X_{2[t+1:n]} ; X_{3[t]} |  X_{2[t]}  \right)  
  \end{IEEEeqnarray}
  \begin{IEEEeqnarray}{rCl}
 &=& \sum_{t=1}^n I\left(\mathcal{J}_{1}^{[1:K]} \mathcal{J}_{2}^{[1:K]}  X_{2[t+1:n]} X_{3[1:t-1]}  ; X_{1[t]} | X_{2[t]} X_{3[t]} \right) \nonumber \\
  &+&  \sum_{t=1}^n\Big[ I\left( X_{3[t+1:n]} X_{1[1:t-1]}  ; X_{1[t]} | \mathcal{J}_{1}^{[1:K]} \mathcal{J}_{2}^{[1:K]}  X_{2[t:n]} X_{3[1:t]} \right)\nonumber \\
  \IEEEeqnarraymulticol{3}{c}{+ I\left(\mathcal{J}_{1}^{[1:K]} \mathcal{J}_{2}^{[1:K]}   X_{3[1:t-1]}  X_{2[t+1:n]} ; X_{3[t]} |  X_{2[t]}  \right)  \Big]} \\  
&=& \sum_{t=1}^n I\left(\mathcal{J}_{1}^{[1:K]} \mathcal{J}_{2}^{[1:K]}  X_{2[t+1:n]} X_{3[1:t-1]}  ; X_{1[t]} X_{3[t]} | X_{2[t]} \right) \nonumber \\
&+&  \sum_{t=1}^n I\left( X_{3[t+1:n]} X_{1[1:t-1]}  ; X_{1[t]} | \mathcal{J}_{1}^{[1:K]} \mathcal{J}_{2}^{[1:K]}  X_{2[t:n]} X_{3[1:t]} \right)\\
 &\stackrel{(f)}{=}&  \sum_{t=1}^n I\left(\mathcal{J}_{1}^{[1:K]} \mathcal{J}_{2}^{[1:K]}  X_{2[t+1:n]} X_{3[1:t-1]}  ; X_{1[t]} X_{3[t]} | X_{2[t]} \right) \nonumber \\
 &+&  \sum_{t=1}^n I\left( X_{3[t+1:n]} X_{1[1:t-1]}  ; X_{1[t]} X_{2[t]}| \mathcal{J}_{1}^{[1:K]} \mathcal{J}_{2}^{[1:K]}  X_{2[t+1:n]} X_{3[1:t]} \right)\\ 
   &\stackrel{(g)}{\geq}&  \sum_{t=1}^n I\left(\mathcal{J}_{1}^{[1:K]} \mathcal{J}_{2}^{[1:K]}  X_{2[t+1:n]} X_{3[1:t-1]}  ; X_{1[t]} X_{3[t]} | X_{2[t]} \right) \nonumber \\
    &+&  \sum_{t=1}^n I\left( X_{3[t+1:n]}  ; X_{1[t]}| \mathcal{J}_{1}^{[1:K]} \mathcal{J}_{2}^{[1:K]}  X_{2[t+1:n]} X_{3[1:t]} \right)\\ 
    &\stackrel{(h)}{=}&     \sum_{t=1}^n  I\left( U_{1\rightarrow 23,[1:K],[t]} U_{2\rightarrow 13,[1:K],[t]}  ; X_{1[t]}X_{3[t]} | X_{2[t]}  \right)\nonumber \\
&+&  \sum_{t=1}^n I\left( U_{1\rightarrow 3,K,[t]} ; X_{1[t]}| U_{1\rightarrow 23,[1:K],[t]} U_{2\rightarrow 13,[1:K],[t]}X_{3[t]} \right)\\ 
 &=&     \sum_{t=1}^n  I\left( U_{1\rightarrow 23,[1:K],[Q]} U_{2\rightarrow 13,[1:K],[Q]}  ; X_{1[Q]}X_{3[Q]} | X_{2[Q]}  ,Q=t\right)\nonumber \\
&+&  \sum_{t=1}^n I\left( U_{1\rightarrow 3,K,[Q]} ; X_{1[Q]}| U_{1\rightarrow 23,[1:K],[Q]} U_{2\rightarrow 13,[1:K],[Q]}X_{3[Q]}, Q=t \right)\\ 
 &\stackrel{(i)}{=}&    n\Big[  I\left( \widetilde{U}_{1\rightarrow 23,[1:K]} \widetilde{U}_{2\rightarrow 13,[1:K]}  ; X_{1}X_{3} | X_{2}\right)\!+ \! I\left( \widetilde{U}_{1\rightarrow 3,K} ; X_{1}| \widetilde{U}_{1\rightarrow 23,[1:K]} \widetilde{U}_{2\rightarrow 13,[1:K]}X_{3}\right)\!\Big]\nonumber\\
 &=& n\Big[  I\left( \widetilde{\mathcal{W}}_{[1,K+1]}  ; X_{1}X_{3} | X_{2}\right)\!+ \! I\left( \widetilde{U}_{1\rightarrow 3,K} ; X_{1}| \widetilde{\mathcal{W}}_{[1,K+1]} X_{3}\right)\!\Big]
   \end{IEEEeqnarray}
  
where
\begin{itemize}
\item step~$(a)$ follows from the fact that $\mathcal{J}_{1}^{[1:K]}$ is a function of the sources $(X_1^n,X_2^n,X_3^n)$, 
\item step~$(b)$ follows  from the non-negativity of mutual information, 
\item step~$(c)$ follows  from the fact that $\mathcal{J}_{2}^{[1:K]}$ is a function of $\mathcal{J}_{1}^{[1:K]}$ and the source $X_2^n$, 
\item step~$(d)$ follows  from the chain rule for conditional mutual information and the memoryless property across time of the sources $(X_1^n,X_2^n,X_3^n)$,
\item step~$(e)$ follows  from the non-negativity of mutual information,
\item step~$(f)$  follows from the Markov chain $X_{2[t]} \mkv  ( \mathcal{J}_{1}^{[1:K]} \mathcal{J}_{2}^{[1:K]}  X_{2[t+1:n]} X_{3[1:t]} ) \mkv  (X_{3[t+1:n]}X_{1[1:t-1]})$ (Corollary \ref{coro:Markov_interactive_encoding_1} in the appendices), for all $t=[1:n]$ which follows from $X_1\mkv X_3\mkv X_2$.,
\item step~$(g)$ follows from the non-negativity of mutual information, 
\item step~$(h)$ follows from defintions (\ref{eq-def-theo5-5}).
\item step~$(i)$ follows from the  standard time-sharing arguments and the definition of new random variables, (i.e. $\widetilde{U}_{1\rightarrow 23,[1:K]} \triangleq (U_{1\rightarrow 23,[1:K],[Q]},Q)$ and $X_1\triangleq (X_{1[Q]},Q)$). 
\end{itemize}
and the last step follows from the definition of the past shared common descriptions $\mathcal{W}_{[1,l]}\ \forall l$. It is also immediate to show that $ (\widetilde{U}_{1\rightarrow 23,l}, \widetilde{U}_{2\rightarrow 13,l})$ satisfies the Markov chains in (\ref{eq:two_dist_crit_mark5-1})-(\ref{eq:two_dist_crit_mark5-3}) for all $l\in[1:K]$.

\subsubsection{Rate at node 2}

For the second rate, by following the same steps as before we have
\begin{IEEEeqnarray}{rCl}
n(R_2+\varepsilon)
	&\geq& H\left(\mathcal{J}_{2}^{[1:K]}\right) \\
		&\stackrel{(a)}{=}&		I\left(\mathcal{J}_{2}^{[1:K]}; X_1^nX_2^nX_3^n\right)\\
		&\stackrel{(b)}{\geq }&		I\left(\mathcal{J}_{2}^{[1:K]}; X_2^n|X_1^n X_3^n\right)\\	
	 	&\stackrel{(c)}{= }&		I\left(\mathcal{J}_{1}^{[1:K]}\mathcal{J}_{2}^{[1:K]}; X_2^n|X_1^n X_3^n\right)\\	
	 	&\stackrel{(d)}{= }&		I\left(\mathcal{J}_{1}^{[1:K]}\mathcal{J}_{2}^{[1:K]}X_1^n; X_2^n|X_3^n\right)\\
	 	&\stackrel{(e)}{\geq }&		I\left(\mathcal{J}_{1}^{[1:K]}\mathcal{J}_{2}^{[1:K]}; X_2^n|X_3^n\right)\\
	 	&\stackrel{(f)}{=}&		\sum_{t=1}^n I\left(\mathcal{J}_{1}^{[1:K]}\mathcal{J}_{2}^{[1:K]}X_{3[1:t-1]}  X_{3[t+1:n]} X_{2[t+1:n]}; X_{2[t]}|X_{3[t]}\right)\\
	 	&\stackrel{(g)}{\geq}&		\sum_{t=1}^n I\left(\mathcal{J}_{1}^{[1:K]}\mathcal{J}_{2}^{[1:K]}X_{3[1:t-1]} X_{2[t+1:n]}; X_{2[t]}|X_{3[t]}\right)\\
	 	&\stackrel{(h)}{=}&		\sum_{t=1}^n I\left(U_{1\rightarrow 23,[1:K],[t]} U_{2\rightarrow 13,[1:K],[t]} ; X_{2[t]}|X_{3[t]}\right)\\
	  &\stackrel{(i)}{=}&	nI\left(\widetilde{\mathcal{W}}_{[1,K+1]} ; X_{2}|X_{3}\right)
       \end{IEEEeqnarray}
where
\begin{itemize}
\item step~$(a)$ follows from the fact that $\mathcal{J}_{2}^{[1:K]}$ is a function of the sources $(X_1^n,X_2^n,X_3^n)$, 
\item step~$(b)$ follows  from the non-negativity of mutual information, 
\item step~$(c)$ follows  from the fact that $\mathcal{J}_{1}^{[1:K]}$ is a function of $\mathcal{J}_{2}^{[1:K]}$ and the source $(X_1^n,X_3^n)$, 
\item step~$(d)$ follows  from the Markov chain $X_1\mkv X_3\mkv X_2$.
\item step~$(e)$ follows  from the non-negativity of mutual information,
\item step~$(f)$  follows  from the chain rule for conditional mutual information and the memoryless property across time of the sources $(X_1^n,X_2^n,X_3^n)$,
\item step~$(g)$ follows from the non-negativity of mutual information, 
\item step~$(h)$ follows from definitions (\ref{eq-def-theo5-5}).
\item step~$(i)$ follows  the definition for $\mathcal{W}_{[1,l]}\ \forall l$ and from standard time-sharing arguments similar to the ones for rate at node 1.
\end{itemize}

\subsubsection{Distortion at nodes 1 and 2}

Node $1$ reconstructs  an estimate $\hat{X}_{12}^n \equiv g_{12}\left(\mathcal{J}_{1}^{[1:K]}, \mathcal{J}_{2}^{[1:K]},X_1^n,X_3^n\right)$ while node $2$ reconstructs  $\hat{X}_{21}^n \equiv g_{21}\left(\mathcal{J}_{1}^{[1:K]}, \mathcal{J}_{2}^{[1:K]},X_2^n\right)$. For each $t\in\{1,\dots,n\}$, define functions $\hat{X}_{12[t]}$ and $\hat{X}_{21[t]}$  as being the $t$-th coordinate of the corresponding estimates  of $\hat{X}_{12}^n$ and $\hat{X}_{21}^n$, respectively:
\begin{IEEEeqnarray}{rCl}
\hat{X}_{12[t]}\left(\mathcal{W}_{[1,K+1],[t]} , U_{1\rightarrow 3,K,[t]},X_{1[t]},X_{3[t]},X_{1[1:t-1]},X_{1[t+1:n]} \right)
& \triangleq &   g_{12[t]} \left(\mathcal{J}_{1}^{[1:K]} , \mathcal{J}_{2}^{[1:K]},X_1^n \right), \\
\hat{X}_{21[t]}\left(\mathcal{W}_{[1,K+1],[t]} , X_{2[t]} \right) &\triangleq&   g_{21[t]} \left(\mathcal{J}_{1}^{[1:K]} , \mathcal{J}_{2}^{[1:K]},X_2^n \right) \ .
\label{eq:definition_dist_12}
\end{IEEEeqnarray}
The component-wise mean distortions thus verify
\begin{IEEEeqnarray}{rCl}
D_{12} + \varepsilon
	&\geq&	\bE\left[ d\left (X_2,g_{12}\big(\mathcal{J}_{1}^{[1:K]}, \mathcal{J}_{2}^{[1:K]},X_1^n \big) \right) \right] \\
	&\stackrel{(a)}{=}&   	\frac1n \sum_{t=1}^n \bE\left[ d\left (X_{2[t]},\hat{X}_{12[t]}\left(\mathcal{W}_{[1,K+1],[t]}, U_{1\rightarrow 3,K,[t]},X_{1[t]},X_{3[t]},X_{1[1:t-1]},X_{1[t+1:n]} \right)\right) \right] \,\,\,\,\,\,\,\,\,\,\,\,\,\\
	&\stackrel{(b)}{\geq}&   \frac1n \sum_{t=1}^n \bE\left[ d\left( X_{2[t]},\hat{X}^{*}_{12[t]}\left(\mathcal{W}_{[1,K+1],[t]}, U_{1\rightarrow 3,K,[t]},X_{1[t]},X_{3[t]} \right)\right) \right] \\
	&=&  \frac1n \sum_{t=1}^n \bE\left[ d\left (X_{2[Q]}, \hat{X}^{*}_{12[Q]}\left(\mathcal{W}_{[1,K+1],[Q]}, U_{1\rightarrow 3,K,[Q]},X_{1[Q]},X_{3[Q]}\right) \right)\ \middle|\ Q=t \right] \\
	&=& 	\bE\left[ d\left (X_{2[Q]}, \hat{X}^{*}_{12[Q]}\left(\mathcal{W}_{[1,K+1],[Q]}, U_{1\rightarrow 3,K,[Q]}X_{1[Q]},X_{3[Q]}\right) \right) \right] \\
	&\stackrel{(c)}{=}& \bE\left[ d\left (X_{2}, \widetilde{X}_{12}\left(\widetilde{\mathcal{W}}_{[1,K+1]}, \widetilde{U}_{1\rightarrow 3,K},X_{1},X_{3}\right) \right) \right] \ ,
\end{IEEEeqnarray}
where 
\begin{itemize}
\item step~$(a)$ follows from (\ref{eq:definition_dist_12}),
\item step~$(b)$ follows from Markov chain $ X_{2[t]}  \mkv \big( X_{1[t]},X_{3[t]},\mathcal{J}_{1}^{[1:K]}, \mathcal{J}_{2}^{[1:K]}, X_{3[1:t-1]},X_{3[t+1:n]},$ \\ $X_{2[t+1:n]}\big)  \mkv \left(X_{1[1:t-1]},X_{1[t+1:n]}\right)$ $\forall$ $t=[1:n]$ (which can be obtained from Corollary \ref{coro:Markov_interactive_encoding_1} in the appendices) and Lemma \ref{lemma:Markov_distortion}.
\item step~$(c)$ follows from the following relations:
\end{itemize}
\begin{IEEEeqnarray*}{rCl}
\widetilde{X}_{12}\left(\widetilde{\mathcal{W}}_{[1,K+1]}, \widetilde{U}_{1\rightarrow 3,K},X_{1},X_{3}\right) & =& \widetilde{X}_{12}\left(Q,\mathcal{W}_{[1,K+1],[Q]}, U_{1\rightarrow 3,K,[Q]},X_{1[Q]},X_{3[Q]} \right) \nonumber\\
&\triangleq& \hat{X}^{*}_{12[Q]}\left(\mathcal{W}_{[1,K+1],[Q]}, U_{1\rightarrow 3,K,[Q]},X_{1[Q]},X_{3[Q]}\right) \ .
\end{IEEEeqnarray*}
By following the very same steps, we can also show that: 
\begin{IEEEeqnarray}{rCl}
D_{21} + \varepsilon
	&\geq&	\bE\left[ d\left (X_1,g_{21}\big(\mathcal{J}_{1}^{[1:K]}, \mathcal{J}_{2}^{[1:K]},X_2^n \big) \right) \right] \\
	&=& \bE\left[ d\left(X_1,\widetilde{X}_{21}\left( \widetilde{\mathcal{W}}_{[1,K+1]},X_{2}\right) \right) \right] \ ,
\end{IEEEeqnarray}
and where we used  the Markov chain $ X_{2[1:t-1]}  \mkv \big( X_{2[t]}, \mathcal{J}_{1}^{[1:K]}, \mathcal{J}_{2}^{[1:K]} , X_{3[1:t-1]},X_{2[t+1:n]} \big)  \mkv X_{1[t]}$ $\forall$ $t=[1:n]$ (which can be obtained from Corollary \ref{coro:Markov_interactive_encoding_1} in the appendices) and Lemma \ref{lemma:Markov_distortion} and where we define the function $\widetilde{X}_{21}$:
\begin{IEEEeqnarray*}{rCl}
\widetilde{X}_{21}\left( \widetilde{\mathcal{W}}_{[1,K+1]},X_{2}\right) = \widetilde{X}_{21}\left(Q,  \widetilde{\mathcal{W}}_{[1,K+1],[Q]},X_{2[Q]}\right) \triangleq \hat{X}_{21[Q]}^{*}\left( \widetilde{\mathcal{W}}_{[1,K+1],[Q]},U_{2\rightarrow 13[Q]},X_{2[Q]}\right) \ .
\end{IEEEeqnarray*}

\subsubsection{Distortions at node 3}

Node $3$ compute lossy reconstructions $\hat{X}_{31}^n \equiv g_{31}\left(\mathcal{J}_{1}^{[1:K]}, \mathcal{J}_{2}^{[1:K]},X^n_3\right)$ and $\hat{X}_{32}^n \equiv g_{32}\left(\mathcal{J}_{1}^{[1:K]}, \mathcal{J}_{2}^{[1:K]},X^n_3\right)$. For each $t\in\{1,\dots,n\}$, define functions $\hat{X}_{31[t]}$ and $\hat{X}_{32[t]}$  as being the $t$-th coordinate of the corresponding estimates  of $\hat{X}_{31}^n$ and $\hat{X}_{32}^n$, respectively:
\begin{IEEEeqnarray}{rCl}
\hat{X}_{31[t]}\left(\mathcal{W}_{[1,K+1],[t]}, U_{1\rightarrow 3,K,[t]}, X_{3[t]}\right) \triangleq   g_{31[t]} \left(\mathcal{J}_{1}^{[1:K]} , \mathcal{J}_{2}^{[1:K]},X^n_3\right) \ ,\\
\hat{X}_{32[t]}\left(\mathcal{W}_{[1,K+1],[t]}, U_{1\rightarrow 3,K,[t]} ,X_{3[t]}\right) \triangleq   g_{32[t]} \left(\mathcal{J}_{1}^{[1:K]} , \mathcal{J}_{2}^{[1:K]},X^n_3\right) \ .
\end{IEEEeqnarray}
The component-wise mean distortions thus verify
\begin{IEEEeqnarray}{rCl}
D_{31} + \varepsilon
	&\geq&	\bE\left[ d\left (X_1,g_{31}\big(\mathcal{J}_{1}^{[1:K]}, \mathcal{J}_{2}^{[1:K]},X^n_3\big) \right) \right] \\
 &=&  \frac1n \sum_{t=1}^n \bE\left[ d\left (X_{1[t]},\hat{X}_{31[t]}\left(\mathcal{W}_{[1,K+1],[t]}, U_{1\rightarrow 3,K,[t]}, X_{3[t]}\right)\right) \right] \\
 &=&	\frac1n \sum_{t=1}^n \bE\left[ d\left (X_{1[Q]}, \hat{X}_{31[Q]}\left(\mathcal{W}_{[1,K+1],[Q]}, U_{1\rightarrow 3,K,[Q]}, X_{3[Q]}\right) \right)\ \middle|\ Q=t \right]  \\
	&=& 	\bE\left[ d\left (X_{1[Q]},\hat{X}_{31[Q]}\left(\mathcal{W}_{[1,K+1],[Q]}, U_{1\rightarrow 3,K,[Q]}, X_{3[Q]}\right)\right) \right] \\
	&=& 	\bE\left[ d\left (X_1,\widetilde{X}_{31}\left(\widetilde{\mathcal{W}}_{[1,K+1]}, \widetilde{U}_{1\rightarrow 3,K}, X_{3}\right) \right) \right] \ ,
\end{IEEEeqnarray}
where the last step follows by defining the function $\widetilde{X}_{31}$ by
\begin{IEEEeqnarray*}{rCl}
\widetilde{X}_{31}\left(\widetilde{\mathcal{W}}_{[1,K+1]}, \widetilde{U}_{1\rightarrow 3,K}, X_{3}\right) &=& \widetilde{X}_{31}\left(Q,\mathcal{W}_{[1,K+1],[Q]}, U_{1\rightarrow 3,K,[Q]}, X_{3[Q]}\right) \nonumber\\
&\triangleq& \hat{X}_{31[Q]}\left(\mathcal{W}_{[1,K+1],[Q]}, U_{1\rightarrow 3,K,[Q]}, X_{3[Q]}\right) \ .
\end{IEEEeqnarray*}
By following the very same steps, we can also show that: 
\begin{IEEEeqnarray}{rCl}
D_{32} + \varepsilon
	&\geq&	\bE\left[ d\left (X_2,g_{32}\big(\mathcal{J}_{1}^{[1:K]}, \mathcal{J}_{2}^{[1:K]},X^n_3\big) \right) \right] \\
 &=&  \bE\left[ d\left (X_2, \widetilde{X}_{32}\left(\widetilde{\mathcal{W}}_{[1,K+1]},\widetilde{U}_{1\rightarrow 3,K},X_{3}\right) \right) \right] \ ,
 \end{IEEEeqnarray}
and where we define the function $\widetilde{X}_{32}$ by 
\begin{IEEEeqnarray*}{rCl}
\widetilde{X}_{32}\left(\widetilde{\mathcal{W}}_{[1,K+1]}, \widetilde{U}_{1\rightarrow 3,K}, X_{3}\right) &=& \widetilde{X}_{32}\left(Q,\mathcal{W}_{[1,K+1],[Q]}, U_{1\rightarrow 3,K,[Q]}, X_{3[Q]}\right) \nonumber\\
&\triangleq& \hat{X}_{32[Q]}\left(\mathcal{W}_{[1,K+1],[Q]}, U_{1\rightarrow 3,K,[Q]}, X_{3[Q]}\right) \ .
\end{IEEEeqnarray*}
Cooperative distributed source coding with two distortion criteria under reversal delivery and side information
This concludes the proof of the converse and thus that of the theorem. 

\end{IEEEproof}

\subsection{Three encoders and three decoders subject to  lossless/lossy reconstruction constraints with degraded side information}

\begin{figure}[th]
\centering
\includegraphics[angle=0,width=0.6\columnwidth,keepaspectratio,trim= 0mm 0mm 0mm 0mm,clip]{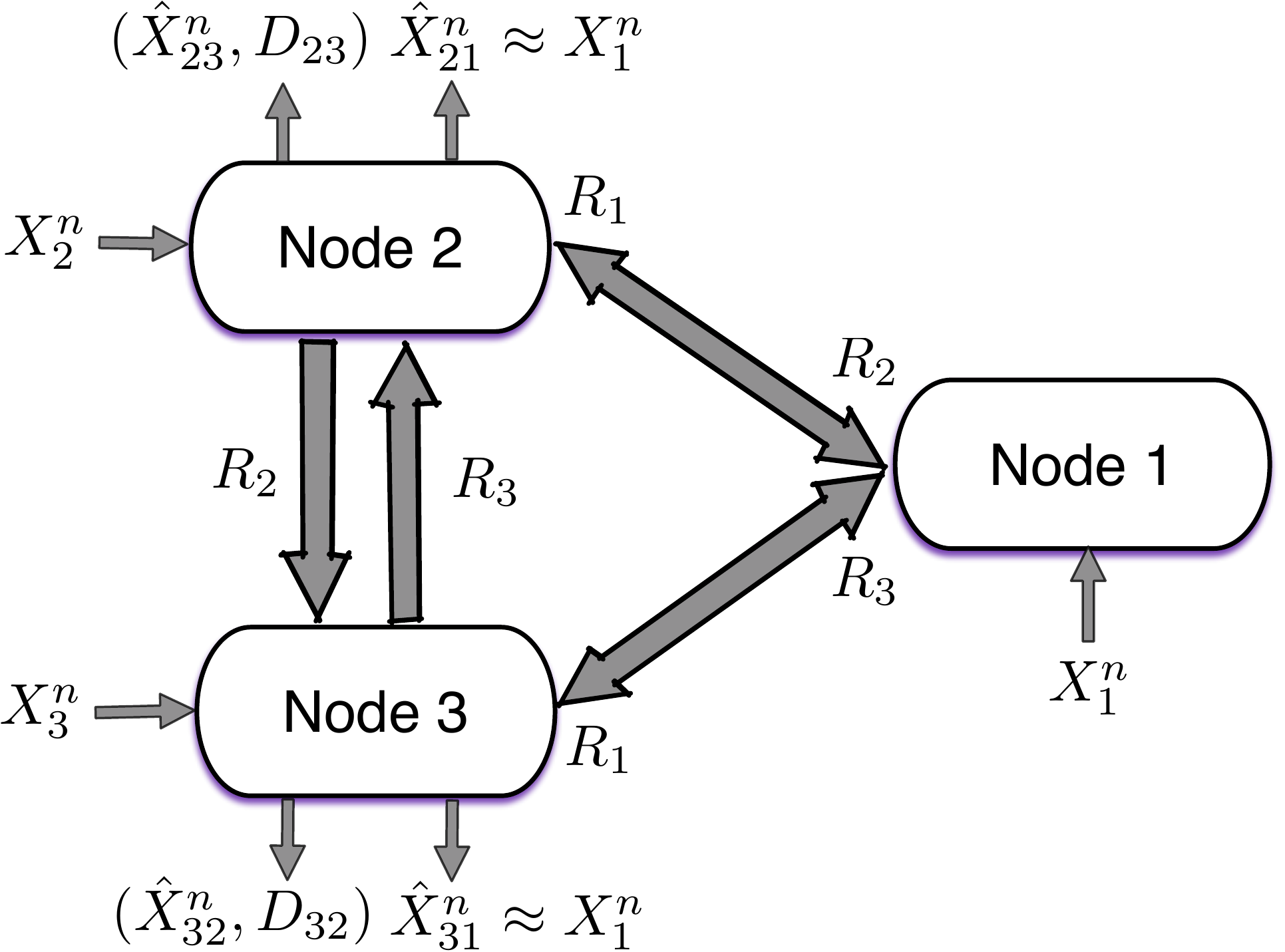} 

\caption{Three encoders and three decoders subject to  lossless/lossy reconstruction constraints with degraded side information.}
\label{fig:model-Coop-BT4}
\end{figure}

Consider now the problem described in Fig.~\ref{fig:model-Coop-BT4} where encoder $1$ wishes to communicate the \emph{lossless} the source $X_1^n$ to nodes $2$ and $3$ while encoder $2$ wishes to send a \emph{lossy} description of its source $X_2^n$ to node $3$ with distortion constraints $D_{32}$ and encoder $3$ wishes to send a \emph{lossy} description of its source $X_3^n$ to node $2$ with distortion constraints $D_{23}$.  In addition to this, the encoders perfom the communication using $K$ communication rounds. This problem can be seen as a generalization of the settings previously investigated in~\cite{heegard_rate_1985}.  This setting can model a problem in which node 1 generate a process $X_1$. This process physically propagates to the locations where nodes 2 and 3 are. These nodes measure $X_2$ and $X_3$ respectively. If node 2 is closer to node 1 than node 3, we can assume that $ X_1 \mkv X_2  \mkv X_3$. Nodes 2 and 3 then interact between them and with node 1, in order to reconstruct $X_1$ in lossless fashion and $X_2$ and $X_3$ with some distortion level. 

\vspace{1mm}
 \begin{theorem}\label{theo-Coop-BT4-B}
The rate-distortion region of the setting described in Fig.~\ref{fig:model-Coop-BT4} where $ X_1 \mkv X_2  \mkv X_3$ form a Markov chain is given by the union over all joint probability measures $p_{X_1X_2X_3 U_{3 \rightarrow 2,[1:K]}  U_{2 \rightarrow 3,[1:K]} }$ satisfying the Markov chains
\begin{IEEEeqnarray}{rcl}
U_{2 \rightarrow 3,l} \mkv (X_1,X_2, \mathcal{V}_{[23,l,2]}) \mkv X_3\, ,\\
U_{3 \rightarrow 2,l} \mkv (X_1,X_3, \mathcal{V}_{[23,l,3]}) \mkv X_2 \, ,
\label{eq:two_dist_crit_markB}
\end{IEEEeqnarray}
 $\forall l \in [1:K]$, and such that there exist  reconstruction mappings:
\begin{IEEEeqnarray*}{rCl}
g_{23} \left(X_1, X_2,   \mathcal{V}_{[23,K+1,2]}   \right)&=&\hat{X}_{23}\,\,\,\, \textrm{ with } \,\,\,\,\mathds{E} \left[d(X_3,\hat{X}_{23})  \right] \leq  D_{23}\ ,\\
g_{32} \left(X_1, X_3, \mathcal{V}_{[23,K+1,2]}  \right)&=&\hat{X}_{32}\,\,\,\, \textrm{ with } \,\,\,\,\mathds{E} \left[d(X_2,\hat{X}_{32})  \right] \leq  D_{32}\ ,
\end{IEEEeqnarray*}
of the set of all tuples satisfying: 
\begin{IEEEeqnarray}{rCl}
R_1 &\geq& H(X_1|X_2) \ ,\\
R_2 &\geq& I(\mathcal{V}_{[23,K+1,2]} ;X_2| X_1 X_3) \ ,\\
R_3 &\geq& I(\mathcal{V}_{[23,K+1,2]} ;X_3| X_1 X_2) \ ,\\
R_1+R_2 &\geq& H(X_1|X_3) +I(\mathcal{V}_{[23,K+1,2]};X_2| X_1 X_3)\ ,
\end{IEEEeqnarray}
The auxiliary random variables have cardinality bounds: 
\begin{equation}
\| \mathcal{U}_{2\rightarrow 3,l}\|\leq \|\mathcal{X}_1\|\|\mathcal{X}_2\|\prod_{i=1}^{l-1}\| \mathcal{U}_{2\rightarrow 3,i}\|\| \mathcal{U}_{3\rightarrow 2,i}\|+1,\ \ l\in[1:K]
\end{equation}
\begin{equation}
\| \mathcal{U}_{3\rightarrow 2,l}\|\leq \|\mathcal{X}_1\|\|\mathcal{X}_3\|\| \mathcal{U}_{2\rightarrow 3,l}\|\prod_{i=1}^{l-1}\| \mathcal{U}_{2\rightarrow 3,i}\|\| \mathcal{U}_{3\rightarrow 2,i}\|+1,\ \ l\in[1:K]
\end{equation}
\end{theorem}
\vspace{1mm}
\begin{remark}
Theorem \ref{theo-Coop-BT4-B}  shows that several exchanges between nodes 2 and 3 can be helpful. Node 1 transmit only once at the beginning its full source. 
\end{remark}
\begin{IEEEproof}
The direct part of the proof follows according to Theorem~\ref{theo-main-theorem} by choosing:
\begin{IEEEeqnarray*}{rCl}
U_{1\rightarrow 3,l} &=&U_{1\rightarrow 2,l} = U_{3\rightarrow 1,l}=U_{2\rightarrow 1,l}=U_{3\rightarrow 12,l} = \varnothing \,\,\,\, \textrm{ $\forall l \in [1:K]$:}\\ 
U_{1\rightarrow 23,l} &=& U_{2\rightarrow 13,l} = \varnothing,\,\,\, \textrm{ $\forall l \in [2:K]$}
\end{IEEEeqnarray*}
and $U_{1\rightarrow 23,1} = U_{2\rightarrow 13,1} =X_1$. The remanding auxiliary random variables  satisfy $\forall l \in [1:K]$:
\begin{IEEEeqnarray}{rcl}
U_{2\rightarrow 3,l}\mkv (X_1,X_2,\mathcal{V}_{[23,l,2]})\mkv X_3\, ,\label{eq:Markov_heegard_berger_coop_1}\\
U_{3\rightarrow 2,l}\mkv (X_1,X_3,\mathcal{V}_{[23,l,3]})\mkv X_2 \, .\label{eq:Markov_heegard_berger_coop_2}
\end{IEEEeqnarray}

If a pair of tuple $(R_1,R_2,R_3)$  and distortions $(D_{23},D_{32})$ are admissible for the $K$-steps interactive cooperative distributed source coding setting described in Fig.~\ref{fig:model-Coop-BT4}, then  for all $\varepsilon >0$ there exists $n_0(\varepsilon,K)$, such that $\forall\,n>n_0(\varepsilon,K)$  there exists a $K$-steps interactive source code $(n,K,\mathcal{F},\mathcal{G})$ with intermediate rates satisfying:  
\begin{equation}
\frac{1}{n}\sum_{l=1}^K\log{\|\mathcal{J}_{i}^l\|}\leq R_i+\varepsilon\ ,\ i\in\{1,2,3\}
 \end{equation} 
and with average per-letter distortions  with respect to the source $2$ and perfect reconstruction with respect to the source $1$ at all nodes: 
\begin{IEEEeqnarray}{rcl}
&&\mathds{E} \left[d(X_2^n,\hat{X}^n_{32})\right]\leq D_{32}+\varepsilon\ ,\label{eq-converse-BT-distortion2B}\\
&&\mbox{Pr}\left(  X_1^n\neq \hat{X}_{21}^n\right)\leq \varepsilon\ ,\label{eq-converse-BT-error3B}\\
&&\mathds{E} \left[d(X_3^n,\hat{X}^n_{23})\right]\leq D_{23}+\varepsilon\ ,\label{eq-converse-BT-distortion3B}\\
&&\mbox{Pr}\left(  X_1^n\neq \hat{X}_{31}^n\right)\leq \varepsilon\ ,\label{eq-converse-BT-error2B}
 \end{IEEEeqnarray}
where 
\begin{IEEEeqnarray}{rCl}
\hat{X}_{32}^n &\equiv & g_{32} \left(\mathcal{J}_{1}^{[1:K]}, \mathcal{J}_{2}^{[1:K]},\mathcal{J}_{3}^{[1:K]}, X_3^n\right)\ ,\,\,\,\,\,
\hat{X}_{23}^n  \equiv  g_{23}\left( \mathcal{J}_{1}^{[1:K]},\mathcal{J}_{2}^{[1:K]},\mathcal{J}_{3}^{[1:K]}, X_2^n\right)\ ,\label{eq-BT2-12B} \\
\hat{X}_{31}^n & \equiv & g_{31}\left(\mathcal{J}_{1}^{[1:K]},\mathcal{J}_{2}^{[1:K]},\mathcal{J}_{3}^{[1:K]}, X_3^n\right)\ ,\,\,\,\,\,
\hat{X}_{21}^n  \equiv  g_{21}\left( \mathcal{J}_{1}^{[1:K]},\mathcal{J}_{2}^{[1:K]},\mathcal{J}_{3}^{[1:K]}, X_2^n\right)\ . \label{eq-BT2-21B}
\end{IEEEeqnarray}

For each $t\in\{1,\dots,n\}$ and $l \in [1:K]$, we define random variables $U_{2\rightarrow 3,l,[t]}$ and $U_{3\rightarrow 2,l,[t]}$ as follows:
\begin{IEEEeqnarray}{rCl}
U_{2\rightarrow 3,1,[t]} & \triangleq & \left(\mathcal{J}_{1}^1, \mathcal{J}_{2}^1,  X_{1[1:t-1]},X_{1[t+1:n]}, X_{2[t+1:n]} , X_{3[1:t-1]} \right) \label{eq:MarkovBT2-1B} , \\
U_{2\rightarrow 3,l,[t]} & \triangleq & \left(\mathcal{J}_{1}^l, \mathcal{J}_{2}^l\right), \ l\in[2:K] \label{eq:MarkovBT2-2B} \\
U_{3\rightarrow 2,l,[t]} & \triangleq & \mathcal{J}_{3}^l,\ l\in[1:K]\ \label{eq:MarkovBT2-3B} .
\end{IEEEeqnarray}
These auxiliary random variables satisfy the Markov conditions (\ref{eq:Markov_heegard_berger_coop_1}) and (\ref{eq:Markov_heegard_berger_coop_2}), which can be verified from Lemma \ref{lemma:Markov_interactive_encoding_2} in the appendices. By the conditions~\eqref{eq-converse-BT-error2B} and~\eqref{eq-converse-BT-error3B}, and Fano's inequality~\cite{cover2006elements}, we have 
\begin{IEEEeqnarray}{rCl}
H(X_1^n| \hat{X}_{31}^n)  &\leq &\mbox{Pr}\left(  X_1^n\neq \hat{X}_{31}^n\right) \log_2 (\|\mathcal{X}_1^n \|-1) + H_2\left( \mbox{Pr}\big(  X_1^n\neq \hat{X}_{31}^n\big)  \right)  \triangleq    n \epsilon_n   \ , \label{eq-converse-BT-fano2-1B}\\
H(X_1^n| \hat{X}_{21}^n)  &\leq &\mbox{Pr}\left(  X_1^n\neq \hat{X}_{21}^n\right) \log_2 (\|\mathcal{X}_1^n \|-1) + H_2\left( \mbox{Pr}\big(  X_1^n\neq \hat{X}_{21}^n\big)  \right)  \triangleq    n \epsilon_n   \ ,\label{eq-converse-BT-fano2-2B}
\end{IEEEeqnarray}
where $\epsilon_n(\varepsilon) \rightarrow 0$ provided that $\varepsilon\rightarrow 0$ and $n \rightarrow \infty$. 

\subsubsection{Rate at node 1}

For the first rate, we have
\begin{IEEEeqnarray}{rCl}
n(R_1+\varepsilon)
	&\geq& 					H\left(\mathcal{J}_{1}^{[1:K]}\right) \\
	&\geq&	H\left(\mathcal{J}_{1}^{[1:K]}| X_2^nX_3^n \right)   \\
	&\stackrel{(a)}{=}&	 H\left(\mathcal{J}_{1}^{[1:K]}\mathcal{J}_{2}^{[1:K]}\mathcal{J}_{3}^{[1:K]}| X_2^nX_3^n \right)   \\
	&\stackrel{(b)}{=}&	I\left(\mathcal{J}_{1}^{[1:K]}\mathcal{J}_{2}^{[1:K]}\mathcal{J}_{3}^{[1:K]}; X_1^n| X_2^nX_3^n  \right)  \\
	&\geq & n H(X_1| X_2X_3) -  H\left(X_1^n | X_2^n \mathcal{J}_{1}^{[1:K]}\mathcal{J}_{2}^{[1:K]}\mathcal{J}_{3}^{[1:K]}  \right) \\
	&\stackrel{(c)}{\geq }&  n H(X_1| X_2X_3) -  H(X_1^n| \hat{X}_{21}^n) \\
	&\stackrel{(d)}{\geq }& n\left[ H(X_1| X_2X_3) -  \epsilon_n\right]\ ,\\
	&\stackrel{(e)}{= }& n\left[ H(X_1| X_2) -  \epsilon_n\right]\ ,
\end{IEEEeqnarray}
where
\begin{itemize}
\item step~$(a)$ follows from the fact that by definition of the code the sequence $\mathcal{J}_{2}^{[1:K]},\mathcal{J}_{3}^{[1:K]}$ are functions of $(\mathcal{J}_{1}^{[1:K]},X_2^n,X_3^n)$,
\item step~$(b)$ follows from the fact that by definition of the code the sequences $(\mathcal{J}_{1}^{[1:K]},\mathcal{J}_{2}^{[1:K]},\mathcal{J}_{3}^{[1:K]})$ are functions of the sources$(X_1^n,X_2^n,X_3^n)$,
\item step~$(c)$ follows from the code assumption in~\eqref{eq-BT2-21B} that guarantees the existence of a reconstruction function $\hat{X}_{21}^n \equiv g_{21}\left(\mathcal{J}_{1}^{[1:K]},\mathcal{J}_{2}^{[1:K]},\mathcal{J}_{3}^{[1:K]},X_2^n\right)$, 
\item step~$(d)$ follows from Fano's inequality in~\eqref{eq-converse-BT-fano2-2B},
\item step~$(e)$ follows from the assumption that $ X_1 \mkv X_2  \mkv X_3$ form a Markov chain.
\end{itemize}

\subsubsection{Rate at nodes 2 and 3}
For the second rate, we have
\begin{IEEEeqnarray}{rCl}
n(R_2+\varepsilon)
	&\geq& 				H\left(\mathcal{J}_{2}^{[1:K]}\right) \\
	&\stackrel{(a)}{=}&		I\left(\mathcal{J}_{2}^{[1:K]}; X_1^nX_2^nX_3^n\right)\\
	&\stackrel{(b)}{\geq }&		I\left(\mathcal{J}_{2}^{[1:K]}; X_2^n|X_1^nX_3^n\right)\\	
 	&\stackrel{(c)}{= }&		I\left(\mathcal{J}_{1}^{[1:K]}\mathcal{J}_{2}^{[1:K]}\mathcal{J}_{3}^{[1:K]}; X_2^n|X_1^nX_3^n\right)\\	
		&\stackrel{(d)}{=}&  
	 \sum_{t=1}^n  I\left(\mathcal{J}_{1}^{[1:K]}\mathcal{J}_{2}^{[1:K]}  \mathcal{J}_{3}^{[1:K]}; X_{2[t]} | X_1^nX_3^nX_{2[t+1:n]} \right) \\
	  	  &\stackrel{(e)}{\geq }&      \sum_{t=1}^n  I\left(\mathcal{V}_{[23,K+1,2][t]}; X_{2[t]} | X_{1[t]}X_{3[t]} \right)\\
			  &\stackrel{(f)}{=}&   \sum_{t=1}^n  I\left( \mathcal{V}_{[23,K+1,2][Q]} ; X_{2[Q]} | X_{1[Q]}X_{3[Q]}  ,Q=t \right)	  \\
&\stackrel{(g)}{\geq }&  n I\left( \widetilde{\mathcal{V}}_{[23,K+1,2]}; X_{2} | X_{1}X_{3}\right)\ ,
\end{IEEEeqnarray}
where
\begin{itemize}
\item step~$(a)$ follows from the fact that $\mathcal{J}_{2}^{[1:K]}$ is a function of the sources $(X_1^n,X_2^n,X_3^n)$, 
\item step~$(b)$ follows  from the non-negativity of mutual information, 
\item step~$(c)$ follows  from the fact that $(\mathcal{J}_{1}^{[1:K]},\mathcal{J}_{2}^{[1:K]},\mathcal{J}_{3}^{[1:K]})$ are functions of  the sources $(X_1^n,X_2^n,X_3^n)$, 
\item step~$(d)$  follows  from the chain rule for conditional mutual information,
\item step~$(e)$ follows from the definitions~\eqref{eq:MarkovBT2-1B}, ~\eqref{eq:MarkovBT2-2B} and \eqref{eq:MarkovBT2-3B}, the memoryless property of the sources and the non-negativity of mutual information,  
\item step~$(f)$ follows from the use of a time sharing random variable $Q$ uniformly distributed over the set $\{1,\dots,n\}$, 
\item step~$(g)$ follows by letting a new random variable $\widetilde{\mathcal{V}}_{[23,K+1,2]}\triangleq (\mathcal{V}_{[23,K+1,2][Q]} ,Q)$.
\end{itemize}
By following similar steps, it is not difficult to check that
\begin{IEEEeqnarray}{rCl}
n(R_3+\varepsilon)
	&\geq& 	 \sum_{t=1}^n  I\left(\mathcal{V}_{[23,K+1,2][t]}; X_{3[t]} | X_{1[t]}X_{2[t]} \right)\\
			  &=&   \sum_{t=1}^n  I\left( \mathcal{V}_{[23,K+1,2][Q]} ; X_{3[Q]} | X_{1[Q]}X_{2[Q]}  ,Q=t \right)	  \\
&\geq &  n I\left( \widetilde{\mathcal{V}}_{[23,K+1,2]}; X_{3} | X_{1}X_{2}\right)\ .
\end{IEEEeqnarray}	

\subsubsection{Sum-rate of nodes $1$ and $2$}
For the sum-rate, we have
\begin{IEEEeqnarray}{rCl}
n(R_1 + R_2+2\varepsilon)
	&\geq& 				H\left(\mathcal{J}_{1}^{[1:K]}\right)+ H\left(\mathcal{J}_{2}^{[1:K]}\right)\\
&\geq & H\left(\mathcal{J}_{1}^{[1:K]}\mathcal{J}_{2}^{[1:K]} \right) \\
	&\stackrel{(a)}{=}&	I\left(\mathcal{J}_{1}^{[1:K]}\mathcal{J}_{2}^{[1:K]}; X_1^nX_3^nX_2^n \right) \\
& \stackrel{(b)}{\geq } & I\left(\mathcal{J}_{1}^{[1:K]}\mathcal{J}_{2}^{[1:K]}; X_1^nX_2^n | X_3^n \right)\\
&=& I\left(\mathcal{J}_{1}^{[1:K]}\mathcal{J}_{2}^{[1:K]}; X_1^n | X_3^n \right)+I\left(\mathcal{J}_{1}^{[1:K]}\mathcal{J}_{2}^{[1:K]}\mathcal{J}_{3}^{[1:K]}; X_2^n | X_1^nX_3^n \right)\\
&=& H\left(X_1^n | X_3^n \right) - H\left(X_1^n | \mathcal{J}_{1}^{[1:K]}\mathcal{J}_{2}^{[1:K]}X_3^n \right)\nonumber\\
  \IEEEeqnarraymulticol{3}{c}{   +I\left(\mathcal{J}_{1}^{[1:K]}\mathcal{J}_{2}^{[1:K]}\mathcal{J}_{3}^{[1:K]}; X_2^n | X_1^nX_3^n \right)}\\
  & \stackrel{(c)}{\geq } & H\left(X_1^n | X_3^n \right) - H(X_1^n | \hat{X}_{31}^n ) +I\left(\mathcal{J}_{1}^{[1:K]}\mathcal{J}_{2}^{[1:K]}\mathcal{J}_{3}^{[1:K]}; X_2^n | X_1^nX_3^n \right)\\
    & \stackrel{(d)}{\geq } & n\left[H\left(X_1 | X_3 \right) - \epsilon_n \right]+I\left(\mathcal{J}_{1}^{[1:K]}\mathcal{J}_{2}^{[1:K]}\mathcal{J}_{3}^{[1:K]}; X_2^n | X_1^nX_3^n \right)\\
   & \stackrel{(e)}{\geq } &   \sum_{t=1}^n  I\left(\mathcal{J}_{1}^{[1:K]}\mathcal{J}_{2}^{[1:K]}\mathcal{J}_{3}^{[1:K]} X_{1[1:t-1]}X_{1[t+1:n]}X_{3[1:t-1]}X_{2[t+1:n]}; X_{2[t]} |  X_{1[t]}X_{3[t]} \right) \nonumber\\
    &+& n\left[H\left(X_1 | X_3 \right) - \epsilon_n \right] \\
        & \stackrel{(f)}{= } & n\left[H\left(X_1 | X_3 \right) - \epsilon_n \right] +  \sum_{t=1}^n  I\left( \mathcal{V}_{[23,K+1,2][t]}; X_{2[t]} |  X_{1[t]}X_{3[t]} \right) \\
                & \stackrel{(g)}{= } & n\left[H\left(X_1 | X_3 \right) - \epsilon_n+ I\left( \mathcal{V}_{[23,K+1,2][Q]}; X_{2[Q]} |  X_{1[Q]}X_{3[Q]},Q \right)  \right] \\
                & \stackrel{(h)}{= } & n\left[H\left(X_1 | X_3 \right) - \epsilon_n  + I\left( \widetilde{\mathcal{V}}_{[23,K+1,2]}; X_{2} |  X_{1}X_{3} \right) \right]   \ ,              
\end{IEEEeqnarray}
where
\begin{itemize}
\item step~$(a)$ follows from the fact that $\mathcal{J}_{1}^{[1:K]}$ and $\mathcal{J}_{2}^{[1:K]}$ are functions of the sources $(X_1^n,X_2^n,X_3^n)$, 
\item step~$(b)$ follows non-negativity of mutual information, 
\item step~$(c)$ follows  from the code assumption in~\eqref{eq-BT2-21B} that guarantees the existence of reconstruction function $\hat{X}_{31}^n \equiv g_{31}\left(\mathcal{J}_{1}^{[1:K]},\mathcal{J}_{2}^{[1:K]},\mathcal{J}_{3}^{[1:K]},X^n_3\right)$,  
\item step~$(d)$ follows from Fano's inequality in~\eqref{eq-converse-BT-error2B},  
\item step~$(e)$ follows  from the chain rule of conditional mutual information and the memoryless property across time of the sources $(X_1^n,X_2^n,X_3^n)$, and non-negativity of mutual information, 
\item step~$(f)$ from follows from the definitions~\eqref{eq:MarkovBT2-1B} and ~\eqref{eq:MarkovBT2-2B}, 
\item step~$(g)$ follows from the use of a time sharing random variable $Q$ uniformly distributed over the set $\{1,\dots,n\}$,
\item step~$(h)$ follows from $\widetilde{\mathcal{V}}_{[23,K+1,2]} \triangleq (\widetilde{\mathcal{V}}_{[23,K+1,2][Q]},Q)$.
\end{itemize}

\subsubsection{Distortion at node 2}

Node $2$ reconstructs a lossy  $\hat{X}_{23}^n \equiv g_{23}\left(\mathcal{J}_{1}^{[1:K]},\mathcal{J}_{2}^{[1:K]},\mathcal{J}_{3}^{[1:K]},X_2^n\right)$. For each $t\in\{1,\dots,n\}$, define a function $\hat{X}_{23[t]}$ as beging the $t$-th coordinate of this estimate:
\begin{IEEEeqnarray}{rCl}
\hat{X}_{23[t]}\left({\mathcal{V}}_{[23,K+1,2][t]}, X_{2[t]} \right) \triangleq   g_{23[t]} \left(\mathcal{J}_{1}^{[1:K]} , \mathcal{J}_{2}^{[1:K]},\mathcal{J}_{3}^{[1:K]},X_2^n \right) \ .
\end{IEEEeqnarray}
The component-wise mean distortion thus verifies
\begin{IEEEeqnarray}{rCl}
D_{23} + \varepsilon
	&\geq&	\bE\left[ d\left (X_3,g_{23}\big(\mathcal{J}_{1}^{[1:K]}, \mathcal{J}_{2}^{[1:K]},\mathcal{J}_{3}^{[1:K]},X_2^n \big) \right) \right] \\
	&=& 	\frac1n \sum_{t=1}^n \bE\left[ d\left (X_{3[t]},\hat{X}_{23[t]}\left(  {\mathcal{V}}_{[23,K+1,2][t]} ,X_{2[t]}\right) \right) \right] \\
	&=& 	\frac1n \sum_{t=1}^n \bE\left[ d\left (X_{3[Q]}, \hat{X}_{23[Q]}\left({\mathcal{V}}_{[23,K+1,2][Q]},X_{2[Q]}\right) \right)\ \middle|\ Q=t \right] \\
	&=& 	\bE\left[ d\left (X_{3[Q]}, \hat{X}_{23[Q]}\left({\mathcal{V}}_{[23,K+1,2][Q]},X_{2[Q]}\right) \right) \right] \\
	&=& 	\bE\left[ d\left (X_3, \widetilde{X}_{23}\left(\widetilde{\mathcal{V}}_{[23,K+1,2]},X_{2}\right) \right) \right] \ ,
\end{IEEEeqnarray}
where we defined function $\widetilde{X}_{23}$ by 
\begin{IEEEeqnarray}{rCl}
\widetilde{X}_{23}\left(\widetilde{\mathcal{V}}_{[23,K+1,2]},X_{2}\right) = \widetilde{X}_{23}\left(Q, \mathcal{V}_{[23,K+1,2][Q]},X_{2[Q]}\right) \triangleq \hat{X}_{23[Q]}\left(\mathcal{V}_{[23,K+1,2][Q]},X_{2[Q]}\right) \ .
\end{IEEEeqnarray}

\subsubsection{Distortion at node 3}
Node $3$ reconstructs a  lossy description $\hat{X}_{32}^n \equiv g_{32}\left(\mathcal{J}_{1}^{[1:K]},\mathcal{J}_{2}^{[1:K]}, \right.$ $\left.\mathcal{J}_{3}^{[1:K]},X^n_3\right)$. For each $t\in\{1,\dots,n\}$, define a function $\hat{X}_{32[t]}$ as beging the $t$-th coordinate of this estimate:
\begin{IEEEeqnarray}{rCl}
\hat{X}_{32[t]}\left({\mathcal{V}}_{[23,K+1,2][t]},X_{3[t]}\right) \triangleq   g_{32[t]} \left(\mathcal{J}_{1}^{[1:K]} , \mathcal{J}_{2}^{[1:K]},\mathcal{J}_{3}^{[1:K]},X^n_3\right) \ .
\end{IEEEeqnarray}
The component-wise mean distortion thus verifies
\begin{IEEEeqnarray}{rCl}
D_{32} + \varepsilon
	&\geq&	\bE\left[ d\left (X_2,g_{32}\big(\mathcal{J}_{1}^{[1:K]}, \mathcal{J}_{2}^{[1:K]},\mathcal{J}_{3}^{[1:K]},X^n_3\big) \right) \right] \\
	&=& 	\frac1n \sum_{t=1}^n \bE\left[ d\left (X_{2[t]},\hat{X}_{32[t]}\left({\mathcal{V}}_{[23,K+1,2][t]},X_{3[t]}\right) \right) \right] \\
	&=& 	\frac1n \sum_{t=1}^n \bE\left[ d\left (X_{2[Q]}, \hat{X}_{32[Q]}\left({\mathcal{V}}_{[23,K+1,2][Q]},X_{3[Q]}\right) \right)\ \middle|\ Q=t \right] \\
	&=& 	\bE\left[ d\left (X_{2[Q]}, \hat{X}_{32[Q]}\left({\mathcal{V}}_{[23,K+1,2][Q]},X_{3[Q]}\right) \right) \right] \\
	&=& 	\bE\left[ d\left (X_2, \widetilde{X}_{32}\left(\widetilde{\mathcal{V}}_{[23,K+1,2]},X_3\right) \right) \right] \ ,
\end{IEEEeqnarray}
where we defined function $\widetilde{X}_{32}$ by 
\begin{IEEEeqnarray}{rCl}
\widetilde{X}_{32}\left(\widetilde{\mathcal{V}}_{[23,K+1,3]},X_3\right)& =& \widetilde{X}_{32}\left(Q, {\mathcal{V}}_{[23,K+1,3][Q]}, X_{3[Q]}\right)\nonumber  \\
&\triangleq & \hat{X}_{32[Q]}\left({\mathcal{V}}_{[23,K+1,3][Q]},X_{3[Q]} \right) \ .
\end{IEEEeqnarray}

This concludes the proof of the converse and thus that of the theorem. 
\end{IEEEproof}

\section{Discussion}
\label{sec:discuss}

\subsection{Numerical example}
\label{sec:num_results}

In order to obtain further insight into the gains obtained from cooperation, we consider the case of two encoders and one decoder subject to lossy/lossless reconstruction constraints without side information in which the sources are distributed according to:
\vspace{2mm}
\begin{IEEEeqnarray}{rCl}
\label{eq:num5}
p_{X_1X_2}(x_1,x_2)&=&\alpha\mathds{1}\left\{x_1=1\right\}\frac{1}{\sqrt{2\pi}\sigma_1}\exp{\left(-\frac{x_2^2}{2\sigma_1^2}\right)}\nonumber\\
&+&(1-\alpha)\mathds{1}\left\{x_1=0\right\}\frac{1}{\sqrt{2\pi}\sigma_0}\exp{\left(-\frac{x_2^2}{2\sigma_0^2}\right)}.
\end{IEEEeqnarray}
This model yields a mixed between discrete and continuous components. We observe  that $X_1$ follows a  \emph{Bernoulli} distribution with parameter $\alpha\in [0:1]$ while $X_2$ given $X_1$ follows a Gaussian distribution with different variance according to the value of $X_1\in\{0,1\}$. In this sense, $X_2$ follows  a Gaussian mixture distribution\footnote{Although the inner bound region in Theorem 1 is strictly valid for discrete sources with finite alphabets, the Gaussian distribution is sufficiently well-behaved to apply a uniform quantization procedure prior to the application of the results of Theorem 1. Then, a limiting argument using a sequence of decreasing quantization step-sizes will deliver the desired result. See chapter 3 in \cite{ELGamal-Kim-book}.}. 

The optimal rate-distortion region for this case was characterized  in Theorem~\ref{theo-Coop-BT} and  can be alternatively written as:
\begin{IEEEeqnarray}{rCl}
\label{eq:num1}
\mathcal{R}(D)=\bigcup_{p\in\mathcal{L}}\Big\{(R_1,R_2): R_1&>&H(X_1|X_2) \ ,\nonumber\\
 R_2&>&I(X_2;U|X_1)\ ,\nonumber\\
 R_1+R_2&>& H(X_1)+I(X_2;U|X_1)\Big\}\ ,
\end{IEEEeqnarray}
where
\begin{equation}
\label{eq:num2}
\mathcal{L}=\left\{p_{U|X_1X_2}: \textrm{ there exists }  (x_1,u)\mapsto g(x_1,u)\ \mbox{such that}\ \bE[d(X_2,g(X_1,U))]\leq D\right\}.
\end{equation}
The corresponding non-cooperative region for the same problem was characterized  in \cite{32119}:
\begin{IEEEeqnarray}{rCl}
\label{eq:num3}
\mathcal{R}(D)=\bigcup_{p\in\mathcal{L}^{\star}}\Big\{(R_1,R_2): R_1&>&H(X_1|U)\ ,\\ R_2&>&I(X_2;U|X_1)\ ,\\ R_1+R_2&>&H(X_1)+I(X_2;U|X_1)\Big\}\ ,
\end{IEEEeqnarray}
where
\begin{equation}
\label{eq:num4}
\mathcal{L}^{\star}=\left\{p_{U|X_2}: \textrm{ there exists }  (x_1,u)\mapsto g(x_1,u)\ \mbox{such that}\ \bE[d(X_2,g(X_1,U))]\leq D\right\}.
\end{equation}
From the previous expressions, it is evident that the cooperative case offers some gains with respect to the non-cooperative setup. This is clearly evidenced from the lower limit in $R_1$ and the fact that $\mathcal{L}^{\star}\subseteq\mathcal{L}$. We have the following result. \vspace{1mm}

\begin{theorem}[Cooperative region for mixed discrete/continuous source]
\label{theo:num1}
Assume the source distribution is given by (\ref{eq:num5}) and that, without loss of generality,  $\sigma_0^2\leq\sigma_1^2$. The rate-distortion region from Theorem~\ref{theo-Coop-BT} can be written as:
\begin{IEEEeqnarray}{rCl}
R_1 &>&\displaystyle\frac{1-\alpha}{\sqrt{2\pi}\sigma_0}\int_{-\infty}^{\infty}\exp{\left(-\frac{x_2^2}{2\sigma_0^2}\right)}H_2(g(x_2))dx_2\nonumber \\
\IEEEeqnarraymulticol{3}{r}{+\frac{\alpha}{\sqrt{2\pi}\sigma_1}\int_{-\infty}^{\infty}\exp{\left(- \frac{x_2^2}{2\sigma_1^2}\right)}H_2(g(x_2))dx_2}\ ,\nonumber\\
R_2 &>& \left\{\begin{array}{ll} \displaystyle\frac{1}{2}\log{\left(\displaystyle\frac{\sigma_0^{2(1-\alpha)}\sigma_1^{2\alpha}}{D}\right)} & D\leq \sigma_0^2\\
\displaystyle\frac{\alpha}{2}\left[\log{\left(\displaystyle\frac{\alpha\sigma_1^2}{D-(1-\alpha)\sigma_0^2}\right)}\right]^{+} & D>\sigma_0^2
\end{array}\right.,\nonumber\\
R_1+R_2 &>& \left\{\begin{array}{ll} H_2(\alpha)+\displaystyle\frac{1}{2}\log{\left(\displaystyle\frac{\sigma_0^{2(1-\alpha)}\sigma_1^{2\alpha}}{D}\right)} & D\leq \sigma_0^2\\
H_2(\alpha)+\displaystyle\frac{\alpha}{2}\left[\log{\left(\displaystyle\frac{\alpha\sigma_1^2}{D-(1-\alpha)\sigma_0^2}\right)}\right]^{+} & D>\sigma_0^2
\end{array}\right.,\nonumber
\end{IEEEeqnarray}
where $H_2(z)\equiv -z\log{z}-(1-z)\log{(1-z)}$  for $z\in[0,1]$, $\left[x\right]^{+}=\max{\left\{0,x\right\}}$ and 
\begin{equation}
\label{eq:num7}
g(x_2)=
\displaystyle\frac{\displaystyle\frac{\alpha}{\sqrt{2\pi}\sigma_1}\exp{\left(\displaystyle-\frac{x_2^2}{2\sigma_1^2}\right)}}{\displaystyle\frac{\alpha}{\sqrt{2\pi}\sigma_1}\exp{\left(-\displaystyle\frac{x_2^2}{2\sigma_1^2}\right)}+\displaystyle\frac{1-\alpha}{\sqrt{2\pi}\sigma_0}\exp{\left(-\displaystyle\frac{x_2^2}{2\sigma_0^2}\right)}}\ .
\end{equation}  
\end{theorem}
\vspace{1mm}
\begin{IEEEproof}
The converse proof  is straightforward by observing that when $D\leq \sigma_0^2$:
\begin{equation}
I(X_2;U|X_1)=h(X_2|X_1)-h(X_2|U,X_1)\geq h(X_2|X_1)-\frac{1}{2}\log{(2\pi e D) }
\label{eq:num8}
\end{equation}
and $h(X_2|X_1)=\frac{\alpha}{2}\log{(2\pi e\sigma_1^2)}+\frac{1-\alpha}{2}\log{(2\pi e\sigma_0^2)}$. For the case when $\sigma_0^2<D\leq \alpha\sigma_1^2+(1-\alpha)\sigma_0^2$ we can write:
\begin{IEEEeqnarray}{rCl}
I(X_1;U|X_1)&=& h(X_2|X_1)-h(X_2|U,X_1)\nonumber\\
&\geq & h(X_2|X_1)-\alpha h(X_2|U,X_1=1)-(1-\alpha)h(X_2|X_1=0)\nonumber\\
&=& \frac{\alpha}{2}\log{(2\pi e\sigma_1^2)}-\alpha \log{\left(\frac{2\pi e (D-(1-\alpha)\sigma_0^2)}{\alpha}\right)}\\
&=& \frac{\alpha}{2}\log{\left(\frac{\alpha\sigma_1^2}{D-(1-\alpha)\sigma_0^2}\right)}\ . 
\label{eq:num8b}
\end{IEEEeqnarray}
When $D>\alpha\sigma_1^2+(1-\alpha)\sigma_0^2$ we can lower bound the mutual information by zero.

The achievability follows from the choice:
\begin{equation}
\label{eq:num8c}
g(U,X_1)=\left\{\begin{array}{cc}
\left(\frac{\sigma_0^2}{\sigma_0^2+\sigma_{Z_0}^2}\right) U & \mbox{if}\ X_1=0\\
\left(\frac{\sigma_1^2}{\sigma_1^2+\sigma_{Z_1}^2}\right) U & \mbox{if}\ X_1=1\ ,
\end{array}\right. 
\end{equation}
and by setting the auxiliary random variable:
\begin{equation}
U=\left\{\begin{array}{cc}
X_2+Z_0 & \mbox{if}\ X_1=0\ ,\\
X_2+Z_1 & \mbox{if}\ X_1=1\ ,
\end{array}\right.
\label{eq:num9}
\end{equation}
where $Z_0,Z_1$ are zero-mean Gaussian random variables, independent from $X_2$ and $X_1$ and with variances given by:
\begin{equation}
\label{eq:num10}
\sigma_{Z_0}^2=\frac{D\sigma_0^2}{\sigma_0^2-D}\ ,\ \ \sigma_{Z_1}^2=\frac{D\sigma_1^2}{\sigma_1^2-D}\ ,
\end{equation}
for $D\leq \sigma_0^2$ while for  $\sigma_0^2<D\leq \alpha\sigma_1^2+(1-\alpha)\sigma_0^2$, we choose:
\begin{equation}
\label{eq:num10b}
\sigma_{Z_0}^2\rightarrow \infty\ ,\ \ \sigma_{Z_1}^2=\frac{\left[D-(1-\alpha)\sigma_0^2\right]\sigma_1^2}{\alpha\sigma_1^2-\left[D-(1-\alpha)\sigma_0^2\right]}\ .
\end{equation}
Finally, for $D>\alpha\sigma_1^2+(1-\alpha)\sigma_0^2$, we let  $\sigma_{Z_0}^2\rightarrow \infty$ and $\sigma_{Z_1}^2\rightarrow \infty$.
\end{IEEEproof}

Unfortunately, the non-cooperative region is hard to evaluate for the assumed source model\footnote{However, there are cases where an exact characterization is possible. This is the case, for example,  when $X_1$ and $X_2$ are the input and output of binary channel with crossover probability $\alpha$ and the distortion function is the Hamming distance \cite{32119}.}. In order to present some comparison between the cooperative and non-cooperative case let us fix the same value for the rate $R_1$ in both cases and compare the rate $R_2$ that can be obtained in each case. Clearly, in this way, we are not taking into account the gain in $R_1$ that could be obtained by the cooperative scheme (as $H(X_1|X_2)\leq H(X_1|U)$ for every $U\mkv X_2\mkv X_1$). For both schemes, it follows that for fixed $R_1$: 
\begin{equation}
R_2>\max{\left\{I(X_2;U|X_1),H(X_1)+I(X_2;U|X_1)-R_1\right\}}\ .
\label{eq:num11}
\end{equation}
From Theorem~\ref{theo:num1}, we can compute (\ref{eq:num11}) for the cooperative case. For the non-cooperative case we need to obtain a lower bound on $I(X_2;U|X_1)$ for $p_{U|X_1}\in\mathcal{L}^{\star}$. It is easy to check that:
\begin{equation}
I(X_2;U|X_1)\geq \frac{1-\alpha}{2}\log{\left(\frac{\sigma_0^2}{\beta_0}\right)}+\frac{\alpha}{2}\log{\left(\frac{\sigma_1^2}{\beta_1}\right)}\ ,
\label{eq:num12}
\end{equation}
where
\begin{equation}
\beta_0=\mathbb{E}_{X_2U|X_1}\left[\left(X_2-\mathbb{E}_{X2|UX_1}\left[X_2|U,X_1=0\right]\right)^2|X_1=0\right],
\label{eq:num13}
\end{equation}
\begin{equation}
\beta_1=\mathbb{E}_{X_2U|X_1}\left[\left(X_2-\mathbb{E}_{X2|UX_1}\left[X_2|U,X_1=1\right]\right)^2|X_1=1\right].
\label{eq:num14}
\end{equation}
The distortion constraint imposes the condition:
\begin{equation}
(1-\alpha)\beta_0+\alpha\beta_1\leq D\ .
\label{eq:num15}
\end{equation}
In order to guarantee that (\ref{eq:num12}) and (\ref{eq:num13}) are achievable, under the Markov constraint  $U\mkv X_2\mkv X_1$, the following conditions on $p_{U|X_2}(u|x_2)$ should be satisfied:
\begin{equation}
\frac{p_{U|X_2}(u|x_2)\displaystyle \frac{1}{\sqrt{2\pi}\sigma_0}\exp{\left(-\frac{x_2^2}{2\sigma_0^2}\right)}}{\int_{-\infty}^\infty p_{U|X_2}(u|x_2)\displaystyle \frac{1}{\sqrt{2\pi}\sigma_0}\exp{\left(-\frac{x_2^2}{2\sigma_0^2}\right)}dx_2}=\frac{1}{\sqrt{2\pi\beta_0}}\exp{\left\{-\frac{\left(x_2-f_0(u)\right)^2}{2\beta_0}\right\}}
\label{eq:num16}
\end{equation}
and 
\begin{equation}
\frac{p_{U|X_2}(u|x_2)\displaystyle \frac{1}{\sqrt{2\pi}\sigma_1}\exp{\left(-\frac{x_2^2}{2\sigma_1^2}\right)}}{\int_{-\infty}^\infty p_{U|X_2}(u|x_2)\displaystyle\frac{1}{\sqrt{2\pi}\sigma_1}\exp{\left(-\frac{x_2^2}{2\sigma_1^2}\right)}dx_2}=\frac{1}{\sqrt{2\pi\beta_1}}\exp{\left\{-\frac{\left(x_2-f_1(u)\right)^2}{2\beta_1}\right\}}\ ,
\label{eq:num17}
\end{equation}
where 
\begin{equation}
f_0(U)\equiv\mathbb{E}_{X2|UX_1}\left[X_2|U,X_1=0\right],\ f_1(U)\equiv\mathbb{E}_{X2|UX_1}\left[X_2|U,X_1=1\right]\ .
\label{eq:num18}
\end{equation}
The characterization of all distributions $p_{U|X_2}(u|x_2)$ that satisfies (\ref{eq:num17}) and (\ref{eq:num18}) appears to be a difficult  problem. In order to show a numerical example,  we shall simply assume that:
\begin{equation}
p_{U|X_2}(u|x_2)=\frac{1}{\sqrt{2\pi}\sigma_w}\exp{\left\{-\frac{\left(u-x_2\right)^2}{2\sigma_w^2}\right\}}\ .
\label{eq:num19}
\end{equation}
Indeed, this choice satisfies simultaneously expressions (\ref{eq:num17}) and (\ref{eq:num18}). In this way, we can calculate the corresponding values of $\beta_0$ and $\beta_1$ obtaining the parametrization of $I(X_2;U|X_1)$ as function of $\sigma_w^2$:
\begin{equation}
I(X_2;U|X_1)=\frac{1-\alpha}{2}\log{\left(\frac{\sigma_0^2+\sigma_w^2}{\sigma_w^2}\right)}+\frac{\alpha}{2}\log{\left(\frac{\sigma_1^2+\sigma_w^2}{\sigma_w^2}\right)}
\label{eq:num20}
\end{equation}
with the following constraint:
\begin{equation}
(1-\alpha)\frac{\sigma_w^2\sigma_0^2}{\sigma_0^2+\sigma_w^2}+\alpha\frac{\sigma_w^2\sigma_1^2}{\sigma_1^2+\sigma_w^2}=D\ .
\label{eq:num21}
\end{equation}
We can replace (\ref{eq:num21}) in (\ref{eq:num11}) to obtain an indication of the performance of the non-cooperative case when $R_1$ is fixed. 

We present now some numerical evaluations. As equation (\ref{eq:num11}) is valid for both the cooperative and the non-cooperative setups, it is sufficient to compare the mutual information term $I(X_2;U|X_1)$ for each of them.  Let us consider the next scenarios: 
\begin{enumerate}
\item $\alpha=0.1$, $\sigma_0^2=0.01$, $\sigma_1^2=2$, 
\item $\alpha=0.1$, $\sigma_0^2=0.5$, $\sigma_1^2=2$.
\end{enumerate}
 
 \begin{figure}[th]
 \centering
 \ifpdf\includegraphics[angle=0,width=1\columnwidth,keepaspectratio,trim= 0mm 0mm 0mm 0mm,clip]{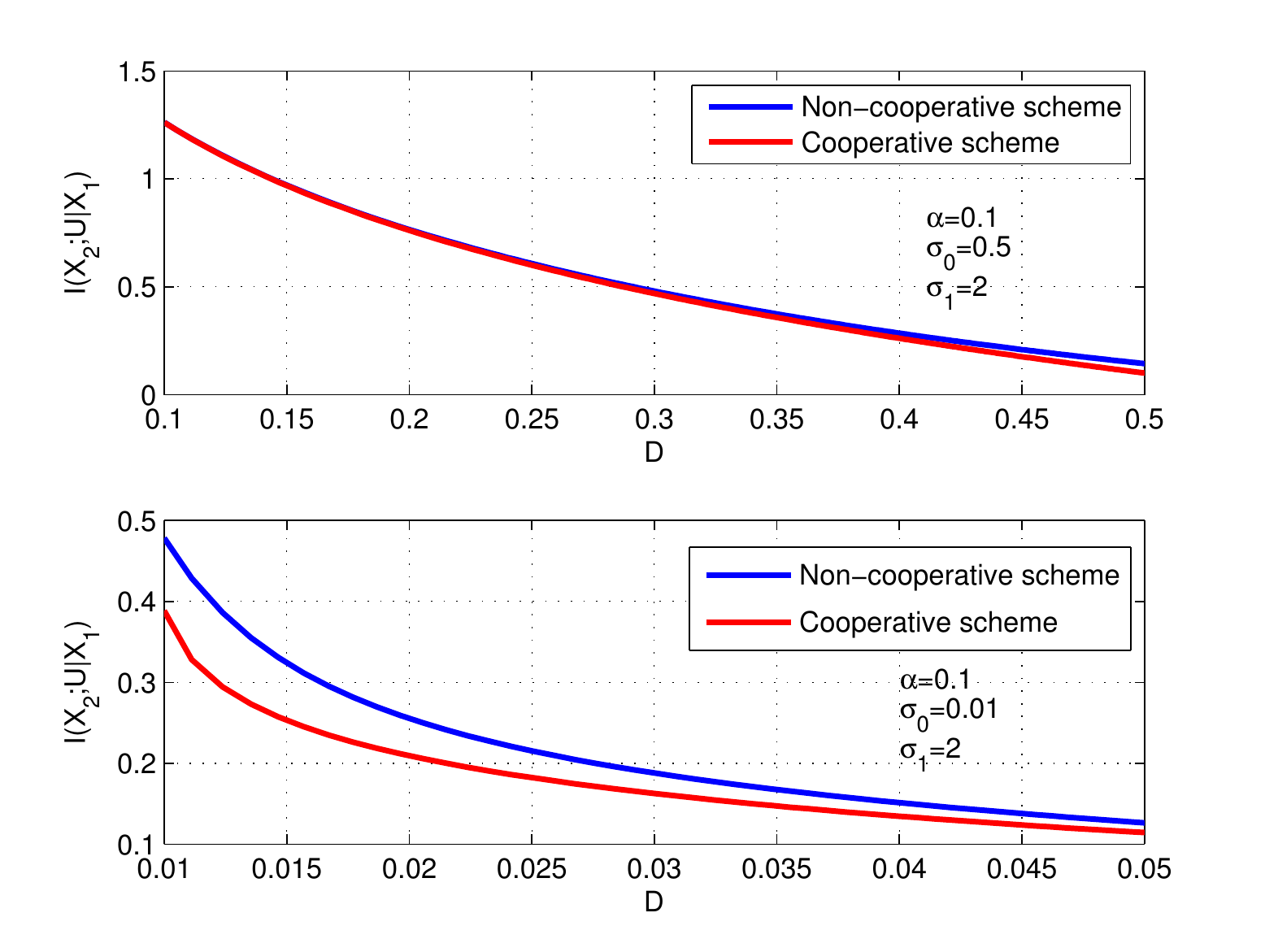} 
 \else
 	\includegraphics[angle=270,width=0.65\columnwidth,keepaspectratio,trim= 0mm 0mm 0mm 0mm,clip]{model.eps} 
 \fi
 \caption{Comparison between the cooperative and the non-cooperative schemes.}
 \label{fig:small_sigma_0}
 \end{figure}
 
%

From Fig.  \ref{fig:small_sigma_0} we see that in the case $\sigma_0^2\ll\sigma_1^2$ the gain of the cooperative scheme is pretty noticeable. However, as $\sigma_0^2$ becomes comparable to $\sigma_1^2$ the gains are reduced. This was expected from the fact that as $\sigma_0^2\rightarrow\sigma_1^2$, the random variable $X_2$ converges to a Gaussian distribution. In that case, the reconstruction of $X_2$ at Node $3$ is equivalent for the cooperative scenario to a lossy source coding problem with side information $X_1$ at both the encoder and the decoder while for the non-cooperative setting to the standard Wyner-Ziv problem. It is known that in this case there is no gains that can be expected~\cite{1055508}. 

\subsection{Interactive Lossless  Source Coding}

\begin{figure}[th]
\centering
\ifpdf\includegraphics[angle=0,width=0.65\columnwidth,keepaspectratio,trim= 0mm 0mm 0mm 0mm,clip]{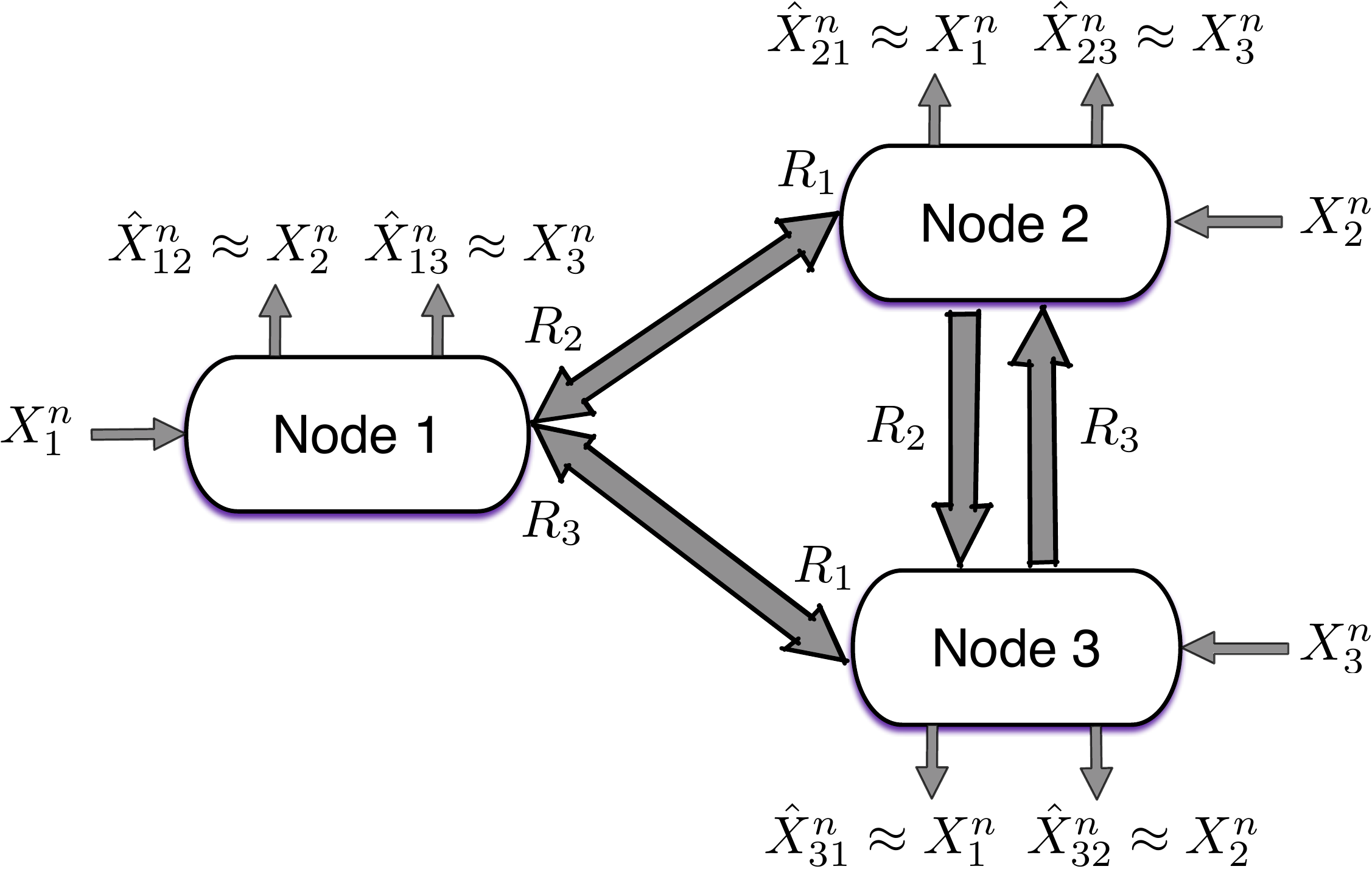} 
\else
	\includegraphics[angle=270,width=0.65\columnwidth,keepaspectratio,trim= 0mm 0mm 0mm 0mm,clip]{model.eps} 
\fi
\caption{Interactive lossless  source coding.}
\label{fig:model-Coop-HB}
\end{figure}

Consider now the problem described in Fig.~\ref{fig:model-Coop-HB} where encoder $1$ wishes to communicate \emph{lossless} the source $X_1^n$ to two decoders which observe sources $X_2^n$ and $X_3^n$. At the same time node 1 wishes to recover $X_2^n$ and $X_3^n$ lossless.  Similarly the other encoders want to communicate \emph{lossless} their sources and recover the sources from the rest. It is wanted to do this through $K$ rounds of exchanges.  


\begin{theorem}[Interactive lossless  source coding]\label{theo-Coop-HB}
The rate region of the setting described in Fig.~\ref{fig:model-Coop-HB} is given by the set of all tuples satisfying: 
\begin{IEEEeqnarray}{rcl}
R_1&>&H(X_1|X_2X_3)\ ,\\
R_2&>&H(X_2|X_1X_3)\ ,\\
R_3&>&H(X_3|X_1X_2)\ ,\\
R_1+R_2&>&H(X_1X_2|X_3)\ ,\\
R_1+R_3&>&H(X_1X_3|X_2)\ ,\\
R_2+R_3&>&H(X_2X_3|X_1)\ .
\end{IEEEeqnarray}
\end{theorem}
\vspace{1mm}
\begin{remark}
It is worth observing that the multiple exchanges of descriptions between all nodes cannot increase the rate compared to standard Slepian-Wolf coding~\cite{Slepian_1973}. 
\end{remark}
\vspace{1mm}

\begin{IEEEproof}
The achievability part is a standard exercise. The converse proof is straightforward from cut-set arguments. For these reasons both proofs are ommited.
\end{IEEEproof}
We should note that for this important case, Theorem~\ref{theo-main-theorem} does not provide the optimal rate-region. That is, the coding scheme used is not optimal for this case. In fact, from Theorem~\ref{theo-main-theorem} and for this problem we can obtain the following achievable region\footnote{Consider $U_{1\rightarrow 23,1}\equiv X_1$, $U_{2\rightarrow 13,1}\equiv X_2$ and $U_{3\rightarrow 12,1}\equiv X_3$ and the other auxiliary random variables to be constants for all $l\in[1:K]$.}:
\begin{IEEEeqnarray}{rCl}
R_1&>&H(X_1|X_2)\ ,\\
R_2&>&H(X_2|X_1X_3)\ ,
\end{IEEEeqnarray}
\begin{IEEEeqnarray}{rCl}
R_3&>&H(X_3|X_1X_2)\ ,\\
R_1+R_2&>&H(X_1X_2|X_3)\ ,\\
R_2+R_3&>&H(X_2X_3|X_1)\ .
\end{IEEEeqnarray}
It is easily seen that in this region the node 2 is not performing joint decoding of the descriptions generated at node 1 and 3. Because of the encoding ordering assumed ($1\rightarrow 2\rightarrow 3$) and the fact that the common description generated in node 2 should be conditionally generated on the common description generated at node 1, node 2 has to recover first this common description first. At the end, it recovers the common description generated at node 3. On the other hand, nodes 1 and 3 perform joint decoding of the common information generated at nodes 2 and 3, and at nodes 1 and 2, respectively. Clearly, this is consequence of the \emph{sequential} encoding and decoding structure imposed between the nodes in the network and which is the basis of the interaction. 

If all nodes would be allowed to perform a joint decoding procedure in order to recover all the exchanged descriptions only at the end of each  round,  this problem would not appear. However, this would destroy the sequential encoding-decoding structure assumed in the paper which seems to be optimal in other situations.
\section{Summary}
\label{sec:summary}
The three-node multiterminal lossy source coding problem was investigated. This problem is not a straightforward generalization of the original problem posed by Kaspi in 1985. As this general problem encompasses several open problems in multiterminal rate distortion the mathematical complexity of it is formidable. For that reason we only provided a general inner bound for the rate distortion region. It is shown that this (rather involved) inner bound contains several rate-distortion regions of some relevant source coding settings. It In this way, besides the non-trivial extension of the interactive two terminal problem, our results can be seen as a generalization and hence unification of several previous works in the field. We also showed, that our inner bound provides definite answers to special cases of the general problem. It was shown that in some cases the cooperation induced by the interaction can be helpful while in others not.  It is clear that further study is needed on the topic of multiple terminal cooperative coding, including a proper generalization to larger networks and to the problem of interactively estimating arbitrary functions of the sources sensed at the nodes.

  \newpage
\appendices

 \section{Strongly typical sequences and related results}
  \label{app:strongly}
  In this appendix we introduce standard notions in information theory but suited for the mathematical developments and proof needed in this work. The results presents can be easily derived from the standard formulations provided in~\cite{ELGamal-Kim-book} and~\cite{Csiszar81}. Be $\mathcal{X}$ and $\mathcal{Y} $  finite alphabets and $(x^n,y^n)\in\mathcal{X}^n\times \mathcal{Y}^n$.  With $\mathcal{P}(\mathcal{X}\times \mathcal{Y})$ we denote the set of all probability distributions on $\mathcal{X}\times\mathcal{Y}$. We define the \emph{strongly $\delta$-typical} sets as:
 
 \begin{definition}[Strongly typical set]
 Consider $p\in\mathcal{P}(\mathcal{X})$ and $\delta>0$. We say that $x^n\in\mathcal{X}^n$ is $p\delta$- strongly typical if $x^n\in\mathcal{T}_{[p]\delta}^n$ with:
 \begin{equation}
 \mathcal{T}_{[p]\delta}^n=\left\{x^n\in\mathcal{X}^n:\Big|\frac{N(a|x^n)}{n}-p(a)\Big|\leq\frac{\delta}{\|\mathcal{X}\|},\ \forall a\in\mathcal{X}\ \mbox{such that}\ p(a)\neq 0\right\}\ ,
 \label{eq:strong_delta_typ}
 \end{equation}
 where $N(a|x^n)$ denotes de number of occurrences of $a\in\mathcal{X}$ in $x^n$ and $p\in\mathcal{P}(\mathcal{X})$.  When $X\sim p_X(x)$ we can denote the corresponding set of strongly typical sequences as $\mathcal{T}_{[X]\delta}^n$. 
  \end{definition}
 Similarly, given $p_{XY}\in\mathcal{P}\left(\mathcal{X}\times\mathcal{Y}\right)$ we can construct the set of $\delta$-jointly typical sequences as:
  \begin{gather}
  \mathcal{T}_{[XY]\delta}^n=\left\{(x^n,y^n)\in\mathcal{X}^n\times\mathcal{Y}^n:\Big|\frac{N(a,b|x^n,y^n)}{n}-p_{XY}(a,b)\Big|\leq\frac{\delta}{\|\mathcal{X}\|\|\mathcal{Y}\|},\right.\nonumber\\
  \left.\ \forall (a,b)\in\mathcal{X}\times\mathcal{Y}\ \mbox{such that}\ p_{XY}(a,b)\neq 0\right\}\ .
  \label{eq:joint_strong_delta_typ}
  \end{gather}
 We also define the \emph{conditional} typical sequences. In precise terms, given $x^n\in\mathcal{X}^n$ we consider the set:
 \begin{gather}
 \mathcal{T}_{[Y|X]\delta}^n(x^n)=\Big\{y^n\in\mathcal{Y}^n:\Big|\frac{N(a,b|x^n,y^n)}{n}-p_{XY}(a,b)\Big|\leq\frac{\delta}{\|\mathcal{X}\|\|\mathcal{Y}\|},\nonumber\\
\ \forall (a,b)\in\mathcal{X}\times\mathcal{Y}\ \mbox{such that}\ p_{XY}(a,b)\neq 0\Big\}\ .
\label{eq:cond_strong_delta_typ}
 \end{gather} 
 Notice that we the following is an alternative writing of this set:
 \begin{equation} 
 \mathcal{T}_{[Y|X]\delta}^n(x^n)=\left\{y^n\in\mathcal{Y}^n:(x^n,y^n)\in\mathcal{T}_{[XY]\delta}^n\right\} \ .
 \end{equation}
 
 We have several useful and standard Lemmas, which will be presented without proof (except for the last one):
 
 \begin{lemma}[Properties of typical sets~\cite{Csiszar81}] 
 The following are true:
 \begin{enumerate}
 \item  Consider $(x^n,y^n)\in\mathcal{T}_{[XY]\epsilon}^n$. Then, $x^n\in\mathcal{T}_{[X]\epsilon}^n$, $y^n\in\mathcal{T}_{[Y]\epsilon}$, $x^n\in\mathcal{T}_{[X|Y]\epsilon}^n(y^n)$ and $y^n\in\mathcal{T}_{[Y|X]\epsilon}^n(x^n)$ \ .
 \item Be $T_{[Y|X]\epsilon}^n(x^n)$ with $x^n\notin\mathcal{T}^n_{[X]\epsilon}$. Then $T_{[Y|X]\epsilon}^n(x^n)=\varnothing$ \ .
 \item Be $(X^n,Y^n)\sim\prod_{t=1}^np_{XY}(x_t,y_t)$. If $x^n\in\mathcal{T}_{[X]\epsilon}^n$ we have
 \[2^{-n(H(X)+\delta(\epsilon))}\leq p_{X^n}(x^n)\leq 2^{-n(H(X)-\delta(\epsilon))} \]
 with $\delta(\epsilon)\rightarrow 0$ when $\epsilon\rightarrow 0$. Similarly, if $y^n\in\mathcal{T}_{[Y|X]\epsilon}^n(x^n)$:
  \[2^{-n(H(Y|X)+\delta'(\epsilon))}\leq p_{Y^n|X^n}(y^n|x^n)\leq 2^{-n(H(Y|X)-\delta'(\epsilon))} \]
  with $\delta'(\epsilon)\rightarrow 0$ when $\epsilon\rightarrow 0$\ .
 \label{lemma:cont}
 \end{enumerate}
 \label{lemma:useful}
 \end{lemma}
 
 \begin{lemma}[Conditional typicality lemma~\cite{Csiszar81}]
 Consider de product measure $\prod_{t=1}^np_{XY}(x_t,y_t)$. Using that measure, we have the following 
 \[\operatorname{Pr}\left\{\mathcal{T}_{[X]\epsilon}^n\right\}\geq 1-\mathcal{O}\left(c_1^{-nf(\epsilon)}\right),\ \ c_1>1\]
 where $f(\epsilon)\rightarrow 0$ when $\epsilon\rightarrow 0$. In addition, for every  $x^n\in\mathcal{T}^n_{[X]\epsilon'}$ with $\epsilon'<\frac{\epsilon}{\|\mathcal{Y}\|}$ we have:
 \[\operatorname{Pr}\left\{\mathcal{T}_{[Y|X]\epsilon}^n(x^n)|x^n\right\}\geq 1-\mathcal{O}\left(c_2^{-ng(\epsilon,\epsilon')}\right),\ \ c_2>1\]
 where $g(\epsilon,\epsilon')\rightarrow 0$ when $\epsilon,\epsilon'\rightarrow 0$.
 \label{lemma:prob_lim}
 \end{lemma}
 
 \begin{lemma}[Size of typical sets~\cite{Csiszar81}]
Given $p_{XY}\in\mathcal{P}\left(\mathcal{X}\times\mathcal{Y}\right)$ we have
  \[\frac{1}{n}\|\mathcal{T}_{[X]\epsilon}^n\|\leq H(X)+\delta(\epsilon),\ \  \frac{1}{n}\|\mathcal{T}_{[X]\epsilon}^n\|\geq H(X)-\delta'(\epsilon,n)\]
  where $\delta(\epsilon),\delta'(\epsilon,n)\rightarrow 0$ when $\epsilon\rightarrow 0$ and $n\rightarrow\infty$. Similarly for every $x^n\in\mathcal{X}^n$ we have:
  \[\frac{1}{n}\|\mathcal{T}_{[Y|X]\epsilon}^n(x^n)\|\leq H(Y|X)+\delta(\epsilon)\]
  with $\delta(\epsilon)\rightarrow 0$ with $\epsilon\rightarrow 0$. In addition, for every $x^n\in\mathcal{T}_{[X]\epsilon'}^n$ with $\epsilon'<\frac{\epsilon}{\|\mathcal{Y}\|}$ we have:
    \[\frac{1}{n}\|\mathcal{T}_{[Y|X]\epsilon}^n(x^n)\|\geq H(Y|X)-\delta'(\epsilon,\epsilon',n)\]
 where $\delta'(\epsilon,\epsilon',n)\rightarrow 0$ when $\epsilon,\epsilon'\rightarrow 0$ and $n\rightarrow\infty$.
 \label{lemma:size}
 \end{lemma}
 
 \begin{lemma}[Joint typicality lemma~\cite{Csiszar81}]
 \label{lemma:joint_typ_strong}
 Consider $(X,Y,Z)\sim p_{XYZ}(x,y,z)$ and $(x^n,y^n)\in\mathcal{T}_{[XY]\epsilon'}^n$ with $\epsilon'<\frac{\epsilon}{\|\mathcal{Z}\|}$ and $\epsilon<\epsilon''$. If $p_{Z^n|X^n}(z^n|x^n)=\frac{\mathds{1}\left\{z^n\in\mathcal{T}_{[Z|X]\epsilon''}^n(x^n)\right\}}{\|\mathcal{T}_{[Z|X]\epsilon''}(x^n)\|}$ there exists $\delta'(\epsilon,\epsilon',\epsilon'',n)$ which goes to zero  as $\epsilon,\epsilon',\epsilon''\rightarrow 0$ and $n\rightarrow\infty$ and: 
 \[2^{-n(I(Y;Z|X)+\delta')}\leq\operatorname{Pr}\left\{\tilde{Z}^n\in\mathcal{T}_{[Z|XY]\epsilon}^n(x^n,y^n)\right\}\leq 2^{-n(I(Y;Z|X)-\delta')}\]
 \end{lemma}

\begin{lemma}[Covering Lemma~\cite{ELGamal-Kim-book}]
\label{lemma:covering}
Be $(U,V,X)\sim p_{UVX}$ and $(x^n,u^n)\in\mathcal{T}_{[XU]\epsilon'}^n$, $\epsilon'<\frac{\epsilon}{\|\mathcal{V}\|}$ and $\epsilon<\epsilon''$. Consider also $\left\{V^n(m)\right\}_{m=1}^{2^{nR}}$ random vectors  which are independently generated according to $\frac{\mathds{1}\left\{v^n\in\mathcal{T}_{[V|U]\epsilon''}^n(u^n)\right\}}{\|\mathcal{T}_{[V|U]\epsilon''}(u^n)\|}$. Then:
\begin{equation}
\mbox{Pr}\left\{V^n(m)\notin\mathcal{T}_{[V|UX]\epsilon}^n(x^n,u^n)\ \mbox{for all}\ m\right\}\xrightarrow[n\rightarrow\infty]{}0
\end{equation}
uniformly for every  $(x^n,u^n)\in\mathcal{T}^n_{[XU]\epsilon'}$  if: 
\begin{equation}
R>I\left(V;X|U\right)+\delta(\epsilon,\epsilon',\epsilon'',n)
\label{eq:covering_lemma}
\end{equation}
where $\delta(\epsilon,\epsilon',\epsilon'',n)\rightarrow 0$ when $\epsilon,\epsilon',\epsilon''\rightarrow 0$ and $n\rightarrow \infty$.
\end{lemma}
\begin{corollary}
\label{coro:covering}
Assume the conditions in Lemma~\ref{lemma:covering}, and  also:
\begin{equation}
\mbox{Pr}\left\{\left(X^n,U^n\right)\in\mathcal{T}^n_{[XU]\epsilon'}\right\}\xrightarrow[n\rightarrow\infty]{}1\ .
\end{equation}
Then:
\begin{equation}
\mbox{Pr}\left\{\left(U^n,X^n,V^n(m))\right)\notin\mathcal{T}_{[UXV]\epsilon}^n\ \mbox{for all}\ m\right\}\xrightarrow[n\rightarrow\infty]{}0
\end{equation}
when (\ref{eq:covering_lemma}) is satisfied.
\end{corollary}

\begin{lemma}[Packing Lemma~\cite{ELGamal-Kim-book}]
\label{lemma:packing}
Be $(U_1U_2WV_1V_2X)\sim p_{U_1U_2WV_1V_2X}$, $(x^n,w^n,v_1^n,v_2^n)\in\mathcal{T}^n_{[XWV_1V_2]\epsilon'}$ and  $\epsilon'<\frac{\epsilon}{\|\mathcal{U}_1\|\|\mathcal{U}_2\|}$ and $\epsilon<\min{\left\{\epsilon_1,\epsilon_2\right\}}$. Consider random vectors $\left\{U_1^n(m_1)\right\}_{m_1=1}^{\mathcal{A}_1}$ and $\left\{U_2^n(m_2)\right\}_{m_2=1}^{\mathcal{A}_2}$ which are independently generated according to 
\begin{equation*}
\frac{\mathds{1}\left\{u_i^n\in\mathcal{T}_{[U_i|V_iW]\epsilon_i}^n(v_i^n,w^n)\right\}}{\|\mathcal{T}_{[U_i|V_iW]\epsilon_i}(w^n,v_i^n)\|},\ i=1,2\ ,
\end{equation*}
and $\mathcal{A}_1,\mathcal{A}_2$ are positive random variables independent of everything else. Then
\begin{equation}
\mbox{Pr}\left\{\left(U_1^n(m_1),U_2^n(m_2)\right)\in\mathcal{T}_{[U_1U_1|XWV_1V_2]\epsilon}^n(x^n,w^n,v_1^n,v_2^n)\ \mbox{for some}\ (m_1,m_2)\right\}\xrightarrow[n\rightarrow\infty]{}0
\end{equation}
uniformly for every  $(x^n,w^n,v_1^n,v_2^n)\in\mathcal{T}^n_{[XWV_1V_2]\epsilon'}$  provided that: 

\begin{equation}
\frac{\log{\mathbb{E}\left[\mathcal{A}_1\mathcal{A}_2\right]}}{n}<I\left(U_1;XV_2U_2|WV_1\right)+I\left(U_2;XV_1U_1|WV_2\right)-I\left(U_1;U_2|XWV_1V_2\right)-\delta
\label{eq:packing_lemma}
\end{equation}
where $\delta\equiv\delta(\epsilon,\epsilon',\epsilon_1,\epsilon_2,n)\rightarrow 0$ when $\epsilon,\epsilon',\epsilon_1,\epsilon_2\rightarrow 0$ and $n\rightarrow \infty$.
\end{lemma}

\begin{corollary}
\label{coro:packing}
Assume the conditions in Lemma~\ref{lemma:packing}, and  also:
\begin{equation}
\mbox{Pr}\left\{\left(X^n,W^n,V_1^n,V_2^n\right)\in\mathcal{T}^n_{[XWV_1V_2]\epsilon'}\right\}\xrightarrow[n\rightarrow\infty]{}1
\end{equation}
Then:
\begin{equation}
\mbox{Pr}\left\{\left(U_1^n(m_1),U_2^n(m_2),X^n,W^n,V_1^n,V_2^n)\right)\in\mathcal{T}_{[U_1U_1XWV_1V_2]\epsilon}^n\ \mbox{for some}\ (m_1,m_2)\right\}\xrightarrow[n\rightarrow\infty]{}0
\end{equation}
when (\ref{eq:packing_lemma}) is satisfied.
\end{corollary}

\begin{lemma}[Generalized Markov Lemma~\cite{GML_ours_2014} ]
Consider a pmf $p_{UXY}$ belonging to $\mathcal{P}\left(\mathcal{X}\times\mathcal{Y}\times\mathcal{U}\right)$ and that satisfies de following:
\[Y\mkv X\mkv U\]
Consider $(x^n,y^n)\in\mathcal{T}^n_{[XY]\epsilon'}$ and random vectors $U^n$  generated according to:

\begin{equation}
\label{eq:u_dist}
\mbox{Pr}\left\{U^n=u^n\Big|x^n,y^n,U^n\in\mathcal{T}^n_{[U|X]\epsilon''}(x^n)\right\}=\frac{\mathds{1}\left\{u^n\in\mathcal{T}_{[U|X]\epsilon''}^n(x^n)\right\}}{\|\mathcal{T}_{[U|X]\epsilon''}^n(x^n)\|}
\end{equation}
For sufficiently small $\epsilon,\epsilon',\epsilon''$ the following holds uniformly for every  $(x^n,y^n)\in\mathcal{T}^n_{[XY]\epsilon'}$:

\begin{equation}
\mbox{Pr}\left\{U^n\notin\mathcal{T}^n_{[U|XY]\epsilon}(x^n,y^n)\Big|x^n,y^n,U^n\in\mathcal{T}^n_{[U|X]\epsilon''}(x^n)\right\}=\mathcal{O}\left(c^{-n}\right)
\end{equation}
where $c>1$.
  \label{lemma:markov}
  \end{lemma}

\begin{corollary}
\label{coro:markov}
Assume the conditions in Lemma~\ref{lemma:markov}, and also:
\begin{equation}
\mbox{Pr}\left\{\left(X^n,Y^n\right)\in\mathcal{T}^n_{[XY]\epsilon'}\right\}\xrightarrow[n\rightarrow\infty]{}1
\end{equation}
and that uniformly for every $(x^n,y^n)\in\mathcal{T}^n_{[XY]\epsilon'}$:
\begin{equation}
\mbox{Pr}\left\{U^n\notin\mathcal{T}^n_{[U|X]\epsilon''}(x^n)\Big| x^n,y^n\right\}\xrightarrow[n\rightarrow\infty]{}0
\end{equation}
we obtain:
\begin{equation}
\mbox{Pr}\left\{(U^n,X^n,Y^n)\in\mathcal{T}^n_{[UXY]\epsilon}\right\}\xrightarrow[n\rightarrow\infty]{}1\ .
\end{equation}
\end{corollary}
 Lemma~\ref{lemma:markov} and Corollary~\ref{coro:markov} will be central for us. They will guarantee the joint typicality of the descriptions generated in different encoders considering the pmf of the chosen descriptions induced by the coding scheme used. 
 The original proof of this result is given in~\cite{PhD-Tung} and involves a combination of rather sophisticated algebraic and combinatorial arguments over finite alphabets. Alternative  proof was also provided  in~\cite{ELGamal-Kim-book}, 
 \begin{equation}
 \mbox{Pr}\left\{(U^nX^n,Y^n)\in\mathcal{T}^n_{[UXY]\epsilon}\right\}\xrightarrow[n\rightarrow\infty]{}1
 \end{equation}
which strongly relies on a rather obscure result by Uhlmann~\cite{Uhlmann} on combinatorics. In~\cite{GML_ours_2014} a short and more general proof of this result is given.

We next present a result which will be useful for proving Theorem  \ref{theo-main-theorem}. In order to use the Markov Lemma we need to show that the descriptions induced by the encoding procedure in each node satisfies (\ref{eq:u_dist}).
   \begin{lemma}[Encoding induced distribution]   \label{lemma:induced}
 Consider a pmf $p_{UXW}$ belonging to $\mathcal{P}\left(\mathcal{U}\times\mathcal{X}\times\mathcal{W}\right)$ and $\epsilon'\geq\epsilon$.  Be $\left\{U^n(m)\right\}_{m=1}^{S}$ random vectors independently generated according to 
 $$
 \frac{\mathds{1}\left\{u^n\in\mathcal{T}_{[U|W]\epsilon'}^n(w^n)\right\}}{\|\mathcal{T}_{[U|W]\epsilon'}(w^n)\|}
 $$ 
 and where $(W^n,X^n)$ are generated with an arbitrary distribution. Once these vectors are generated, and given $x^n$ and $w^n$, we choose one of them if:
  \begin{equation}
  \left(u^n(m),w^n,x^n\right)\in\mathcal{T}^n_{[UWX]\epsilon},\ \mbox{for some}\ m\in[1:S] \ .\end{equation}
 If there are various vectors $u^n$ that satisfies this we choose the one with smallest index. If there are none we choose an arbitrary one. Let $M$ denote the index chosen. 
 Then we have that:
  \begin{equation}
 \mbox{Pr}\left\{U^n(M)=u^n\Big|x^n,w^n,U^n(M)\in\mathcal{T}^n_{[U|XW]\epsilon}(x^n,w^n)\right\}=
 \frac{\mathds{1}\left\{u^n\in\mathcal{T}_{[U|XW]\epsilon}^n(x^n,w^n)\right\}}{\|\mathcal{T}_{[U|XW]\epsilon}(x^n,w^n)\|} \ .
 \end{equation}
   \end{lemma}
   \begin{IEEEproof}
   From the selection procedure for $M$ we know that:
  \begin{equation}
  M=f\left(\mathds{1}\left\{(U^n(m),X^n,W^n)\in\mathcal{T}_{[UXW]\epsilon}^n\right\}, m\in[1:S]\right)\ , \end{equation}
   where $f(\cdot)$ is an appropriate function. Moreover, because of this and the way in which the random vectors $U^n$ are generated we have:
\begin{equation}
U^n(M)\mkv\left(\mathds{1}\left\{(U^n(M),X^n,W^n)\in\mathcal{T}_{[UXW]\epsilon}^n\right\},W^n,X^n\right)\mkv M\ .
\label{eq:mkv}
\end{equation}
We can write:
\begin{IEEEeqnarray}{lCl}
\mbox{Pr}\left\{U^n(M)=u^n\Big|x^n,w^n,U^n(M)\in\mathcal{T}^n_{[U|XW]\epsilon}(x^n,w^n)\right\}=\nonumber\\
\sum_{m=1}^S\mbox{Pr}\left\{M=m\Big|x^n,w^n,U^n(m)\in\mathcal{T}^n_{[U|XW]\epsilon}(x^n,w^n)\right\}\times\nonumber\\
\mbox{Pr}\left\{U^n(m)=u^n\Big|x^n,w^n,U^n(m)\in\mathcal{T}^n_{[U|XW]\epsilon}(x^n,w^n),M=m\right\}.
\label{eq:mkv2}
\end{IEEEeqnarray}
From (\ref{eq:mkv}), the second probability term in the RHS of (\ref{eq:mkv2}) can be written as:
\begin{equation}
\mbox{Pr}\left\{U^n(m)=u^n\Big|x^n,w^n,U^n(m)\in\mathcal{T}^n_{[U|XW]\epsilon}(x^n,w^n)\right\}\ \forall m\ .
\end{equation}
We are going to analyze this term. It is clear that we can write:
\begin{IEEEeqnarray}{rcl}
\mbox{Pr}\left\{U^n(m)=u^n,U^n(m)\in\mathcal{T}^n_{[U|XW]\epsilon}(x^n,w^n)\Big| x^n,w^n\right\} &=&
\mathds{1}\left\{u^n\in\mathcal{T}_{[U|WX]\epsilon}^n(x^n,w^n)\right\}\nonumber\\
& &\times\mbox{Pr}\left\{U^n(m)=u^n\Big| x^n,w^n\right\}\ \forall m\ .\,\,\, \,\, \, \,\, 
\end{IEEEeqnarray}
\end{IEEEproof}
This means that
\begin{IEEEeqnarray}{rCl}
&&\mbox{Pr}\left\{U^n(m)=u^n\Big|x^n,w^n,U^n(m)\in\mathcal{T}^n_{[U|XW]\epsilon}(x^n,w^n)\right\}\nonumber\\
&=&\frac{\mathds{1}\left\{u^n\in\mathcal{T}_{[U|WX]\epsilon}^n(x^n,w^n)\cap\mathcal{T}_{[U|W]\epsilon'}^n(w^n)\right\}}{
\mbox{Pr}\left\{U^n(m)\in\mathcal{T}^n_{[U|XW]\epsilon}(x^n,w^n)\Big| x^n,w^n\right\}\|\mathcal{T}_{[U|W]\epsilon'}(w^n)\|}\ \forall m\ .
\label{eq:uniform_1}
\end{IEEEeqnarray}
From (\ref{eq:uniform_1}) and the fact that for $\epsilon'\geq\epsilon$, we have that $\mathcal{T}^n_{[U|XW]\epsilon}(x^n,w^n)\subseteq
\mathcal{T}^n_{[U|W]\epsilon'}(w^n)$ we obtain:
\begin{equation}
\mbox{Pr}\left\{U^n(m)=u^n\Big|x^n,w^n,U^n(m)\in\mathcal{T}^n_{[U|XW]\epsilon}(x^n,w^n)\right\}
=\frac{\mathds{1}\left\{u^n\in\mathcal{T}_{[U|WX]\epsilon}^n(x^n,w^n)\right\}}
{\|\mathcal{T}_{[U|XW]\epsilon}(x^n,w^n)\|}\ \forall m\ .
\label{eq:uniform_2}
\end{equation}
From this equation and (\ref{eq:mkv2}) we easily obtain the desired result.

%
%

We present, without proof, a useful result about reconstruction functions for lossy source coding problems:
\begin{lemma}[Reconstruction functions for degraded random variables \cite{kaspi_two-way_1985}]
\label{lemma:Markov_distortion}
Consider random variables $(X,Y,Z)$ such that $X\mkv Y\mkv Z$. Consider an arbitrary function $\hat{X}=f(Y,Z)$ and an arbitrary positive distortion function $d(\cdot,\cdot)$. Then $\exists\ g^{*}(Y)$ such that
\begin{equation}
E\left[d(X,g^{*}(Y))\right]\leq E[d(X,f(Y,Z))]\ .
\label{eq:best_dist_function}
\end{equation}
\end{lemma}

Finally we present two lemmas about Markov chains induced by the interactive encoding schemes which will be relevant for the paper converse results 
\begin{lemma}[Markov chains induced by interactive encoding of two nodes]
\label{lemma:Markov_interactive_encoding_1}
Consider a set of three sources $(X^n,Y^n,Z^n)\sim\prod\limits_{t=1}^np_{XYZ}(x_t,y_t,z_t)$ and integer $K\in\mathbb{N}$. For each $l\in[1:K]$ consider arbitrary message sets $\mathcal{I}_{x}^l$, $\mathcal{I}_y^l$ and arbitrary functions
\begin{IEEEeqnarray}{rCl}
f_x^l\left(X^n,\mathcal{J}_{x}^{[1:l-1]},\mathcal{J}_y^{[1:l-1]}\right)&=&\mathcal{J}_x^l\ ,\\
f_y^l\left(Y^n,\mathcal{J}_{x}^{[1:l]},\mathcal{J}_y^{[1:l-1]}\right)&=&\mathcal{J}_y^l
\end{IEEEeqnarray}
with $\mathcal{J}_x^l\in\mathcal{I}_{x}^l$ and $\mathcal{J}_y^l\in\mathcal{I}_{y}^l$. The following Markov chain relations are valid for each $t\in[1:n]$ and $l\in[1:K]$:
\begin{enumerate}
\item $\left(\mathcal{J}_{x}^1,X_{[1:t-1]},Y_{[t+1:n]}\right)\mkv X_{[t]}\mkv (Y_{[t]},Z_{[t]})$\ ,
\item $\left(\mathcal{J}_{x}^l,X_{[t+1:n]}\right)\mkv \left(\mathcal{ J}_{x}^{[1:l-1]},\mathcal{J}_{y}^{[1:l-1]},X_{[1:t]},Y_{[t+1:n]}\right)\mkv (Y_{[t]},Z_{[t]})$\ ,
\item $\left(\mathcal{J}_{y}^l,Y_{[1:t-1]}\right)\mkv \left( \mathcal{J}_{x}^{[1:l]},\mathcal{J}_{y}^{[1:l-1]},X_{[1:t-1]},Y_{[t:n]}\right)\mkv (X_{[t]},Z_{[t]})$\ ,
\item $X_{[t+1:n]}\mkv \left(\mathcal{J}_{x}^{[1:K]},\mathcal{J}_{y}^{[1:K]},X_{[1:t]},Y_{[t+1:n]}\right)\mkv (Y_{[t]},Z_{[t]})$\ ,
\item $Y_{[1:t-1]}\mkv \left(\mathcal{J}_{x}^{[1:K]},\mathcal{J}_{y}^{[1:K]},X_{[1:t-1]},Y_{[t:n]}\right)\mkv (X_{[t]},Z_{[t]})$\ ,
\item $\left(\mathcal{J}_{x}^{[1:K]},\mathcal{J}_{y}^{[1:K]},X_{[1:t-1]},X_{[t+1:n]},Z_{[1:t-1]},Z_{[t+1:n]},Y_{[1:t-1]}\right)\mkv (X_{[t]},Y_{[t]}) \mkv Z_{[t]}$\ .
\end{enumerate} 
\end{lemma}
\vspace{1mm}
\begin{IEEEproof}
Relations 1), 2) and 3) where obtained in \cite{kaspi_two-way_1985}. For completeness we present here a short proof of 2). The proof of 1) and 3) are similar. For simplicity let us consider  $A=I\left(\mathcal{J}_{x}^lX_{[t+1:n]};Y_{[t]}Z_{[t]}\Big|\mathcal{ J}_{x}^{[1:l-1]}\mathcal{J}_{y}^{[1:l-1]}X_{[1:t]}Y_{[t+1:n]}\right)$. We can write the following:
\begin{IEEEeqnarray}{rCl}
A
&\stackrel{(a)}{\leq}&
I\left(\mathcal{J}_{x}^lX_{[t+1:n]};Y_{[1:t]}Z_{[t]}\Big|\mathcal{ J}_{x}^{[1:l-1]}\mathcal{J}_{y}^{[1:l-1]}X_{[1:t]}Y_{[t+1:n]}\right)\nonumber\\
&\stackrel{(b)}{=}& I\left(X_{[t+1:n]};Y_{[1:t]}Z_{[t]}\Big|\mathcal{ J}_{x}^{[1:l-1]}\mathcal{J}_{y}^{[1:l-1]}X_{[1:t]}Y_{[t+1:n]}\right)\nonumber\\
&=& H\left(X_{[t+1:n]}\Big|\mathcal{ J}_{x}^{[1:l-1]}\mathcal{J}_{y}^{[1:l-1]}X_{[1:t]}Y_{[t+1:n]}\right)-H\left(X_{[t+1:n]}\Big|\mathcal{ J}_{x}^{[1:l-1]}\mathcal{J}_{y}^{[1:l-1]}X_{[1:t]}Z_{[t]}Y^n\right)\nonumber\\
&\stackrel{(c)}{\leq}&  I\left(X_{[t+1:n]};Y_{[1:t]}Z_{[t]}\Big|\mathcal{ J}_{x}^{[1:l-1]}\mathcal{J}_{y}^{[1:l-2]}X_{[1:t]}Y_{[t+1:n]}\right)\nonumber\\
&=& H\left(Y_{[1:t]}Z_{[t]}\Big|\mathcal{ J}_{x}^{[1:l-1]}\mathcal{J}_{y}^{[1:l-2]}X_{[1:t]}Y_{[t+1:n]}\right)-H\left(Y_{[1:t]}Z_{[t]}\Big|\mathcal{ J}_{x}^{[1:l-1]}\mathcal{J}_{y}^{[1:l-2]}X^nZ_{[t]}Y_{[t+1:n]}\right)\nonumber\\
&\stackrel{(d)}{\leq}&  I\left(X_{[t+1:n]};Y_{[1:t]}Z_{[t]}\Big|\mathcal{J}_{x}^{[1:l-2]}\mathcal{J}_{y}^{[1:l-2]}X_{[1:t]}Y_{[t+1:n]}\right)
\end{IEEEeqnarray}
where
\begin{itemize}
\item step (a) follows from non-negativity of mutual information,
\item step (b) follows from the fact that $\mathcal{J}_x^l=f_x^l\left(X^n,\mathcal{J}_{x}^{[1:l-1]},\mathcal{J}_y^{[1:l-1]}\right)$,
\item step (c) follows from the fact that $\mathcal{J}_y^{l-1}=f_y^l\left(Y^n,\mathcal{J}_{x}^{[1:l-1]},\mathcal{J}_y^{[1:l-2]}\right)$ and that conditioning reduces entropy,
\item step (d) follows from the fact that $\mathcal{J}_x^{l-1}=f_x^l\left(X^n,\mathcal{J}_{x}^{[1:l-2]},\mathcal{J}_y^{[1:l-2]}\right)$ and that conditioning reduces entropy. 
\end{itemize}
Continuing this procedure we obtain:
\begin{equation}
A\leq I\left(X_{[t+1:n]};Y_{[1:t]}Z_{[t]}\Big|X_{[1:t]}Y_{[t+1:n]}\right)=0.
\label{eq:Markov_proof_chain_1}
\end{equation}
This shows that 2) is true. 4) and 5) are straightforward consequences of 2) and 3). Just consider $\mathcal{J}_{x}^{K+1}=\mathcal{J}_{y}^{K+1}=\varnothing$. The proof for 6) is straightforward from the fact that $\left(\mathcal{J}_{x}^{[1:K]}\mathcal{J}_{y}^{[1:K]}\right)$ is only function of $(X^n,Y^n)$. 
\end{IEEEproof}

\begin{corollary}
\label{coro:Markov_interactive_encoding_1}
Consider the setting in Lemma~\ref{lemma:Markov_interactive_encoding_1} with the following modifications:
\begin{itemize}
\item $X\mkv Z\mkv Y$\ ,
\item $f_x^l\left(X^n,Z^n,\mathcal{J}_{x}^{[1:l-1]},\mathcal{J}_y^{[1:l-1]}\right)=\mathcal{J}_x^l$\ .
\end{itemize}
The following are true:
\begin{enumerate}
\item $\left(\mathcal{J}_{x}^1,Z_{[1:t-1]},Y_{[t+1:n]}\right) \mkv Z_{[t]} \mkv Y_{[t]}$\ ,
\item $\left(\mathcal{J}_{x}^l,Z_{[t+1:n]}\right) \mkv \left(\mathcal{ J}_{x}^{[1:l-1]},\mathcal{J}_{y}^{[1:l-1]},Z_{[1:t]},Y_{[t+1:n]}\right) \mkv Y_{[t]}$\ ,
\item $\left(\mathcal{J}_{y}^l,Y_{[1:t-1]}\right) \mkv \left( \mathcal{J}_{x}^{[1:l]},\mathcal{J}_{y}^{[1:l-1]},Z_{[1:t-1]} ,Y_{[t:n]}\right)\mkv (X_{[t]},Z_{[t]})$\ ,
\item $\left(Z_{[t+1:n]},X^n\right)\mkv \left(\mathcal{J}_{x}^{[1:K]},\mathcal{J}_{y}^{[1:K]},Z_{[1:t]},Y_{[t+1:n]}\right)\mkv Y_{[1:t]}$\ .
\end{enumerate} 
\end{corollary}
\begin{IEEEproof}
The proof follows the same lines of Lemma~\ref{lemma:Markov_interactive_encoding_1}.
\end{IEEEproof}
We consider next some Markov chains that arises naturally when we have 3 interacting nodes and which will be needed for Theorem \ref{theo-Coop-BT4-B}.
\begin{lemma}[Markov chains induced by interactive encoding of three nodes]
\label{lemma:Markov_interactive_encoding_2}
Consider a set of three sources $(X^n,Y^n,Z^n)\sim\prod_{t=1}^np_{XYZ}(x_t,y_t,z_t)$ and integer $K\in\mathbb{N}$. For each $l\in[1:K]$ consider arbitrary message sets $\mathcal{I}_{x}^l$, $\mathcal{I}_y^l$, $\mathcal{I}_z^l$  and arbitrary functions
\begin{IEEEeqnarray}{rCl}
f_x^l\left(X^n,\mathcal{J}_{x}^{[1:l-1]},\mathcal{J}_y^{[1:l-1]},\mathcal{J}_z^{[1:l-1]}\right)&=&\mathcal{J}_x^l\ ,\\
f_y^l\left(Y^n,\mathcal{J}_{x}^{[1:l]},\mathcal{J}_y^{[1:l-1]},\mathcal{J}_z^{[1:l-1]}\right)&=&\mathcal{J}_y^l\ ,\\
f_z^l\left(Z^n,\mathcal{J}_{x}^{[1:l]},\mathcal{J}_y^{[1:l]},\mathcal{J}_z^{[1:l-1]}\right)&=&\mathcal{J}_z^l
\end{IEEEeqnarray}
with $\mathcal{J}_x^l\in\mathcal{I}_{x}^l$, $\mathcal{J}_y^l\in\mathcal{I}_{y}^l$ and $\mathcal{J}_z^l\in\mathcal{I}_{z}^l$. The following Markov chain relations are valid for each $t\in[1:n]$ and $l\in[1:K]$:
\begin{enumerate}
\item $\left(\mathcal{J}_{x}^1,\mathcal{J}_{y}^1,X_{[1:t-1]},X_{[t+1:n]},Y_{[t+1:n]},Z_{[1:t-1]}\right)\mkv \left(X_{[t]},Y_{[t]}\right)\mkv Z_{[t]}$\ ,
\item $\left(\mathcal{J}_{x}^l,\mathcal{J}_{y}^l,Y_{[1:t-1]}\right)\mkv \left(\mathcal{ J}_{x}^{[1:l-1]},\mathcal{J}_{y}^{[1:l-1]},\mathcal{J}_{z}^{[1:l-1]},X^n,Y_{[t:n]},Z_{[1:t-1]}\right)\mkv Z_{[t]}$\ ,
\item $\left(\mathcal{J}_{z}^l,Z_{[t+1:n]}\right)\mkv \left( \mathcal{J}_{x}^{[1:l]},\mathcal{J}_{y}^{[1:l]},\mathcal{J}_{z}^{[1:l-1]},X^n,Y_{[t+1:n]},Z_{[1:t]}\right)\mkv Y_{[t]}$\ ,
\item $Z_{[t+1:n]}\mkv \left(\mathcal{J}_{x}^{[1:K]},\mathcal{J}_{y}^{[1:K]},\mathcal{J}_z^{[1:K]},X^n,Y_{[t+1:n]},Z_{[1:t]}\right)\mkv Y_{[t]}$\ ,
\item $Y_{[1:t-1]}\mkv \left(\mathcal{J}_{x}^{[1:K]},\mathcal{J}_{y}^{[1:K]},\mathcal{J}_z^{[1:K]},X^n,Y_{[t:n]},Z_{[1:t-1]}\right)\mkv Z_{[t]}$\ .
\end{enumerate} 
\end{lemma}
\begin{IEEEproof}
Along the same lines of Lemma\ref{lemma:Markov_interactive_encoding_1} and for that reason omitted. 
\end{IEEEproof}

\section{Cooperative Berger-Tung Problem with Side Information at the Decoder}
\label{app:coop_berger}

We derive an inner bound on the rate region of the setup described in Fig.~\ref{fig:coop_berger}. It should be emphasize that we will not consider distortion measures, we only focus is on the exchange  of descriptions.  Encoders 1 and 2 observe source sequences $X_1^n$ and $X_2^n$, and also have access to a common side information $V_1^n$.  Whereas, the decoder  has access to side informations $(X^n_3,V_1^n,V_2^n)$. Upon observing $X_1^n$  and $V_1^n$, Encoder $1$ generates a message $M_1$ which is transmitted to Encoder $2$ and the decoder. Encoder $2$, upon observing $(X_2^n,V_1^n)$ and the message $M_1$, generates a message $M_2$ which is transmitted only to the decoder. Finally, the decoder uses messages $(M_1,M_2)$ and the side informations $(X_3^n,V_1^n,V_2^n)$ to reconstruct  two sequences $(\hat{U}_1^n,\hat{U}_2^n)$ which are jointly typical with $(X_1^n,X_2^n,X_3^n,V_1^n,V_2^n)$. In precise terms, we will assume the following:
\begin{figure}[t]
\centering
\ifpdf\includegraphics[angle=0,width=0.6\columnwidth,keepaspectratio,trim= 0mm 0mm 0mm 0mm,clip]{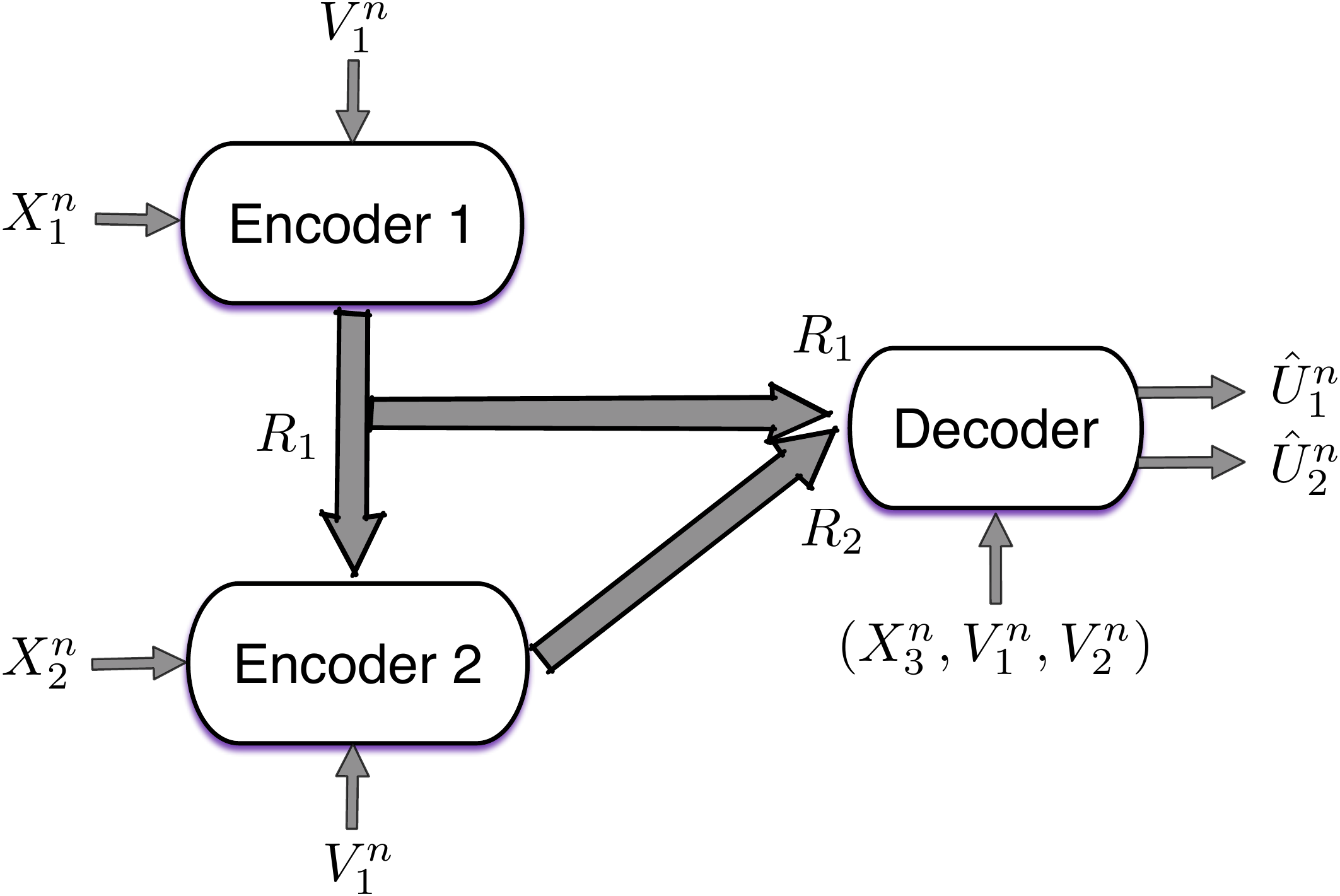} 
\else
	\includegraphics[angle=270,width=0.45\columnwidth,keepaspectratio,trim= 0mm 0mm 0mm 0mm,clip]{dependence_2.eps} 
\fi
\caption{Cooperative Berger-Tung problem.}
\label{fig:coop_berger}
\end{figure}

\begin{itemize}
\item A probability mass function $p_{X_1X_2X_3U_1U_2V_1V_2}$ which takes values on cartesian product  finite alphabets $\mathcal{X}_1\times\mathcal{X}_2\times\mathcal{X}_3\times\mathcal{U}_1\times\mathcal{U}_2\times\mathcal{V}_1\times\mathcal{V}_2$, and that satisfies the following Markov chains:
\begin{equation}
U_1\mkv (X_1,V_1) \mkv (X_2,X_3,V_2)\ ,\ \ U_2\mkv (U_1,X_2,V_1)\mkv (X_1,X_3,V_2)\ .\label{eq:PD-coop-BT}
\end{equation}
\item Five random vectors $(X_1^n,X_2^n,X_3^n,V_1^n,V_2^n)$ (not necessarily independently and identically distributed with $p_{X_1X_2X_3V_1V_2}$ which take values on alphabets  
$\mathcal{X}_1^n\times\mathcal{X}_2^n\times\mathcal{X}_3^n\times\mathcal{V}_1^n\times\mathcal{V}_2^n$
such that, for every $\epsilon>0$,
\begin{equation}
\label{eq:condition_coop}
\lim_{n\rightarrow\infty}\mbox{Pr}\left\{(X_1^n,X_2^n,X_3^n,V_1^n,V_2^n)\in\mathcal{T}^n_{[X_1X_2X_3V_1V_2]\epsilon}\right\}=1\ .
\end{equation}
\end{itemize}
\begin{definition}[Cooperative code]
A code $(n,f_1^n,f_2^n,g^n,\mathcal{M}_1,\mathcal{M}_2)$ for the setup in Fig. \ref{fig:coop_berger} is composed by:
\begin{itemize}
\item Two set of indices $\mathcal{M}_1$, $\mathcal{M}_2$.
\item An encoding function $f_1^n:\mathcal{X}_1^n\times\mathcal{V}_1^n\rightarrow \mathcal{M}_1$, such that $f_1^n(x_1^n,v_1^n)=m_1$.
\item An encoding function  $f_2^n:\mathcal{X}_2^n\times\mathcal{V}_1^n\times\mathcal{M}_1\rightarrow \mathcal{M}_2$, such that  $f_2^n(x_2^n,v_1^n,m_1)=m_2$.
\item A decoding function $g^n:\mathcal{X}_3^n\times\mathcal{V}_1^n\times\mathcal{V}_2^n\times\mathcal{M}_1\times\mathcal{M}_2\rightarrow \mathcal{U}_1^n\times\mathcal{U}_2^n$, such that $g^n(x_3^n,v_1^n,v_2^n,m_1,m_2)=(\hat{u}_1^n,\hat{u}_2^n)$.
\end{itemize}
\end{definition}
\begin{definition}[Achievable rates]
We say that $(R_1,R_2)$ are $\epsilon$-achievable if there exists a code $(n,f_1^n,f_2^n,g^n,\mathcal{M}_1,\mathcal{M}_2)$ such that:
\begin{equation}
\frac{1}{n}\log\|\mathcal{M}_1\|\leq R_1+\epsilon\ ,\ \ \frac{1}{n}\log\|\mathcal{M}_2\|\leq R_2+\epsilon\end{equation}
and 
\begin{equation}
\mbox{Pr}\left\{(\hat{U}_1^n,\hat{U}_2^n,V_1^n,V_2^n,X_1^n,X_2^n,X_3^n)\in\mathcal{T}^n_{[U_1U_2V_1V_2X_1X_2X_3]\epsilon}\right\}\leq \epsilon\ .
\end{equation}
The closure of the set of all achievable rates $(R_1,R_2)$ is denoted by $\mathcal{R}_{CBT}$.
\end{definition}

The following theorem presents an inner bound to $\mathcal{R}_{CBT}$.
\begin{theorem}(Inner bound on the rate region of the cooperative Berger-Tung problem)
\label{theo:coop_berger}
Consider $\mathcal{R}_{ CBT}^{\mbox{inner}}$ the closure of the set of rates satisfying:
\begin{eqnarray*}
R_1 &>& I(X_1;U_1|X_2V_1)\ ,\\
R_2 & >& I(X_2;U_2|X_3V_1V_2U_1)\ , \\
R_1+R_2 &>& I(X_1X_2;U_1U_2|X_3V_1V_2)\ ,
\end{eqnarray*}
where the union is over all probability distributions verifying~\eqref{eq:PD-coop-BT}. Then $\mathcal{R}_{ CBT}^{\mbox{inner}}\subseteq\mathcal{R}_{ CBT}$. 
\end{theorem}
\vspace{1mm}
\begin{remark}
Notice that we are not asking for $(X_1^n,X_2^n,X_3^n,V_1^n,V_2^n)$ to be independently and identically distributed. This is in fact not needed for the result that follows. For us, when trying to use this result, the case of most interest will be when  $(X_1^n,X_2^n,X_3^n)$ is generated using the product measure $\prod_{i=1}^np_{X_1X_2X_3}(x_{1i},x_{2i},x_{3i})$, (that is, when $(X_1X_2X_3)$ is a DMS). However, $(V_1^n,V_2^n)$ will not be independently and identically distributed.  Still, (\ref{eq:condition_coop}) will be satisfied. 
\end{remark}
\vspace{1mm}
\begin{remark}
Notice that unlike the classical rate-distortion problem we are not interested in an average per-symbol distortion constraints at the decoder. We only require that the obtained sequences be jointly typical with the sources. Clearly the problem can be slightly modified to consider the case in which reconstruction distortion constraints are of interest. In fact, case (C) reported in~\cite{Kaspi_1982}, considers a similar setting. Here, given the importance of this result for our interactive scheme, we present a slightly different and more direct proof of the achievability, where we discuss the key points in the encoding and decoding procedures which will be relevant for our extension to the interactive problem. 
\end{remark}

\begin{IEEEproof}
Our proof uses standard ideas from multiterminal source coding. As $V^n_1$ is common to both encoders and decoder we can set  without loss of generality $V_1^n=\varnothing$. Conditioning with respect to $V_1$ the final expressions can take into account the situation in which $V_1^n\neq\varnothing$. 
\subsection{Coding generation}
We randomly generate $2^{n\hat{R}_1}$ codewords $U_1^n(k),\ k\in[1:2^{\hat{R}_1}]$ according to

\begin{equation}
U_1^n(k)\sim\frac{\mathds{1}\left\{u_1^n\in\mathcal{T}_{[U_1]\epsilon_{cd}}^n\right\}}{\left\|\mathcal{T}_{[U_1]\epsilon_{cd}}^n\right\|},\ \epsilon_{cd}>0\ .
\end{equation}
These $2^{n\hat{R}_1}$ codewords are distributed uniformly over $2^{nR_1}$ bins denoted by $\mathcal{B}_1(m_1)$, where $m_1\in[1:2^{nR_1}]$. For each codeword $u_1^n(k)$ with $k\in[1:2^{n\hat{R}_1}]$ , we randomly generate $2^{n\hat{R}_2}$ codewords according to:
\begin{equation}
U_2^n(l,k)\sim\frac{\mathds{1}\left\{u_2^n\in\mathcal{T}_{[U_2|U_1]\epsilon_{cd}}^n(u_1^n(k))\right\}}{\left\|\mathcal{T}_{[U_2|U_1]\epsilon_{cd}}^n(u_1^n(k))\right\|},\ \epsilon_{cd}>0
\end{equation}
with $l\in[1:2^{n\hat{R}_2}]$. The $2^{n(\hat{R}_1+\hat{R}_2)}$ codewords generated are distributed uniformly in $2^{nR_2}$ bins, denoted by $\mathcal{B}_2(m_2)$, $m_2\in[1:2^{nR_2}]$.  It is worth to mention that the codewords $\{U_2^n(l,k)\}$ are not distributed in a different structure of bins for each $k$, but on only one \emph{super-bin} structure of size $2^{n(\hat{R}_1+\hat{R}_2)}/2^{nR_2}$ where $\mathcal{B}_2(m_2)$ does not needed to be indexed with $k$. 

As will be clear, this will not constraint the decoder to use \emph{successive decoding} and instead use \emph{joint decoding} in order to recover the desired codewords $(\hat{U}_1^n,\hat{U}_2^n)$. Finally all codebooks are revealed to all parties.

\subsection{Encoding at node 1}
Given $x_1^n$, the encoder search for $k\in[1:2^{n\hat{R}_1}]$ in such a way that:
\begin{equation}
(x_1^n,u_1^n(k))\in\mathcal{T}^n_{[X_1U_1]\epsilon_{2}},\ \epsilon_{2}>0\ .
\end{equation}
 If more than one index satisfies this condition, then we choose the one with the smallest index. Otherwise, if no such index exists,  we choose an arbitrary one and declare an error. Finally we select $m_1$ as the index of the bin which contains the codeword $u_1^n(k)$ found and transmit it to nodes 2 and 3.
 
 \subsection{Decoding at node 2}
 Given $x_2^n$ and $m_1$, we search in the bin  $\mathcal{B}_{1}(m_1)$ for an index  $k\in[1:2^{n\hat{R}_1}]$ such that:
\begin{equation}
(x_2^n,u_1^n(k))\in\mathcal{T}^n_{[X_2U_1]\epsilon_{3}},\ \epsilon_{3}>0\ .
\end{equation}
If there only one index that satisfies this we declare it as the index generated at node $1$. If there several or none we choose a predefined one and declare an error. The chosen index is denoted as $\hat{k}(2)$.

 \subsection{Encoding at node 2}
 Given $x_2^n$ and $\hat{k}(2)$ we search for $l\in[1:2^{n\hat{R}_2}]$  such that:
\begin{equation}
(x_2^n,u_1^n( \hat{k}(2)),u_2^n(l,\hat{k}(2)))\in\mathcal{T}^n_{[X_2U_1U_2]\epsilon_{4}},\ \epsilon_{4}>0\ .
\end{equation}
  If more than one index satisfies this condition, then we choose the one with the smallest index. Otherwise, if no such index exists,  we choose an arbitrary one and declare an error. Finally we select $m_2$ as the index of the bin which contains the codeword $u_2^n(l,\hat{k}(2))$ selected and transmit it to node $3$.
  
  \subsection{Decoding at node 3}
 Given $x_3^n$, $v_2^n$ and $m_1,m_2$, the decoder search in the bins $\mathcal{B}_1(m_1)$ and $\mathcal{B}_2(m_2)$ for a pair of indices $(k,l)\in[1:2^{n\hat{R}_1}]\times[1:2^{n\hat{R_2}}]$ such that
\begin{equation}
(x_3^n,v_2^n,u_1^n(k),u_2^n(l,k))\in\mathcal{T}^n_{[X_3V_2U_1U_2]\epsilon},\ \epsilon>0\ .
\end{equation}
  If there only one pair of indices that satisfy this we declare it as the indices generated at node 1 and 2. If there several or none we choose a predefined pair and declare an error. The chosen pair is denoted by $(\hat{k}(3),\hat{l}(3))$. Finally, the decoder declares $(\hat{u}_1^n,\hat{u}_2^n)= (u_1^n( \hat{k}(3)),u_2^n(\hat{l}(3),\hat{k}(3)))$.
  
  \subsection{Error probability analysis}
  Consider $(K,L)$  the description indices generated at node 1 and 2, and $(M_1,M_2)$  the corresponding bin indices. With $\hat{K}(2)$ and $(\hat{K}(3),\hat{L}(3))$ we denote the indices recovered at nodes 2 and 3. We want to prove that $\mbox{Pr}\left\{\mathcal{E}\right\}\leq \epsilon'$ when $n$ is sufficiently large, where
\begin{equation}
\mathcal{E}=\left\{\left(X_1^n,X_2^n,X_3^n,V_2^n,U_1^n(\hat{K}(3)),U_2^n(\hat{L}(3),\hat{K}(3))\right)\notin\mathcal{T}_{[X_1X_2X_3V_2U_1U_2]\epsilon}^n\right\}\ .
  \end{equation}
  We consider the following events of error:
  \begin{itemize}
  \item $\mathcal{E}_1=\left\{\left(X_1^n,X_2^n,X_3^n,V_2^n\right)\notin \mathcal{T}_{[X_1X_2X_3V_2]\epsilon_1}^n\right\},\ \epsilon_1>0$.
  
  \item $\mathcal{E}_2=\left\{\left(X_1^n,U_1^n(k)\right)\notin \mathcal{T}_{[X_1U_1]\epsilon_2}^n\ \forall k\in[1:2^{n\hat{R}_1}]\right\},\ \epsilon_2>0$.
  
  \item  $\mathcal{E}_3=\left\{\left(X_1^n,X_2^n,X_3^n,V_2^n,U_1^n(K)\right)\notin \mathcal{T}_{[X_1X_2X_3V_2U_1]\epsilon_3}^n\right\},\ \epsilon_3>0$.
  
  \item  $\mathcal{E}_4=\left\{\exists \hat{k}\neq K, \hat{k}\in\mathcal{B}_1(M_1),\left(X_2^n,U_1^n(\hat{k})\right)\in \mathcal{T}_{[X_1U_1]\epsilon_3}^n\right\},\ \epsilon_3>0$.
  
  \item $\mathcal{E}_5=\left\{\left(X_2^n,U_1^n(\hat{K}(2)),U_2^n(l,\hat{K}(2)))\right)\notin \mathcal{T}_{[X_2U_1U_2]\epsilon_4}^n\ \forall l\in[1:2^{n\hat{R}_2}]\right\},\ \epsilon_4>0$.
  
  \item  $\mathcal{E}_6=\left\{\left(X_1^n,X_2^n,X_3^n,V_2^n,U_1^n(K),U_2^n(L,\hat{K}(2))\right)\notin \mathcal{T}_{[X_1X_2X_3V_2U_1U_2]\epsilon}^n\right\},\ \epsilon>0$.
  
  \item  $\mathcal{E}_7\!=\!\left\{\exists \hat{k}\neq K, \hat{l}\neq L, \hat{k}\in\mathcal{B}_1(M_1), \hat{l}\in\mathcal{B}_2(M_2),\left(X_3^n,V_2^n,U_1^n(\hat{k}),U_2^n(\hat{l},\hat{k})\right)\!\!\in\!\! \mathcal{T}_{[X_3V_2U_1U_2]\epsilon}^n\right\}$.
  \end{itemize}\vspace{1mm}
  
Clearly $\mathcal{E}\subseteq\bigcup_{i=1}^7\mathcal{E}_i$. In fact, it is easy to show that $\left\{(\hat{K}(3),\hat{L}(3))\neq (K,L) , \hat{K}(2)\neq K\right\}\subseteq\bigcup_{i=1}^7\mathcal{E}_i$. From hypothesis, we obtain that $\lim_{n\rightarrow \infty}\mbox{Pr}\left\{\mathcal{E}_1\right\}=0$. Choosing $\epsilon_1<\frac{\epsilon_2}{\|\mathcal{U}_1\|}$ and $\epsilon_2<\epsilon_{cd}$ we can use Lemma~\ref{lemma:covering} and its Corollary (with the following equivalences: $V\equiv U_1$, $X\equiv X_1$, $U\equiv \varnothing$) to obtain $\lim_{n\rightarrow\infty}\mbox{Pr}\left\{\mathcal{E}_2\right\}=0$ if
  \begin{equation}
  \hat{R}_1>I(U_1;X_1)+\delta(\epsilon_1,\epsilon_2,\epsilon_{cd},n)\ .
  \label{eq:rate_1b}
  \end{equation}
  For the analysis of $\mbox{Pr}\left\{\mathcal{E}_3\right\}$ we can use Lemma \ref{lemma:markov}, its Corollary and Lemma \ref{lemma:induced} defining $Y\equiv X_2X_3$, $X\equiv X_1$ and $U\equiv U_1$ and using $\epsilon_2$, $\epsilon_3$ and $\epsilon_{cd}$ sufficiently small\footnote{In the following, we will not indicate anymore the corresponding values of the constants $\epsilon$, the arguments of $\delta$ and the equivalence between the involved random variables in order to use the lemmas from Appendix~\ref{app:strongly}} to obtain $\lim_{n\rightarrow\infty}\mbox{Pr}\left\{\mathcal{E}_3\right\}=0$. For the analysis of $\mbox{Pr}\left\{\mathcal{E}_4\right\}$ we can write:
  \begin{eqnarray}
\mbox{Pr}\left\{\mathcal{E}_4\right\}&= & \mathbb{E}\left[\mbox{Pr}\left\{\mathcal{E}_4|K=k,M_1=m_1\right\}\right]\nonumber\\
&=&\mathbb{E}\left[\mbox{Pr}\left\{\left.\bigcup_{\substack{\hat{k}\neq k\\ \hat{k}\in\mathcal{B}_1(m_1)}}\left\{\left(X_2^n,U_1^n(\hat{k})\right)\in \mathcal{T}_{[X_1U_1]\epsilon_3}^n \right\}\right|\substack{K=k\\ M_1=m_1}\right\}\right]\ .
 \end{eqnarray}
 Using Lemma~\ref{lemma:packing} (with the appropriate equivalences on the involved random variables) and the statistical properties of the codebooks, binning and encoding, we have, that for each $k$, $m_1$:
  \begin{equation}
  \lim_{n\rightarrow \infty}\mbox{Pr}\left\{\left.\bigcup_{\hat{k}\neq k, \hat{k}\in\mathcal{B}_1(m_1)}\left\{\left(X_2^n,U_1^n(\hat{k})\right)\in \mathcal{T}_{[X_1U_1]\epsilon_3}^n \right\}\right|\substack{K=k\\ M_1=m_1}\right\}=0
    \end{equation}
provided that
  \begin{equation}
  \frac{1}{n} \log\, {\mathbb{E} \|\mathcal{B}_1(m_1)\| }< I(X_2;U_1)-\delta(\epsilon_1,\epsilon_3,\epsilon_{cd},n)\ .   
   \end{equation}
As $\mathbb{E}\left[\|\mathcal{B}_1(m_1)\|\right]=2^{n(\hat{R}_1-R_1)}\ \forall m_1$ we have that $\lim_{n\rightarrow\infty}\mbox{Pr}\left\{\mathcal{E}_4\right\}=0$ provided that 
\begin{equation}
\hat{R}_1-R_1< I(X_2;U_1)-\delta\ .
\label{eq:rate_2b}
\end{equation} 
The analysis of $\mbox{Pr}\left\{\mathcal{E}_5\right\}$  follows the same lines of $\mbox{Pr}\left\{\mathcal{E}_2\right\}$. The above analysis implies that 
\begin{equation}
\lim_{n\rightarrow\infty}\mbox{Pr}\left\{\left(X_2^n,U_1^n(\hat{K}(2))\right)\in \mathcal{T}_{[X_2U_1]\epsilon_3}^n\right\}=1\ .
\end{equation} 
Then, by Lemma~\ref{lemma:covering} and its Corollary we have that $\lim_{n\rightarrow\infty}\mbox{Pr}\left\{\mathcal{E}_5\right\}=0$ if :
\begin{equation}
\hat{R}_2>I(X_2;U_2|U_1)+\delta\ .
\label{eq:rate_2bb}
\end{equation}
From Lemmas~\ref{lemma:markov} and~\ref{lemma:induced}, similarly as with $\mbox{Pr}\left\{\mathcal{E}_3\right\}$, we have $\lim_{n\rightarrow\infty}\mbox{Pr}\left\{\mathcal{E}_6\right\}=0$. Let us turn to analyze $\mbox{Pr}\left\{\mathcal{E}_7\right\}$:
  \begin{eqnarray}
\mbox{Pr}\left\{\mathcal{E}_7\right\}&= & \mathbb{E}\left[\mbox{Pr}\left\{\mathcal{E}_7|K=k,L=l,M_1=m_1,M_2=m_2\right\}\right]\nonumber\\
&=&\mathbb{E}\left[\mbox{Pr}\left\{\left.\bigcup_{\substack{(\hat{k},\hat{l})\neq (k,l)\\ \hat{k}\in\mathcal{B}_1(m_1)\nonumber \\ 
(\hat{k},\hat{l})\in\mathcal{B}_2(m_2)}}\left\{\left(X_3^n,V_2^n,U_1^n(\hat{k}),U_2^n(\hat{l},\hat{k})\right)\in \mathcal{T}_{[X_3V_2U_1U_2]\epsilon}^n \right\}\right|\substack{K=k, L=l\\ M_1=m_1, M_2=m_2}\right\}\right]\\
&\leq & \mathbb{E}\left[\alpha_1+\alpha_2+\alpha_3\right]
 \end{eqnarray}
where we have
\begin{equation}
  \alpha_1=\mbox{Pr}\left\{\left.\bigcup_{\substack{\hat{k}\neq k\\ \hat{k}\in\mathcal{B}_1(m_1)\\ (\hat{k},l)\in\mathcal{B}_2(m_2)}}\left\{\left(X_3^n,V_2^n,U_1^n(\hat{k}),U_2^n(l,\hat{k})\right)\in \mathcal{T}_{[X_3V_2U_1U_2]\epsilon}^n \right\}\right|\substack{K=k, L=l\\ M_1=m_1, M_2=m_2}\right\}\ ,
  \end{equation}
\begin{equation}
\alpha_2=\mbox{Pr}\left\{\left.\bigcup_{\substack{\hat{l}\neq l\\ (k,\hat{l})\in\mathcal{B}_2(m_2)}}\left\{\left(X_3^n,V_2^n,U_1^n(k),U_2^n(\hat{l},k)\right)\in \mathcal{T}_{[X_3V_2U_1U_2]\epsilon}^n \right\}\right|\substack{K=k, L=l\\ M_1=m_1, M_2=m_2}\right\}\ ,
\end{equation}
\begin{equation}
\alpha_3=\mbox{Pr}\left\{\left.\bigcup_{\substack{\hat{k}\neq k, \hat{l}\neq l\\ \hat{k}\in\mathcal{B}_1(m_1)\\ (\hat{k},\hat{l})\in\mathcal{B}_2(m_2)}}\left\{\left(X_3^n,V_2^n,U_1^n(\hat{k}),U_2^n(\hat{l},\hat{k})\right)\in \mathcal{T}_{[X_3V_2U_1U_2]\epsilon}^n \right\}\right|\substack{K=k, L=l\\ M_1=m_1, M_2=m_2}\right\}\ .
\end{equation}

We can use the Lemma~\ref{lemma:packing} to obtain:
\begin{equation}
\lim_{n\rightarrow\infty} \alpha_1=0\ , \, \lim_{n\rightarrow\infty}\alpha_2=0\ ,\, \lim_{n\rightarrow\infty}\alpha_3=0
\end{equation}
provided that
\begin{IEEEeqnarray}{rCl}
\frac{1}{n} \log\, {\mathbb{E} \left\|\hat{k}:(\hat{k},l)\in\mathcal{B}_2(m_2), \hat{k}\in\mathcal{B}_1(m_1)\right\| } &<& I(X_3V_2;U_1U_2)-\delta,\\
\frac{1}{n} \log\, {\mathbb{E} \left\|\hat{l}:(k,\hat{l})\in\mathcal{B}_2(m_2)\right\| } & < & I(X_3V_2;U_2|U_1)-\delta\ , \\
\frac{1}{n} \log\, {\mathbb{E} \left\|(\hat{k},\hat{l}):(\hat{k},\hat{l})\in\mathcal{B}_2(m_2), \hat{k}\in\mathcal{B}_1(m_1)\right\| } &<& I(X_3V_2;U_1U_2)-\delta \ .
\end{IEEEeqnarray}
Because on how the binning is performed, we have:
\begin{IEEEeqnarray}{rCl} 
\mathbb{E} \left\|\hat{k}:(\hat{k},l)\in\mathcal{B}_2(m_2), \hat{k}\in\mathcal{B}_1(m_1)\right\|  &= &2^{n(\hat{R}_1-R_1-R_2)}\ , \\
\mathbb{E} \left\|\hat{l}:(k,\hat{l})\in\mathcal{B}_2(m_2)\right\|  &= & 2^{n(\hat{R}_2-R_2)}\ , \\
\mathbb{E} \left\|(\hat{k},\hat{l}):(\hat{k},\hat{l})\in\mathcal{B}_2(m_2), \hat{k}\in\mathcal{B}_1(m_1)\right\| & = & 2^{n(\hat{R}_1+\hat{R}_2-R_1-R_2)}\ ,
\end{IEEEeqnarray}
which give us:
\begin{IEEEeqnarray}{rCl}
(\hat{R}_1-R_1) -R_2 &<& I(X_3V_2;U_1U_2)-\delta\ , \label{eq:rate_3b}\\
\hat{R}_2-R_2 &<& I(X_3V_2;U_2|U_1)-\delta\ ,\label{eq:rate_4} \\
(\hat{R}_1+\hat{R}_2)-(R_1+R_2) &<& I(X_3V_2;U_1U_2)-\delta\ . \label{eq:rate_5}
\end{IEEEeqnarray}
 Notice that equation (\ref{eq:rate_3b}) remains inactive because of (\ref{eq:rate_5}). Equations (\ref{eq:rate_1b}), (\ref{eq:rate_2b}), (\ref{eq:rate_2bb}), (\ref{eq:rate_4}), (\ref{eq:rate_5}) can be combined with:
\begin{IEEEeqnarray}{rCl} 
\hat{R}_1&>&R_1\ ,\\ 
\hat{R_1}+\hat{R}_2&>&R_2\ ,
\end{IEEEeqnarray}
which follow from the binning structure assumed in the generated codebooks. A Fourier-Motzkin elimination procedure can be done to eliminate $\hat{R}_1$ and $\hat{R}_2$ obtaining the desired rate region (conditioning also the mutual information terms on $V_1$).
\end{IEEEproof}
The following Corollary considers the case in which a genie gives node 2 the value of $M_1$. Indeed, this case will be important for our main result.
\vspace{1mm}
\begin{corollary}
\label{coro:coop_berger}
If a genie gives $M_1$ to node 2, the achievable region  $\mathcal{R}_{ CBT}^{\mbox{inner}}$ reduces to:
\begin{eqnarray}
R_2 & >& I(X_2;U_2|X_3V_1V_2U_1)\ ,\\
R_1+R_2 &>& I(X_1X_2;U_1U_2|X_3V_1V_2)\ .
\end{eqnarray}
\end{corollary}
The proof of this result is straightforward and thus it will not be presented.

 \section{Proof of Theorem~\ref{theo-main-theorem}}
 \label{app:proof_theo_1}
 Let us describe the coding generation, encoding and decoding procedures. We will consider the following notation. With $M_{i\rightarrow S,l}$ we will denote the index corresponding to the true description $U^n_{i\rightarrow S,l}$ generated at node $i$ at round $l$ and destined to the group of nodes $S\in\mathcal{C}\left(\mathcal{M}\right)$ with $i\notin S$. With $\hat{M}_{i\rightarrow S,l}(j)$ where $S\in\mathcal{C}\left(\mathcal{M}\right)$, $i\notin S$, $j\in S$ we denote the corresponding estimated index at node $j$.
 
 \subsection{Codebook generation}
 \label{subsec:coding_gen}
 
 Consider the round $l\in [1:K]$. For simplicity let us consider the descriptions at node 1. We generate $2^{n\hat{R}_{1\rightarrow 23}^{(l)}}$ independent and identically distributed $n$-length codewords $U^n_{1\rightarrow 23,l}(m_{1\rightarrow 23,l},m_{\mathcal{W}_{[1,l]}})$ according to:
 \begin{equation}
 U^n_{1\rightarrow 23,l}(m_{1\rightarrow 23,l},m_{\mathcal{W}_{[1,l]}})\sim \frac{\mathds{1}\left\{u^n_{1\rightarrow 23,l}\in\mathcal{T}_{[U_{1\rightarrow 23,l}|\mathcal{W}_{[1,l]}]\epsilon(1,23,l)}^n\left(w_{[1,l]}^n\right)\right\}}{\left\|\mathcal{T}_{[U_{1\rightarrow 23,l}|\mathcal{W}_{[1,l]}]\epsilon(1,23,l)}^n\left(w_{[1,l]}^n\right)\right\|},\  \epsilon(1,23,l)>0
 \label{eq:coding_gen_1}
 \end{equation}
where $m_{1\rightarrow 23,l}\in[1:2^{n\hat{R}_{1\rightarrow 23}^{(l)}}]$ and let $m_{\mathcal{W}_{[1,l]}}$ denote the indices of the common descriptions $\mathcal{W}^n_{[1,l]}$ generated in rounds $t\in[1:l-1]$. For example, $m_{\mathcal{W}_{[1,l]}}=\left\{m_{1\rightarrow 23,t}, m_{2\rightarrow 13,t},m_{3\rightarrow 12,t}\right\}_{t=1}^{l-1}$. With $w_{[1,l]}^n$ we denote the set of $n$-length common information codewords from previous rounds corresponding to the indices  $m_{\mathcal{W}_{[1,l]}}$. For each $m_{\mathcal{W}_{[3,l-1]}}$ consider the set of  $2^{n\left(\hat{R}^{(l)}_{1\rightarrow 23}+\hat{R}^{(l)}_{3\rightarrow 12}\right)}$ codewords $U^{n}_{1\rightarrow 23,l}(m_{1\rightarrow 23,l},m_{3\rightarrow 12,l-1},m_{\mathcal{W}_{[3,l-1]}})$. These  $n$-length codewords are distributed independently and uniformly over $2^{nR_{1\rightarrow 23}^{(l)}}$ bins denoted by $\mathcal{B}_{1\rightarrow 23,l}\left(p_{1\rightarrow 23,l},m_{\mathcal{W}_{[3,l-1]}}\right)$ with $p_{1\rightarrow 23,l}\in[1:2^{nR_{1\rightarrow 23}^{(l)}}]$. Notice that this binning structure is exactly the same we used for the cooperative Berger-Tung problem in Appendix ~\ref{app:coop_berger}. Node 1 distributes codewords $U^{n}_{1\rightarrow 23,l}(m_{1\rightarrow 23,l},m_{3\rightarrow 12,l-1},m_{\mathcal{W}_{[3,l-1]}})$ in a super-binning structure. This will allow node 2 to recover both, $m_{1\rightarrow 23,l}$ and $m_{3\rightarrow 12,l-1}$, using the same procedure as in the Berger-Tung problem described above. Notice that a different super-binning structure is generated for every $m_{\mathcal{W}_{[3,l-1]}}$. This is without loss of generality, because at round $l$ nodes 1, 2 and 3, will have a very good estimated of it (see below). 

We also generate $2^{n\hat{R}_{1\rightarrow 2}^{(l)}}$ and $2^{n\hat{R}_{1\rightarrow 3}^{(l)}}$ independent and identically distributed $n$-length codewords $U^n_{1\rightarrow 2,l}(m_{1\rightarrow 2,l},m_{\mathcal{W}_{[2,l]}},m_{\mathcal{V}_{[12,l,1]}})$, and $U^n_{1\rightarrow 3,l}(m_{1\rightarrow 3,l},m_{\mathcal{W}_{[2,l]}},m_{\mathcal{V}_{[13,l,1]}})$ according to:
\begin{IEEEeqnarray}{rCl}
 U^n_{1\rightarrow 2,l}(m_{1\rightarrow 2,l},m_{\mathcal{W}_{[2,l]}},m_{\mathcal{V}_{[12,l,1]}}) &\sim & \frac{\mathds{1}\left\{u^n_{1\rightarrow 2,l}\in\mathcal{T}_{[U_{1\rightarrow 2,l}|\mathcal{W}_{[2,l]}\mathcal{V}_{[12,l,1]}]\epsilon(1,2,l)}^n\left(w_{[2,l]}^n,v_{[12,l,1]}^n\right)\right\}}{\left\|\mathcal{T}_{[U_{1\rightarrow 2,l}|\mathcal{W}_{[2,l]}\mathcal{V}_{[12,l,1]}]\epsilon(1,2,l)}^n\left(w_{[2,l]}^n,v_{[12,l,1]}^n\right)\right\|}\ ,\,\,\,\,\,\, \,\,\,\,\, \\ 
U^n_{1\rightarrow 3,l}(m_{1\rightarrow 3,l},m_{\mathcal{W}_{[2,l]}},m_{\mathcal{V}_{[13,l,1]}}) &\sim & \frac{\mathds{1}\left\{u^n_{1\rightarrow 3,l}\in\mathcal{T}_{[U_{1\rightarrow 3,l}|\mathcal{W}_{[2,l]}\mathcal{V}_{[13,l,1]}]\epsilon(1,3,l)}^n\left(w_{[2,l]}^n,v_{[13,l,1]}^n\right)\right\}}{\left\|\mathcal{T}_{[U_{1\rightarrow 3,l}|\mathcal{W}_{[2,l]}\mathcal{V}_{[13,l,1]}]\epsilon(1,3,l)}^n\left(w_{[2,l]}^n,v_{[13,l,1]}^n\right)\right\|}\ ,  \,\,\, \,\,\,\,\,\,\,\,
   \end{IEEEeqnarray}
  where $ \epsilon(1,2,l)>0$, $\epsilon(1,3,l)>0$, and $m_{1\rightarrow 2,l}\in[1:2^{n\hat{R}_{1\rightarrow 2}^{(l)}}]$ and  $m_{1\rightarrow 3,l}\in[1:2^{n\hat{R}_{1\rightarrow 3}^{(l)}}]$. These codewords are distributed uniformly on $2^{nR_{1\rightarrow 2}^{(l)}}$ bins denoted by $\mathcal{B}_{1\rightarrow 2,l}\big(p_{1\rightarrow 2,l},m_{\mathcal{W}_{[2,l]}},m_{\mathcal{V}_{[12,l,1]}}\big)$ and indexed with $p_{1\rightarrow 2,l}\in[1:2^{nR_{1\rightarrow 2}^{(l)}}]$ and on $2^{nR_{1\rightarrow 3}^{(l)}}$ bins denoted by $\mathcal{B}_{1\rightarrow 3,l}\big(p_{1\rightarrow 3,l},m_{\mathcal{W}_{[2,l]}},m_{\mathcal{V}_{[13,l,1]}}\big)$ and indexed with $p_{1\rightarrow 3,l} \in[1:2^{nR_{1\rightarrow 3}^{(l)}}]$, respectively. Notice that these codewords (which will be used to generated private descriptions to node 2 and 3) are not distributed in a super-binning structure. This is because there is not explicit cooperation between the nodes at this level. That is, node 2 is not compelled to recover the private description that node 1 generate for node 3, and for that reason the private description that node 2 generate for node 3 is not superimposed over the former. Notice that the binning structure used for the codewords to be utilized by node 1 impose the following relationships:
\begin{IEEEeqnarray}{rCl}
R_{1\rightarrow 23}^{(l)} &<& \hat{R}_{1\rightarrow 23}^{(l)}+\hat{R}_{3\rightarrow 12}^{(l-1)}\ , \\ R_{1\rightarrow 2}^{(l)}& < & \hat{R}_{1\rightarrow 2}^{(l)}\ ,\\
R_{1\rightarrow 3}^{(l)} &<& \hat{R}_{1\rightarrow 3}^{(l)} \ .
   \end{IEEEeqnarray}
The common and private codewords to be utilized in nodes 2 and 3, for every round, are generated by following a similar procedure and theirs corresponding rates have analogous relationships. After this is finished the generated codebooks are revealed to all the nodes in the network.

 \subsection{Encoding technique}
 \label{subsec:encoding}

Consider node $1$ at round $l\in [1:K]$. Upon observing $x_1^n$ and given all of its encoding and decoding history up to round $l$, encoder $1$ first looks for a codeword $u_{1\rightarrow 23,l}^n(m_{1\rightarrow 23,l},\hat{m}_{\mathcal{W}_{[1,l]}}(1))$ such that  $\epsilon_c(1,23,l)>0$,
 \begin{equation}
 \left(x_1^n,w^n_{[1,l]}(\hat{m}_{\mathcal{W}_{[1,l]}}(1)),u_{1\rightarrow 23,l}^n(m_{1\rightarrow 23,l},\hat{m}_{\mathcal{W}_{[1,l]}}(1)) \right)
 \in\mathcal{T}^n_{[U_{1\rightarrow 23,l}X_1\mathcal{W}_{[1,l]}]\epsilon_c(1,23,l)} \ . 
 \end{equation}
 Notice that some components in $\hat{m}_{\mathcal{W}_{[1,l]}}(1)$  are generated at node 1 and are perfectly known. If more than one codeword satisfies this condition, then we choose the one with the smallest index. Otherwise, if no such codeword exists,  we choose an arbitrary index and declare an error. With the chosen index $m_{1\rightarrow 23,l}$, and with $\hat{m}_{\mathcal{W}_{[3,l-1]}}(1)$, we determine the index $p_{1\rightarrow 23,l}$ of the bin $\mathcal{B}_{1\rightarrow 23,l}\big(p_{1\rightarrow 23,l},\hat{m}_{\mathcal{W}_{[3,l-1]}}\big)$ to which $u_{1\rightarrow 23,l}^n(m_{1\rightarrow 23,l},\hat{m}_{3\rightarrow 12,l-1}(1),\hat{m}_{\mathcal{W}_{[3,l-1]}}(1))$ belongs. After this, Encoder $1$ generates the private descriptions looking for codewords $u_{1\rightarrow 2,l}^n(m_{1\rightarrow 2,l},$ $\hat{m}_{\mathcal{W}_{[2,l]}}(1),\hat{m}_{\mathcal{V}_{[12,l,1]}}(1)),u_{1\rightarrow 3,l}^n(m_{1\rightarrow 2,l},\hat{m}_{\mathcal{W}_{[2,l]}}(1),\hat{m}_{\mathcal{V}_{[13,l,1]}}(1))$ such that 
 
\begin{IEEEeqnarray}{lCl}
&& \left(x_1^n,w^n_{[2,l]}(\hat{m}_{\mathcal{W}_{[2,l]}}(1)),v^n_{[12,l,1]}(\hat{m}_{\mathcal{V}_{[12,l,1]}}(1)),u_{1\rightarrow 2,l}^n(m_{1\rightarrow 2,l},\hat{m}_{\mathcal{W}_{[2,l]}}(1),\hat{m}_{\mathcal{V}_{[12,l,1]}}(1))\right)\nonumber\\
  \IEEEeqnarraymulticol{3}{c}{ \in\mathcal{T}^n_{[U_{1\rightarrow 2,l}X_1\mathcal{W}_{[2,l]}\mathcal{V}_{[12,l,1]}]\epsilon_c(1,2,l)},\ ,}\\
&& \left(x_1^n,w^n_{[2,l]}(\hat{m}_{\mathcal{W}_{[2,l]}}(1)),v^n_{[13,l,1]}(\hat{m}_{\mathcal{V}_{[13,l,1]}}(1)),u_{1\rightarrow 3,l}^n(m_{1\rightarrow 2,l},\hat{m}_{\mathcal{W}_{[2,l]}}(1),\hat{m}_{\mathcal{V}_{[13,l,1]}}(1))\right) \nonumber\\
  \IEEEeqnarraymulticol{3}{c}{  \in\mathcal{T}^n_{[U_{1\rightarrow 3,l}X_1\mathcal{W}_{[2,l]}\mathcal{V}_{[13,l,1]}]\epsilon_c(1,3,l)}\ ,}
  \end{IEEEeqnarray}
 respectively, where $ \epsilon_c(1,2,l)>0$ and $ \epsilon_c(1,3,l)>0$. Given $\big(\hat{m}_{\mathcal{W}_{[2,l]}}(1),\hat{m}_{\mathcal{V}_{[12,l,1]}}(1),\hat{m}_{\mathcal{V}_{[13,l,1]}}(1)\big)$,   the encoding procedure continues by determining  the bin indices $p_{1\rightarrow 2,l}$ and  $p_{1\rightarrow 3,l}$ to which the generated private descriptions belong to. Node 1 then transmits to node 2 and 3 the indices $(p_{1 \rightarrow 23,l}, p_{1\rightarrow 2,l}, p_{1\rightarrow 3,l})$. The encoding in nodes $2$ and $3$ follows along the same lines and for that reason are not described.
 
\subsection{Decoding technique}
\label{subsec:decoding}

Consider round $l\in[1: K+1]$ and node 2. During round the present and previous round node 2 receives  $(p_{1\rightarrow 23,l}, p_{3\rightarrow 12,l-1}, p_{1\rightarrow 2,l}, p_{1\rightarrow 3,l}, p_{3\rightarrow 1,l-1},p_{3\rightarrow 2,l-1})$.  However, only the indices $(p_{1\rightarrow 23,l}, p_{3\rightarrow 12,l-1}, p_{1\rightarrow 2,l},p_{3\rightarrow 2,l-1})$ are the ones relevant to him. Knowing this set of indices, node 2 aims to recover the exact values of $(m_{1\rightarrow 23,l}, m_{3\rightarrow 12,l-1}, m_{1\rightarrow 2,l},m_{3\rightarrow 2,l-1})$. This is done through \emph{successive decoding}
   where first, the common information indices are recovered by looking for the unique tuple of codewords $u_{1\rightarrow 23,l}^n(m_{1\rightarrow 23,l},m_{3\rightarrow 12,l-1},\hat{m}_{\mathcal{W}_{[3,l-1]}}(2))$, $u_{3\rightarrow 12,l-1}^n(m_{3\rightarrow 12,l-1},$ $\hat{m}_{\mathcal{W}_{[3,l-1]}}(2))$ that satisfies:
\begin{IEEEeqnarray}{rCl}
&&  \left(x_2^n,w^n_{[3,l-1]}(\hat{m}_{\mathcal{W}_{[3,l-1]}}(2)),v^n_{[12,1,l]}(\hat{m}_{\mathcal{V}_{[12,l,1]}}(2)),v^n_{[23,l-1,3]}(\hat{m}_{\mathcal{V}_{[23,l-1,3]}}(2)),\right.\nonumber\\
   \IEEEeqnarraymulticol{3}{c}{ \left.u_{1\rightarrow 23,l}^n(m_{1\rightarrow 23,l},m_{3\rightarrow 12,l-1},\hat{m}_{\mathcal{W}_{[3,l-1]}}(2)),
  u_{3\rightarrow 12,l-1}^n(m_{3\rightarrow 12,l-1},\hat{m}_{\mathcal{W}_{[3,l-1]}}(2))\right)}\nonumber\\
   \IEEEeqnarraymulticol{3}{c}{ \in  \mathcal{T}^n_{[U_{1\rightarrow 23,l}U_{3\rightarrow 12,l-1}X_2\mathcal{W}_{[3,l-1]}\mathcal{V}_{[23,l-1,3]}\mathcal{V}_{[12,l,1]}]\epsilon_{dc}(2,l)},\ \epsilon_{dc}(2,l)>0}
  \end{IEEEeqnarray}
  and also belong to the bins indicated by $p_{1\rightarrow 23,l}$ and $p_{3\rightarrow 12,l-1}$. If there are more than one pair of codewords, or none that satisfies this, we choose a predefined one and declare an error. After this is done, node 2 can recover the private information indices by looking at codewords  $u^n_{1\rightarrow 2,l}(m_{1\rightarrow 2,l},\hat{m}_{\mathcal{W}_{[2,l]}}(2),\hat{m}_{\mathcal{V}_{[12,l,1]}}(2))$ and  $u^n_{3\rightarrow 2,l-1}(m_{3\rightarrow 2,l-1},\hat{m}_{\mathcal{W}_{[1,l]}}(2),\hat{m}_{\mathcal{V}_{[23,l-1,3]}}(2))$ which satisfy
  
\begin{IEEEeqnarray}{rCl}
&&  \big(x_2^n,w^n_{[2,l]}(\hat{m}_{\mathcal{W}_{[2,l]}}(2)),v^n_{[12,l,1]}(\hat{m}_{\mathcal{V}_{[12,l,1]}}(2)),v^n_{[23,l-1,3]}(\hat{m}_{\mathcal{V}_{[23,l-1,3]}}(2)),\big.\nonumber\\
   \IEEEeqnarraymulticol{3}{c}{   \big. u^n_{1\rightarrow 2,l}(m_{1\rightarrow 2,l},\hat{m}_{\mathcal{W}_{[2,l]}}(2),\hat{m}_{\mathcal{V}_{[12,l,1]}}(2)),
  u^n_{3\rightarrow 2,l-1}(m_{3\rightarrow 2,l-1},\hat{m}_{\mathcal{W}_{[1,l]}}(2),\hat{m}_{\mathcal{V}_{[23,l-1,3]}}(2)) \big)}\nonumber\\
   \IEEEeqnarraymulticol{3}{c}{   \in  \mathcal{T}^n_{[U_{1\rightarrow 2,l}U_{3\rightarrow 2,l-1}X_2\mathcal{W}_{[2,l]}\mathcal{V}_{[23,l-1,3]}\mathcal{V}_{[12,l,1]}]\epsilon_{dp}(2,l)},\ \epsilon_{dp}(2,l)>0}
  \end{IEEEeqnarray}
   and are in the bins given by  $p_{1\rightarrow 2,l}$ and $p_{3\rightarrow 2,l-1}$. If there are more than one pair of codewords, or none that satisfies this, we choose a predefined one and declare an error. The decoding in nodes 1 and 3 is exactly the same and for that reason are not described.

 \subsection{Lossy reconstructions}
\label{subsec:lossy}
 When the exchange of information is completed, each node needs to estimate the other nodes sources. For instance, node $1$ reconstruct the source of node $2$ by computing:
 \begin{equation}
 \hat{x}_{12,i} = g_{12}\left( x_{1i}, v_{[12,K+1,1]i},w_{[1,K+1]i}\right), \, i = 1,2,\dots,n,
 \label{eq:reconstruct_1}
 \end{equation}
and similarly, for the source of node $3$:
 \begin{equation}
 \hat{x}_{13,i} = g_{13}\left( x_{1i}, v_{[13,K+1,1]i},w_{[1,K+1]i}\right),\,  i = 1,2,\dots,n.
 \label{eq:reconstruct_2}
 \end{equation}
Reconstruction at nodes $2$ and $3$ is done in a similar way  using the adequate reconstruction functions. 
\vspace{1mm}

\subsection{Error and distortion analysis}
\label{subsec:error_analysis}

In order to maintain expressions simple, in the following when we denote a description without the corresponding index, i.e. $U^n_{i\rightarrow S,l}$ or $\mathcal{W}_{[1,l]}^n$, we will assume that the corresponding index is the true one generated in the corresponding nodes through the detailed  encoding procedure. Consider round $l$ and the event $\mathcal{D}_l=\mathcal{G}_l\cap\mathcal{F}_l$, where for $\epsilon_l>0$, 
\begin{IEEEeqnarray}{rcl}
\mathcal{G}_l&=&\left\{\left(X_1^n,X_2^n,X_3^n, \mathcal{W}^n_{[1,l]},\mathcal{V}^n_{[12,l,1]},\mathcal{V}_{[13,l,1]}^n,\mathcal{V}_{[23,l,2]}^n\right)\in\mathcal{T}^n_{[X_1X_2X_3\mathcal{W}_{[1,l]}\mathcal{V}_{[12,l,1]}\mathcal{V}_{[13,l,1]}\mathcal{V}_{[23,l,2]}]\epsilon_l}\right\} \ ,\,\,\,\,\,\,\,\, \\
\mathcal{F}_l&=&\left\{\hat{M}_{i\rightarrow S,t}(j)=M_{i\rightarrow S,t},\ S\in\mathcal{C}(\mathcal{M}),\ i\notin S,\ j\in S, \ t\in[1:l-1], \mbox{with exception of}\right. \nonumber\\
& &\left.\ \hat{M}_{3\rightarrow 12, l-1}(2),\hat{M}_{3\rightarrow 2, l-1}(2)\right\}\ .
\end{IEEEeqnarray}
The set $\mathcal{G}_l$ indicates that all the descriptions generated in the network, up to round $l$ are jointly typical with the sources. The ocurrence of this depends mainly on the encoding procedure in the nodes. Set $\mathcal{F}_l$ indicates, that up to round $l$, all nodes were able to recovers the true indices of the descriptions. This clearly implies that there were not errors at the decoding procedures in  all the nodes in the network. The condition in $\mathcal{F}_l$ on $\hat{M}_{3\rightarrow 12, l-1}(2),\hat{M}_{3\rightarrow 2, l-1}(2)$ is due to the fact, that the decoding of those descriptions in node 2 occurs during round $l$. The occurrence of $\mathcal{D}_l$ guarantees that at the beginning of round $l$:
\begin{itemize}
\item Node 1 and 2 share a common path of descriptions $\mathcal{W}^n_{[1,l]}\cup\mathcal{V}_{[12,l,1]}^n$ which are typical with $(X_1^n,X_2^n,X_3^n)$.
\item Node 1 and 3 share a common path of descriptions $\mathcal{W}^n_{[1,l]}\cup\mathcal{V}_{[13,l,1]}^n$ which are typical  with $(X_1^n,X_2^n,X_3^n)$.
\item Node 2 and 3 share a common path of descriptions $\mathcal{W}^n_{[3,l-1]}\cup\mathcal{V}_{[23,l-1,3]}^n$ which are typical  with $(X_1^n,X_2^n,X_3^n)$.
\end{itemize}
Let us also define the event $\mathcal{E}_l$:
\begin{IEEEeqnarray}{rcl}
\label{eq:round_error}
\mathcal{E}_l&=&\left\{\mbox{there exists at least an error at the encoding or decoding in a node during round $l$}\right\}\nonumber\\
&=&\bigcup_{i\in\mathcal{M}}\mathcal{E}_{enc}(i,l)\cup\mathcal{E}_{dec}(i,l)
\end{IEEEeqnarray}
where $\mathcal{E}_{enc}(i,l)$ contains the errors at the encoding in node $i$ during round $l$ and $\mathcal{E}_{dec}(i,l)$ considers the event that at node $i$ during round $l$ there is a failure at recovering an index generated previously in other node. For example, at node 1 and during round $l$:
\begin{equation}
\mathcal{E}_{enc}(i,l)=\mathcal{E}_{enc}(1,l,23)\cup\mathcal{E}_{enc}(1,l,2)\cup\mathcal{E}_{enc}(1,l,3)
\end{equation}
where
\begin{IEEEeqnarray}{rcl}
\mathcal{E}_{enc}(1,l,23)&=&\left\{\left(X_1^n,\mathcal{W}^n_{[1,l]}(\hat{M}_{\mathcal{W}_{[1,l]}}(1))U_{1\rightarrow 23,l}^n(m_{1\rightarrow 23,l},\hat{M}_{\mathcal{W}_{[1,l]}}(1)) \right)
 \notin\mathcal{T}^n_{[U_{1\rightarrow 23,l}X_1\mathcal{W}_{[1,l]}]\epsilon_c(1,l,23)}\right.\nonumber\\
& & \left. \forall m_{1\rightarrow 23,l}\in[1:2^{n\hat{R}^{(l)}_{1\rightarrow 23}}] \right\}\\
\mathcal{E}_{enc}(1,l,2)&=&\left\{\left(X_1^n,\mathcal{W}^n_{[2,l]}(\hat{M}_{\mathcal{W}_{[2,l]}}(1)),\mathcal{V}^n_{[12,l,1]}(\hat{M}_{\mathcal{V}_{[12,l,1]}}(1)),U_{1\rightarrow 2,l}^n(m_{1\rightarrow 2,l},\hat{M}_{\mathcal{W}_{[2,l]}}(1),\hat{M}_{\mathcal{V}_{[12,l,1]}}(1))\right)\right.\nonumber\\
&& \left.\notin\mathcal{T}^n_{[U_{1\rightarrow 2,l}X_1\mathcal{W}_{[1,l]}\mathcal{V}_{[12,l,1]}]\epsilon_c(1,l,2)} \ \forall m_{1\rightarrow 2,l}\in[1:2^{n\hat{R}^{(l)}_{1\rightarrow 2}}] \right\}\\
\mathcal{E}_{enc}(1,l,3)&=&\left\{\left(X_1^n,\mathcal{W}^n_{[2,l]}(\hat{M}_{\mathcal{W}_{[2,l]}}(1)),\mathcal{V}^n_{[13,l,1]}(\hat{M}_{\mathcal{V}_{[13,l,1]}}(1)),U_{1\rightarrow 3,l}^n(m_{1\rightarrow 3,l},\hat{M}_{\mathcal{W}_{[2,l]}}(1),\hat{M}_{\mathcal{V}_{[13,l,1]}}(1))\right)\right.\nonumber\\
&& \left.\notin\mathcal{T}^n_{[U_{1\rightarrow 3,l}X_1\mathcal{W}_{[1,l]}\mathcal{V}_{[13,l,1]}]\epsilon_c(1,l,3)} \ \forall m_{1\rightarrow 3,l}\in[1:2^{n\hat{R}^{(l)}_{1\rightarrow 3}}] \right\}\ .
 \end{IEEEeqnarray}
Event $\mathcal{E}_{dec}(i,l)$ can be decomposed as: 
 \begin{equation}
 \mathcal{E}_{dec}(i,l)=\bigcup_{S\in\mathcal{C}\left(\mathcal{M}\right),i\in S}\bigcup_{j:j\notin S}\left\{\hat{M}_{j\rightarrow S,l}(i)\neq M_{j\rightarrow S,l}\right\}\ .
 \end{equation} 

At the end of the information exchange phase we would expect the occurrence of $\mathcal{D}_{K+1}\cap\bar{\mathcal{E}}_{K+1}$, where $\mathcal{E}_{K+1}$ is the event of an error during round $K+1$. As during round $K+1$ only node 2 tries to recover the descriptions generated during round $K$ in node 3, we have:
\begin{equation}
\mathcal{E}_{K+1}=\mathcal{E}_{dec}(2,K+1)=\left\{ \hat{M}_{3\rightarrow 12, K}(2)\neq M_{3\rightarrow 12,K}\ \mbox{or}\ \hat{M}_{3\rightarrow 2, K}(2)\neq M_{3\rightarrow 2,K}\right\}\ .
\end{equation}
The occurrence of $\mathcal{D}_{K+1}\cap\bar{\mathcal{E}}_{K+1}$ guarantees that all the descriptions generated during the $K$ rounds of information exchange in the network are jointly typical with the sources realizations and that those descriptions can be perfectly recovered in all the nodes. In this way, if we can guarantee that $\mbox{Pr}\left\{\mathcal{D}_{K+1}\cap\bar{\mathcal{E}}_{K+1}\right\}\xrightarrow[n\rightarrow\infty]{}1$, then with probability converging to one we obtain:
\begin{itemize}
\item Node 1 and 2 share a common path of descriptions $\mathcal{W}^n_{[1,K+1]}\cup\mathcal{V}_{[12,K+1,1]}^n$ which are typical with $(X_1^n,X_2^n,X_3^n)$.
\item Node 1 and 3 share a common path of descriptions $\mathcal{W}^n_{[1,K+1]}\cup\mathcal{V}_{[13,K+1,1]}^n$ which are typical with $(X_1^n,X_2^n,X_3^n)$.
\item Node 2 and 3 share a common path of descriptions $\mathcal{W}^n_{[1,K+1]}\cup\mathcal{V}_{[23,K+1,2]}^n$ which are typical with $(X_1^n,X_2^n,X_3^n)$.
\end{itemize}
Using standard analysis ideas, the average distortions (over the codebooks) at the reconstruction stages in all the nodes satisfy the required fidelity constraints. From there is straightforward to prove the existence of  good codebooks for the network. In order to prove that  $\mbox{Pr}\left\{\mathcal{D}_{K+1}\cap\bar{\mathcal{E}}_{K+1}\right\}\xrightarrow[n\rightarrow\infty]{}1$ let us write:
\begin{IEEEeqnarray}{rcl}
 \mbox{Pr}\left\{\overline{\mathcal{D}_{K+1}\cap\bar{\mathcal{E}}_{K+1}}\right\}&=&\mbox{Pr}\left\{\bar{\mathcal{D}}_{K+1}\cup \mathcal{E}_{K+1}\right\}=\mbox{Pr}\left\{\bar{\mathcal{D}}_{K+1}\right\}+\mbox{Pr}\left\{\mathcal{D}_{K+1}\cap\mathcal{E}_{K+1}\right\}\nonumber\\
 &\leq & \mbox{Pr}\left\{\bar{\mathcal{D}}_{K+1}\cap\mathcal{D}_K\right\}+\mbox{Pr}\left\{\bar{\mathcal{D}}_{K}\right\}+\mbox{Pr}\left\{\mathcal{D}_{K+1}\cap\mathcal{E}_{K+1}\right\}\nonumber\\
 &\leq & \mbox{Pr}\left\{\bar{\mathcal{D}}_{K}\right\}+\mbox{Pr}\left\{\bar{\mathcal{D}}_{K+1}\cap\left(\mathcal{D}_K\cap \bar{\mathcal{E}}_K\right)\right\}+\mbox{Pr}\left\{\mathcal{D}_{K}\cap\mathcal{E}_{K}\right\}+\mbox{Pr}\left\{\mathcal{D}_{K+1}\cap\mathcal{E}_{K+1}\right\}\nonumber\\
 &\leq & \mbox{Pr}\left\{\bar{\mathcal{D}}_{1}\right\}+\sum_{l=1}^{K+1}\mbox{Pr}\left\{\mathcal{D}_{l}\cap\mathcal{E}_{l}\right\}+
 \sum_{l=1}^K\mbox{Pr}\left\{\bar{\mathcal{D}}_{l+1}\cap\left(\mathcal{D}_l\cap \bar{\mathcal{E}}_l\right)\right\}\ .
 \end{IEEEeqnarray}
 Notice that
 \begin{equation}
 \mathcal{D}_1=\left\{(X_1^n,X_2^n,X_3^n)\in\mathcal{T}^n_{[X_1X_2X_3]\epsilon_1}\right\}\ , \ \epsilon_1>0\ .
 \end{equation}
 From Lemma~\ref{lemma:prob_lim}, we see that for every $\epsilon_1>0$  $\mbox{Pr}\left\{\bar{\mathcal{D}}_{1}\right\}\xrightarrow[n\rightarrow\infty]{}0$. Then, it is easy to see that $\mbox{Pr}\left\{\mathcal{D}_{K+1}\cap\bar{\mathcal{E}}_{K+1}\right\}\xrightarrow[n\rightarrow\infty]{}1$ will hold if the coding generation, the encoding and decoding procedures described above allow us to have the following:
 \begin{enumerate}
 \item If $\mbox{Pr}\left\{\mathcal{D}_{l}\right\}\xrightarrow[n\rightarrow\infty]{}1$ then $\mbox{Pr}\left\{\mathcal{D}_{l+1}\right\}\xrightarrow[n\rightarrow\infty]{}1$ $\forall l\in[1:K+1]$.
 \item $\mbox{Pr}\left\{\mathcal{D}_{l}\cap\mathcal{E}_{l}\right\}\xrightarrow[n\rightarrow\infty]{}0$ $\forall l\in[1:K+1]$.
 \end{enumerate}
 In the following we will prove these facts. Observe that, at round $l$ the nodes act sequentially:
 \[\mbox{Encoding at node 1}\rightarrow\mbox{Decoding at node 2}\rightarrow\cdots\rightarrow\mbox{Encoding at node 3}\rightarrow\mbox{Decoding at node 1}.\]
Then, using (\ref{eq:round_error}) we can write condition 2) as:
 \begin{IEEEeqnarray}{rcl}
 \mbox{Pr}\left\{\mathcal{D}_{l}\cap\mathcal{E}_{l}\right\}&=&\mbox{Pr}\left\{\mathcal{D}_{l}\cap\mathcal{E}_{enc}(1,l)\right\}+\mbox{Pr}\left\{\mathcal{D}_{l}\cap\mathcal{E}_{dec}(2,l)\cap\bar{\mathcal{E}}_{enc}(1,l)\right\}\nonumber\\
& &+ \mbox{Pr}\left\{\mathcal{D}_{l}\cap\mathcal{E}_{enc}(2,l)\cap\bar{\mathcal{E}}_{enc}(1,l)\cap\bar{\mathcal{E}}_{dec}(2,l)\right\}+\nonumber\\
& &\dots+
\mbox{Pr}\left\{\mathcal{D}_{l}\cap\mathcal{E}_{dec}(1,l)\cap\bar{\mathcal{E}}_{enc}(1,l)\cap\cdots\cap\bar{\mathcal{E}}_{enc}(3,l)\right\}.
 \label{eq:error_prob_2}
 \end{IEEEeqnarray}
  Assume then that at the beginning of round $l$ we have $\mbox{Pr}\left\{\mathcal{D}_{l}\right\}\xrightarrow[n\rightarrow\infty]{}1$. Let us analize the encoding procedure at node 1. Let us consider $\mbox{Pr}\left\{\mathcal{D}_l\cap\mathcal{E}_{enc}(1,l)\right\}$. We can write:
 \begin{IEEEeqnarray}{rcl}
 \mbox{Pr}\left\{\mathcal{D}_l\cap\mathcal{E}_{enc}(1,l)\right\}&\leq& \mbox{Pr}\left\{\mathcal{E}_{enc}(1,l,23)\cap\mathcal{D}_l\right\}+\mbox{Pr}\left\{\mathcal{E}_{enc}(1,l,2)\cap\mathcal{D}_l\cap\bar{\mathcal{E}}_{enc}(1,l,23)\right\}\nonumber\\
 & &+\mbox{Pr}\left\{\mathcal{E}_{enc}(1,l,3)\cap\mathcal{D}_l\cap\bar{\mathcal{E}}_{enc}(1,l,23)\right\}\ .
 \end{IEEEeqnarray}
 From Lemma \ref{lemma:useful} and the fact  that $\lim_{n\rightarrow\infty}\mbox{Pr}\left\{\mathcal{G}_l\right\}=1$ we have that $\lim_{n\rightarrow\infty}\mbox{Pr}\left\{\mathcal{A}_l(1,23)\right\}=1$ where:
 \begin{equation}
 \mathcal{A}_l(1,23)=\left\{(X^n_1,\mathcal{W}_{[1,l]}^n)\in\mathcal{T}^n_{[X_1\mathcal{W}_{[1,l]}]\epsilon_l}\right\}\ .
 \end{equation}
Then, we can use Lemma~\ref{lemma:covering} to obtain:
 \begin{equation}
 \lim_{n\rightarrow \infty}\mbox{Pr}\left\{\mathcal{E}_{enc}(1,l,23)\cap\mathcal{D}_l\right\}=0 \end{equation}
provided that 
\begin{equation}
\label{eq:rate_1}
\hat{R}_{1\rightarrow 23}^{(l)}>I\left(X_1;U_{1\rightarrow 23,l}\Big|\mathcal{W}_{[1,l]}\right)+\delta_c(1,l,23)
\end{equation}
where $\delta_c(1,l,23)$ can be made arbitrarily small. On the other hand we can write:
\begin{equation}
\label{eq:encoding_1_2}\mbox{Pr}\left\{\mathcal{E}_{enc}(1,l,2)\cap\mathcal{D}_l\cap\bar{\mathcal{E}}_{enc}(1,l,23)\right\}\leq \mbox{Pr}\left\{\mathcal{E}_{enc}(1,l,2)\cap\mathcal{G}_l(1,2)\cap\mathcal{F}_l\right\}+\mbox{Pr}\left\{\bar{\mathcal{G}}_{l}(1,2)\right\}
\end{equation}
where
 \begin{equation}
\mathcal{G}_l(1,2)= \left\{(X_1^n,X_2^n,X_3^n,\mathcal{W}_{[2,l]}^n,\mathcal{V}_{[12,l,1]}^n,\mathcal{V}_{[13,l,1]}^n,\mathcal{V}_{[23,l,2]}^n) \in\mathcal{T}^n_{[X_1X_2X_3\mathcal{W}_{[2,l]}\mathcal{V}_{[12,l,1]}\mathcal{V}_{[13,l,1]}\mathcal{V}_{[23,l,2]}]\epsilon_l(1,2)}\right\}  
\end{equation}
where $\epsilon_l(1,2)>0$.  As explained before, $\mbox{Pr}\left\{\mathcal{G}_l\right\}\xrightarrow[n\rightarrow\infty]{}1$. Then,
from condition (\ref{eq:rate_1}) we have
\begin{equation}
\mbox{Pr}\left\{\bar{\mathcal{E}}_{enc}(1,l,23)\cap\mathcal{F}_l\right\}=\mbox{Pr}\left\{(X_1^n,\mathcal{W}_{[2,l]}^n)\in\mathcal{T}_{[X_1\mathcal{W}_{[2,l]}]\epsilon_c(1,l,23)}\right\}\xrightarrow[n\rightarrow\infty]{}1\ .
\end{equation}
Moreover, from the coding generation and the encoding procedure proposed is immediate to use Lemma~\ref{lemma:induced} to show that:
\begin{IEEEeqnarray}{rCl}
\mbox{Pr}\left(U_{1\rightarrow 23,l}^n  = u^n_{1\rightarrow 23,l}\big|x_1^n,w_{[1:l]}^n,\bar{\mathcal{E}}_{enc}(1,l,23)\cap\mathcal{F}_l\right) &=& \nonumber\\
\IEEEeqnarraymulticol{3}{r}{ \frac{\mathds{1}\left\{u^n_{1\rightarrow 23,l}\in\mathcal{T}_{[U_{1\rightarrow 23,l}|X_1\mathcal{W}_{[1,l]}]\epsilon_c(1,23,l)}^n\left(x_1^n,w_{[1,l]}^n\right)\right\}}{\left\|\mathcal{T}_{[U_{1\rightarrow 23,l}|X_1\mathcal{W}_{[1,l]}]\epsilon_c(1,23,l)}^n\left(x_1^n,w_{[1,l]}^n\right)\right\|}\ .}
\end{IEEEeqnarray}
Then, from Markov chain 
 \begin{equation}
U_{1\rightarrow 23,l}\mkv (X_1,\mathcal{W}_{[1,l]})\mkv (X_2,X_3,\mathcal{V}_{[12,l,1]},\mathcal{V}_{[13,l,1]},\mathcal{V}_{[23,l,2]})
 \end{equation}
and the Markov Lemma \ref{lemma:markov}, for sufficiently small $\left(\epsilon_c(1,l,23),\epsilon_l,\epsilon_l(1,2)\right)$ and after some minor manipulations, we can obtain:
 \begin{equation}
 \mbox{Pr}\left\{\mathcal{G}_l(1,2)\right\}\xrightarrow[n\rightarrow\infty]{}1\ .
 \end{equation}
From equation (\ref{eq:encoding_1_2}) it is clear that we need to analyze term $\mbox{Pr}\left\{\mathcal{E}_{enc}(1,l,2)\cap\mathcal{G}_l(1,2)\cap\mathcal{F}_l\right\}$. Similarly as before $\lim_{n\rightarrow\infty}\mbox{Pr}\left\{\mathcal{A}_l(1,2)\right\}$ where:
 \begin{equation}
 \mathcal{A}_l(1,2)=\left\{(X^n_1,\mathcal{W}_{[2,l]}^n,\mathcal{V}^n_{[12,l,1]})\in\mathcal{T}^n_{[X_1\mathcal{W}_{[2,l]}\mathcal{V}_{[12,l,1]}]\epsilon_l(1,2)}\right\}\ ,
 \end{equation}
 which allow us to write:
 \begin{equation}
 \mbox{Pr}\left\{\mathcal{E}_{enc}(1,l,2)\cap\mathcal{G}_l(1,2)\cap\mathcal{F}_l\right\}\leq \mbox{Pr}\left\{\mathcal{E}_{enc}(1,l,2)\cap\mathcal{A}_l(1,2)\cap\mathcal{F}_l\right\}\ .
 \end{equation}
 Using again Lemma~\ref{lemma:covering}, we obtain that $ \mbox{Pr}\left\{\mathcal{E}_{enc}(1,l,2)\cap\mathcal{G}_l(1,2)\cap\mathcal{F}_l\right\}\xrightarrow[n\rightarrow\infty]{}0$ provided that
\begin{equation}
\label{eq:rate_2}
\hat{R}_{1\rightarrow 2}^{(l)}>I\left(X_1;U_{1\rightarrow 2,l}\Big|\mathcal{W}_{[2,l]}\mathcal{V}_{[12,l,1]}\right)+\delta_c(1,l,2)
\end{equation}
where $\delta_c(1,l,2)$ can be made arbitrarly small. For the analysis of $\mbox{Pr}\left\{\mathcal{E}_{enc}(1,l,3)\cap\mathcal{D}_l\cap\bar{\mathcal{E}}_{enc}(1,l,23)\right\}$ we follow the same procedure. We can write
\begin{equation}
\label{eq:encoding_1_3}\mbox{Pr}\left\{\mathcal{E}_{enc}(1,l,3)\cap\mathcal{D}_l\cap\bar{\mathcal{E}}_{enc}(1,l,23)\right\}\leq \mbox{Pr}\left\{\mathcal{E}_{enc}(1,l,3)\cap\mathcal{G}_l(1,3)\cap\mathcal{F}_l\right\}+\mbox{Pr}\left\{\bar{\mathcal{G}}_{l}(1,3)\right\}
\end{equation}
with
\begin{equation}
\mathcal{G}_l(1,3)=\left\{(X_1^n,X_2^n,X_3^n,\mathcal{W}_{[2,l]}^n,\mathcal{V}_{[12,l,2]}^n,\mathcal{V}_{[13,l,1]}^n,\mathcal{V}_{[23,l,2]})\in\mathcal{T}^n_{[X_1X_2X_3\mathcal{W}_{[2,l]}\mathcal{V}_{[12,l,2]}\mathcal{V}_{[13,l,1]}\mathcal{V}_{[23,l,2]}]\epsilon_l(1,3)}\right\}\ .
\end{equation}
Using the Markov chain 
\begin{equation}
U_{1\rightarrow 2,l}\mkv (X_1,\mathcal{W}_{[2,l]},\mathcal{V}_{[12,l,1]})\mkv (X_2,X_3,\mathcal{V}_{[13,l,1]}\mathcal{V}_{[23,l,2]})\ ,
\end{equation}
the fact that $ \mbox{Pr}\left\{\mathcal{G}_l(1,2)\right\}\xrightarrow[n\rightarrow\infty]{}1$ and the Markov Lemma \ref{lemma:markov}, and Lemma \ref{lemma:induced} for appropriately chosen values of  $\left(\epsilon_c(1,l,2),\epsilon_l(1,2),\epsilon_l(1,3)\right)$ we have:
\begin{equation}
\mbox{Pr}\left\{\mathcal{G}_l(1,3)\right\}\xrightarrow[n\rightarrow\infty]{}1\ .
\end{equation}
Following exactly the same reasoning as above, we have that in order to have 
\begin{equation}
\mbox{Pr}\left\{\mathcal{E}_{enc}(1,l,3)\cap\mathcal{D}_l\cap\bar{\mathcal{E}}_{enc}(1,l,23)\right\}\xrightarrow[n\rightarrow\infty]{}0\ ,
\end{equation}
 besides conditions (\ref{eq:rate_1}) and (\ref{eq:rate_2}) we need:
 \begin{equation}
 \label{eq:rate_3}
 \hat{R}_{1\rightarrow 3}^{(l)}>I\left(X_1;U_{1\rightarrow 3,l}\Big|\mathcal{W}_{[2,l]}\mathcal{V}_{[13,l,1]}\right)+\delta_c(1,l,3)
 \end{equation}
 for sufficiently small $\delta_c(1,l,3)$. With these conditions we have proved that the encoding procedure in node 1 during round $l$ permit us to have:
 \begin{equation}
\mbox{Pr}\left\{\mathcal{D}_l\cap\mathcal{E}_{enc}(1,l)\right\}\xrightarrow[n\rightarrow\infty]{}0 \ .
 \end{equation}
Another instance of the Markov lemma, jointly with Markov chain 
 \begin{equation}
U_{1\rightarrow 3,l}\mkv (X_1,\mathcal{W}_{[2,l]},\mathcal{V}_{[13,l,1]})\mkv (X_2,X_3,\mathcal{V}_{[12,l,2]}\mathcal{V}_{[23,l,2]})
 \end{equation}
and Lemma~\ref{lemma:induced} allow us to have:
\begin{equation}
\mbox{Pr}\left\{\mathcal{G}_l(2,13)\right\}\xrightarrow[n\rightarrow\infty]{}1
\label{eq:G_at_2}
\end{equation}
where 
\begin{equation}
\mathcal{G}_l(2,13)=\left\{(X_1^n,X_2^n,X_3^n,\mathcal{W}_{[2,l]}^n,\mathcal{V}_{[12,l,2]}^n,\mathcal{V}_{[13,l,3]}^n,\mathcal{V}_{[23,l,2]})\in\mathcal{T}^n_{[X_1X_2X_3\mathcal{W}_{[2,l]}\mathcal{V}_{[12,l,2]}\mathcal{V}_{[13,l,3]}\mathcal{V}_{[23,l,2]}]\epsilon_l(2,13)}\right\}\ .
\end{equation}
At this point we have to analyze the decoding in node 2. If that decoding if successful, with (\ref{eq:G_at_2}),  the analysis of the encoding at node 2 follows the same lines as above\footnote{See that $\mathcal{G}_l(2,13)$ has, for the encoding at node 2 the same role that $\mathcal{G}_l$ has for the enconding at node 1 during round $l$.}. The same can be said of the encoding at node 3 (after successful decoding). In this way, we terminate round $l$ with 
\begin{equation}
\mbox{Pr}\left\{\mathcal{D}_{l+1}\right\}=\mbox{Pr}\left\{\mathcal{G}_{l+1}\cap\mathcal{F}_{l+1}\right\}\xrightarrow[n\rightarrow\infty]{}1
\end{equation}
which is one the results we wanted. Clearly, analyzing now the decoding at node 2 (from which we can easily extrapolate the analysis to the decoding at node 1 and 3) we will be able to obtain $\mbox{Pr}\left\{\mathcal{D}_{l}\cap\mathcal{E}_{l}\right\}\xrightarrow[n\rightarrow\infty]{}0$ which is the other required result.

The decoding in each of nodes follows the approach of \emph{successive decoding}. Decoder 2 will try to find first the common descriptions $M_{1\rightarrow 23,l}$ and $M_{3\rightarrow 12,l-1}$. Then, it will try to find the private descriptions $M_{1\rightarrow 2,l}$ and $M_{3\rightarrow 2,l-1}$ (using of course the previously obtained common descriptions as side information). Clearly, the use of \emph{joint-decoding} could improve the rate region. However, the analysis of this strategy, besides of being more difficult to analyze, it will give rise to more complex rate region. It can be easily seen, that the joint-decoding region will contain several sum-rate equations that will contains common and private rates. Successive decoding allows for a rate region where the sum-rate equations contains solely common rates or private rates, being more easy to analyze and understand.

In order to analyze the decoding, we can write:
\begin{equation}
\mbox{Pr}\left\{\mathcal{D}_{l}\cap\mathcal{E}_{dec}(2,l)\cap\bar{\mathcal{E}}_{enc}(1,l)\right\}\leq\mbox{Pr}\left\{\mathcal{E}_{dec}(2,l)\cap\mathcal{F}_{l}\cap\mathcal{G}_l(2,13)\right\}+\mbox{Pr}\left\{\bar{\mathcal{G}}_l(2,13)\right\}\ .
\end{equation}
As $\mbox{Pr}\left\{\mathcal{G}_l(2,13)\right\}\xrightarrow[n\rightarrow\infty]{}1$ we can concentrate our effort on the first term. Event $\mathcal{E}_{dec}(2,l)$ can be written as:
\begin{equation}
\mathcal{E}_{dec}(2,l)=\mathcal{H}_{common}(2,l)\cup\mathcal{H}_{private}(2,l)\ ,
\end{equation}
where
\begin{IEEEeqnarray}{rcl}
\mathcal{H}_{common}(2,l)&=&\left\{\left(\hat{M}_{3\rightarrow 12,l-1}(2),\hat{M}_{1\rightarrow 23,l}(2)\right)\neq (M_{3\rightarrow 12,l-1},M_{1\rightarrow 23,l})\right\}\ , \\
\mathcal{H}_{private}(2,l)&=& \left\{\left(\hat{M}_{3\rightarrow 2,l-1}(2),\hat{M}_{1\rightarrow 2,l}(2)\right)\neq (M_{3\rightarrow 2,l-1},M_{1\rightarrow 2,l})\right\}\ .
\end{IEEEeqnarray}
From these definitions, we can easily deduce that:
\begin{IEEEeqnarray}{rcl}
\mbox{Pr}\left\{\mathcal{E}_{dec}(2,l)\cap\mathcal{F}_{l}\cap\mathcal{G}_l(2,13)\right\}&=&\mbox{Pr}\left\{\mathcal{H}_{common}(2,l)\cap\mathcal{F}_{l}\cap\mathcal{G}_l(2,13)\right\}\nonumber\\
& &+\mbox{Pr}\left\{\mathcal{H}_{private}(2,l)\cap\mathcal{F}_{l}\cap\mathcal{G}_l(2,13)\cap \bar{\mathcal{H}}_{common}(2,l)\right\}\nonumber\\
&\leq& \mbox{Pr}\left\{\mathcal{K}_{common}(2,l)\right\}+\mbox{Pr}\left\{\mathcal{K}_{private}(2,l)\right\}
\label{eq:prob_K}
\end{IEEEeqnarray}
where
\begin{IEEEeqnarray}{rcl}
\mathcal{K}_{common}(2,l) &=& 
\Big\{\exists (\tilde{m}_{1\rightarrow 23,l},\tilde{m}_{3\rightarrow 12, l-1})\neq(M_{1\rightarrow 23,l},M_{3\rightarrow 12,l-1}), (\tilde{m}_{1\rightarrow 23,l},\tilde{m}_{3\rightarrow 12, l-1})  \nonumber \\
  \IEEEeqnarraymulticol{3}{r}{\in \mathcal{B}_{1\rightarrow 23,l}(P_{1\rightarrow 23,l})\times\mathcal{B}_{3\rightarrow 12,l-1}(P_{3\rightarrow 12,l-1}): \left(X_2^n,\mathcal{W}_{[3,l-1]}^n,\mathcal{V}^n_{[23,l-1,3]},\mathcal{V}_{[12,l,1]}^n,\right.} \nonumber \\
  \IEEEeqnarraymulticol{3}{r}{ \left. U_{1\rightarrow 23,l}^n(\tilde{m}_{1\rightarrow 23,l},\tilde{m}_{3\rightarrow 12,l-1}),
U_{3\rightarrow 12,l-1}^n(\tilde{m}_{3\rightarrow 12,l-1})\in\mathcal{T}_{[X_1\mathcal{W}_{[2,l]}\mathcal{V}_{[12,l,1]}\mathcal{V}_{[23,l-1,3]}]\epsilon_{dc}(2,l)}^n\right)
\Big\} \ ,}\label{eq:K_common} \\ 
\mathcal{K}_{private}(2,l) &=& 
\Big\{\exists (\tilde{m}_{1\rightarrow 2,l},\tilde{m}_{3\rightarrow 2, l-1})\neq(M_{1\rightarrow 2,l},M_{3\rightarrow 2,l-1}), (\tilde{m}_{1\rightarrow 2,l},\tilde{m}_{3\rightarrow 2, l-1})  \nonumber  \\
  \IEEEeqnarraymulticol{3}{r}{\in \mathcal{B}_{1\rightarrow 2,l}(P_{1\rightarrow 2,l})\times\mathcal{B}_{3\rightarrow 2,l-1}(P_{3\rightarrow 2,l-1}): \left(X_2^n,\mathcal{W}_{[2,l]}^n,\mathcal{V}^n_{[23,l-1,3]},\mathcal{V}_{[12,l,1]}^n,\right.} \nonumber \\
  \IEEEeqnarraymulticol{3}{r}{\left. U_{1\rightarrow 2,l}^n(\tilde{m}_{1\rightarrow 2,l}), U_{3\rightarrow 2,l-1}^n(\tilde{m}_{3\rightarrow 2,l-1})\in\mathcal{T}_{[X_1\mathcal{W}_{[2,l]}\mathcal{V}_{[12,l,2]}\mathcal{V}_{[23,l,2]}]\epsilon_{dp}(2,l)}^n\right)
\Big\}\ , }\label{eq:K_private}
\end{IEEEeqnarray}
where $\epsilon_{dc}(2,l),\epsilon_{dp}(2,l)$ are carefully chosen\footnote{
Using Lemma~\ref{lemma:useful} to have:
\begin{IEEEeqnarray*}{rcl}
\mathcal{G}_{l}(2,13)& \subseteq& \left(X_2^n,\mathcal{W}_{[2,l]}^n,\mathcal{V}^n_{[23,l-1,3]},\mathcal{V}_{[12,l,1]}^n)\in\mathcal{T}_{[X_1\mathcal{W}_{[2,l]}\mathcal{V}_{[12,l,1]}\mathcal{V}_{[23,l-1,3]}]\epsilon_{dc}(2,l)}^n\right)\ ,\\
\mathcal{G}_{l}(2,13)& \subseteq & \left(X_2^n,\mathcal{W}_{[2,l]}^n,\mathcal{V}^n_{[23,l,2]},\mathcal{V}_{[12,l,2]}^n)\in\mathcal{T}_{[X_1\mathcal{W}_{[2,l]}\mathcal{V}_{[12,l,2]}\mathcal{V}_{[23,l,2]}]\epsilon_{dp}(2,l)}^n\right)\ .
\end{IEEEeqnarray*}
}, and for a saving of notation we considered only the indices to be recovered, i.e., 
$ U_{1\rightarrow 23,l}^n(\tilde{m}_{1\rightarrow 23,l},\tilde{m}_{3\rightarrow 12,l-1})\equiv  U_{1\rightarrow 23,l}^n(\tilde{m}_{1\rightarrow 23,l},\tilde{m}_{3\rightarrow 12,l-1}, M_{\mathcal{W}_{[3,l-1]}})$
Consider first the recovering of the common information. Node 2 has to recover two indices from a binning structure as the one in the cooperative Berger-Tung problem described in 
Appendix~\ref{app:coop_berger}. 
\begin{figure}[t]
\centering
\ifpdf\includegraphics[angle=0,width=0.8\columnwidth,keepaspectratio,trim= 0mm 0mm 0mm 0mm,clip]{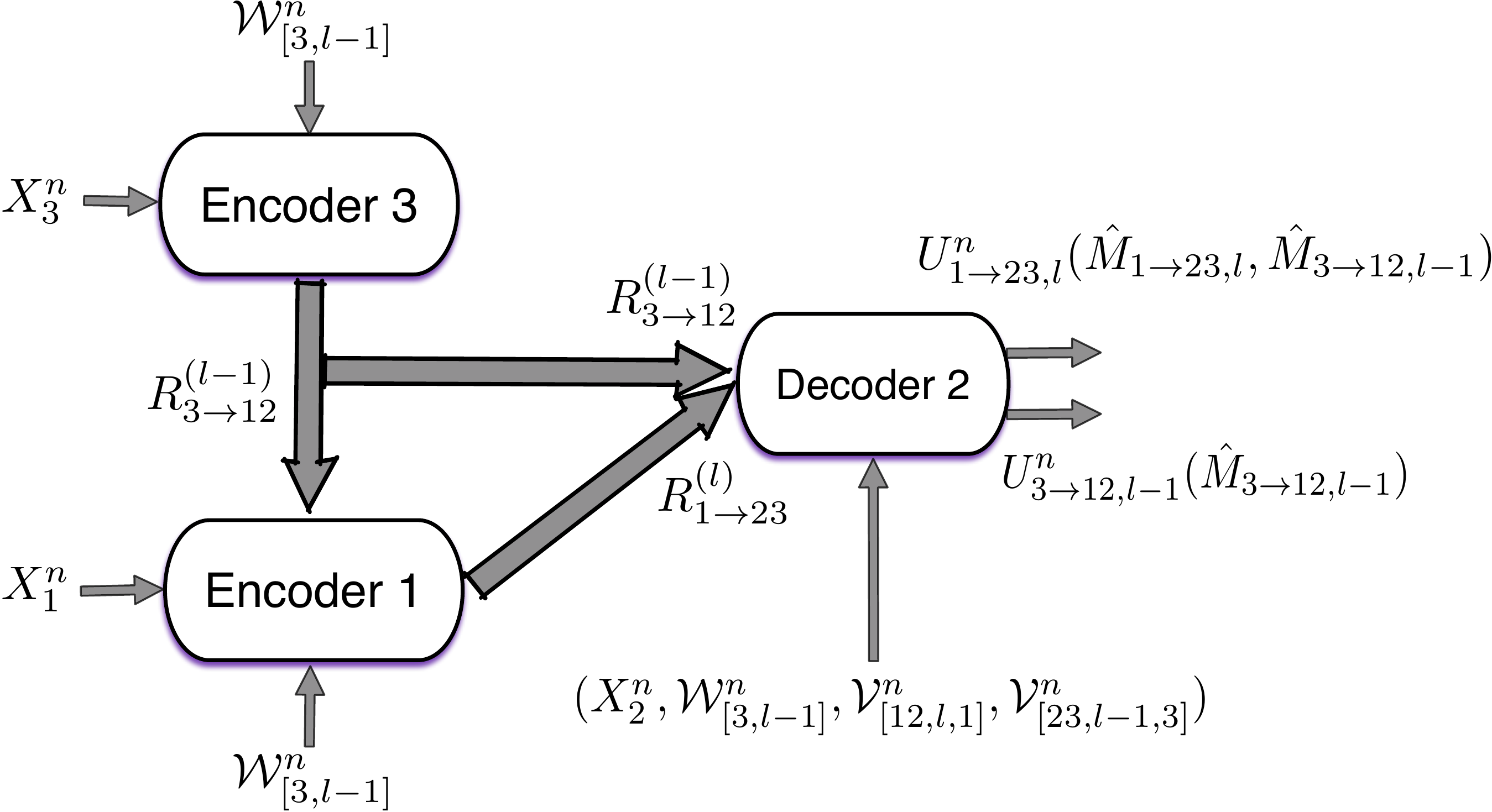} 
\else
	\includegraphics[angle=270,width=0.50\columnwidth,keepaspectratio,trim= 0mm 0mm 0mm 0mm,clip]{coop_berger_interactive.eps}
\fi
\caption{Cooperative Berger-Tung decoding problem for node 2.}
\label{fig:Berger_Tung_1}
\end{figure}

In Fig.~\ref{fig:Berger_Tung_1}, we have a representation of the problem seen at decoder $2$. Node $3$ generate a common description at rate $R_{3\rightarrow 12}^{(l-1)}$ using $\mathcal{W}_{[3,l-1]}^n$ as side information. Similarly node $1$, after decoding the common description from node 3, generate its own description using the recovered one and also $\mathcal{W}_{[3,l-1]}^n$ as side information. All these operations all done using the super-binning structure as in the cooperative Berger-Tung problem in Appendix~\ref{app:coop_berger}.  Then, node $2$, using $(X_3^n,\mathcal{W}_{[3,l-1]}^n,\mathcal{V}_{[12,l,1]}^n,\mathcal{V}_{[23,l-1,3]}^n)$ as side information tries to recover the descriptions generated at node $3$ and $1$. Remember the fact that the encoding procedure at nodes $1$ and $3$ requires:
\begin{IEEEeqnarray}{rCl}
\hat{R}_{3\rightarrow 12}^{(l-1)} &>& I(X_3;U_{3\rightarrow 12,l-1}|\mathcal{W}_{[3,l-1]})+\delta_c(1,l-1,12)\ ,\\ 
\hat{R}_{1\rightarrow 23}^{(l)} &>& I(X_1;U_{1\rightarrow 23,l}|\mathcal{W}_{[1,l]})+\delta_c(1,l,23)\ ,\\ 
R_{1\rightarrow 23}^{(l)} &<& \hat{R}_{1\rightarrow 23}^{(l)}+\hat{R}_{3\rightarrow 12}^{(l-1)}
\end{IEEEeqnarray}
and that the following Markov chains: 
\begin{IEEEeqnarray}{rCl}
U_{3\rightarrow 12,l-1}& \mkv & (X_3,\mathcal{W}_{[3,l-1]}) \mkv  (X_1,X_2,\mathcal{V}_{[12,l,1]},\mathcal{V}_{[23,l-1,3]})\ ,\\
U_{1\rightarrow 23,l} & \mkv &  (X_1,\mathcal{W}_{[1,l]})\mkv (X_2,X_3,\mathcal{V}_{[12,l,1]},\mathcal{V}_{[23,l,2]})/ ,
\end{IEEEeqnarray}
are implied by the Markov chains in the conditions of Theorem \ref{theo-main-theorem}.  In this way, we can use the results in Appendix~\ref{app:coop_berger} to show that the following rates imply  $\mbox{Pr}\left\{\mathcal{K}_{common}(2,l)\right\}\xrightarrow[n\rightarrow\infty]{}0$:
\begin{eqnarray}
R_{1\rightarrow 23}^{(l)}&>&I(X_1;U_{1\rightarrow 23,l}|X_2\mathcal{W}_{[1,l]}\mathcal{V}_{[23,l-1,3]}\mathcal{V}_{[12,l,1]})+\delta_{dc}(2,l)\ ,\\
R_{1\rightarrow 23}^{(l)}+R_{3\rightarrow 12}^{(l-1)}&>&I(X_1X_3;U_{1\rightarrow 23,l}U_{3\rightarrow 12,l-1}|X_2\mathcal{W}_{[3,l-1]}\mathcal{V}_{[23,l-1,3]}\mathcal{V}_{[12,l,1]})+\delta_{dc}'(2,l)\ ,\,\,\,\, \,\,\,\, 
\label{eq:rates_decod_com_2}
\end{eqnarray}
where $\delta_{dc}(2,l),\delta'_{dc}(2,l)$ can be made arbitrarily small\footnote{Here we considered the Corollary to Theorem~\ref{theo:coop_berger}. That is we assumed, that node 1 knows perfectly the value of $M_{3\rightarrow 12, l-1}$. This follows from the assumed fact, that at the beginning of round $l$, the probability of decoding errors at previous rounds in all nodes is goes to zero when $n\rightarrow\infty$. In this way, the constraint on rate $R_{3\rightarrow 12,l-1}$ that should be considered, according to Theorem~\ref{theo:coop_berger} is not needed. In fact, constraints on rate $R_{3\rightarrow 12,l-1}$ will arise when at node 1 we consider the recovering of $M_{3\rightarrow 12,l-1}$ and $M_{2\rightarrow 13,l-1}$.  For that reason, the analysis carried on is valid. Through this analysis we avoid carrying a lengthy and difficult Fourier-Motzkin procedure to eliminate $\hat{R}_{1\rightarrow 23}^{(l)},\hat{R}_{2\rightarrow 13}^{(l)},\hat{R}_{3\rightarrow 12}^{(l)}$ for $l=[1:K]$. }.

\begin{figure}[t]
\centering
\ifpdf\includegraphics[angle=0,width=0.75\columnwidth,keepaspectratio,trim= 0mm 0mm 0mm 0mm,clip]{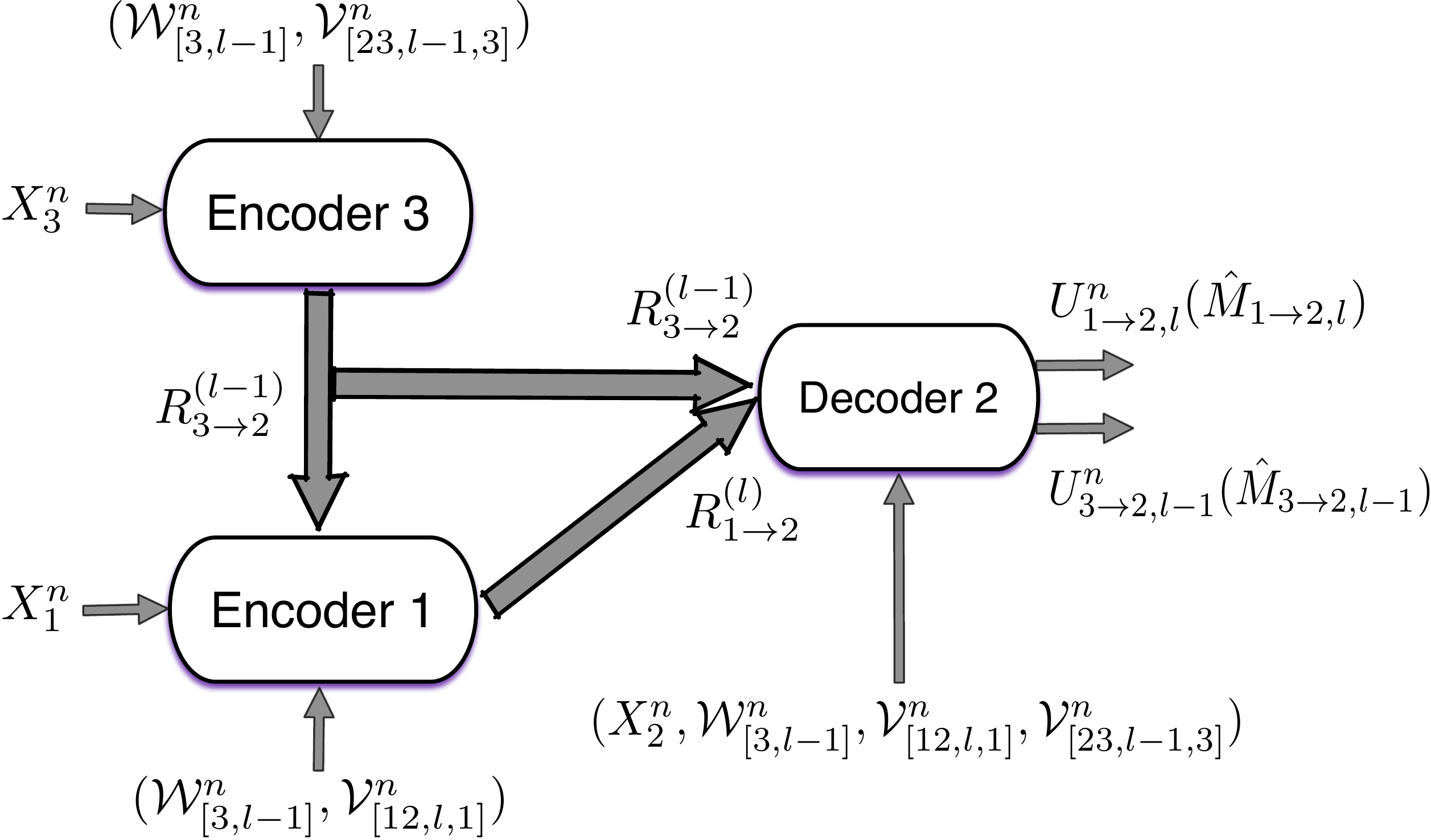} 
\else
	\includegraphics[angle=270,width=0.50\columnwidth,keepaspectratio,trim= 0mm 0mm 0mm 0mm,clip]{berger_interactive.eps}
\fi
\caption{Berger-Tung decoding problem for node 2 when it tries to recover the private descriptions generated in nodes 1 and 3.}
\label{fig:Berger_Tung_std}
\end{figure}

The decoding of the private descriptions can be seen as a standard Berger-Tung decoding problem (see Fig.~\ref{fig:Berger_Tung_std}) where the binning used to transmit the descriptions generated in node 3 and 1 is not cooperative (in the sense of Theorem~\ref{theo:coop_berger}) as in the case of the common descriptions. Lemma~\ref{lemma:packing} can bee easily used to analyze $\mbox{Pr}\left\{\mathcal{K}_{private}(2,l)\right\}$. The  following conditions guarantee that  $\mbox{Pr}\left\{\mathcal{K}_{private}(2,l)\right\}\xrightarrow[n\rightarrow\infty]{}0$:

\begin{IEEEeqnarray}{rcl}
\hat{R}_{1\rightarrow 2}^{(l)} &<& R_{1\rightarrow 2}^{(l)}+I\left(U_{1\rightarrow 2,l};X_2\mathcal{V}_{[23,l,2]}\Big|\mathcal{W}_{[2,l]}\mathcal{V}_{[12,l,1]}\right)-\delta_{dp}(2,l)\\
\hat{R}_{3\rightarrow 2}^{(l-1)} &<& R_{3\rightarrow 2}^{(l-1)}+I\left(U_{3\rightarrow 2,l-1};X_2U_{1\rightarrow 23,l}\mathcal{V}_{[12,l,2]}\Big|\mathcal{W}_{[1,l]}\mathcal{V}_{[23,l-1,3]}\right)-\delta'_{dp}(2,l)\\
\hat{R}_{3\rightarrow 2}^{(l-1)}+\hat{R}_{1\rightarrow 2}^{(l)}&<& R_{3\rightarrow 2}^{(l-1)}+R_{1\rightarrow 2}^{(l)}+I\left(U_{1\rightarrow 2,l};X_2\mathcal{V}_{[23,l,2]}\Big|\mathcal{W}_{[2,l]}\mathcal{V}_{[12,l,1]}\right)\nonumber\\
 \IEEEeqnarraymulticol{3}{c}{ +I\left(U_{3\rightarrow 2,l-1};X_2U_{1\rightarrow 23,l}\mathcal{V}_{[12,l,2]}\Big|\mathcal{W}_{[1,l]}\mathcal{V}_{[23,l-1,3]}\right)}\nonumber\\
 \IEEEeqnarraymulticol{3}{r}{ - I\left(U_{3\rightarrow 2,l-1};U_{1\rightarrow 2,l}\Big|\mathcal{W}_{[2,l]}\mathcal{V}_{[23,l-1,3]}\mathcal{V}_{[12,l,1]}X_2\right)-\delta''_{dp}(2,l)}
\label{eq:rates_decod_priv_2}
\end{IEEEeqnarray}
where $\delta_{dp}(2,l),\delta'_{dp}(2,l),\delta''_{dp}(2,l)$ can be made arbitrarily small. 
Then, combining all the obtained results, we have that:
\begin{equation}
\mbox{Pr}\left\{\mathcal{D}_{l}\cap\mathcal{E}_{dec}(2,l)\cap\bar{\mathcal{E}}_{enc}(1,l)\right\}\xrightarrow[n\rightarrow\infty]{}0\ .
\end{equation}
At this point, the story is as it was at the encoding stage in node $1$ and all the steps can be repeated with minor modifications, proving the desired results at the end of round $l$:
\begin{equation}
\mbox{Pr}\left\{\mathcal{D}_{l+1}\right\}\xrightarrow[n\rightarrow\infty]{}1\, ,\ \mbox{Pr}\left\{\mathcal{D}_{l}\cap\mathcal{E}_{l}\right\}\xrightarrow[n\rightarrow\infty]{}0\ .
\end{equation}
The other rates equations are as follows:
\begin{itemize}
\item Encoding at node 2:
\begin{IEEEeqnarray}{rCl}
\hat{R}_{2\rightarrow 13}^{(l)}&>&I\left(X_2;U_{2\rightarrow 13,l}\Big|\mathcal{W}_{[2,l]}\right)+\delta_c(2,l,13)\\
\hat{R}_{2\rightarrow 1}^{(l)}&>&I\left(X_2;U_{2\rightarrow 1,l}\Big|\mathcal{W}_{[3,l]}\mathcal{V}_{[12,l,2]}\right)+\delta_c(2,l,1)\\
\hat{R}_{2\rightarrow 3}^{(l)}&>&I\left(X_2;U_{2\rightarrow 3,l}\Big|\mathcal{W}_{[3,l]}\mathcal{V}_{[23,l,2]}\right)+\delta_c(2,l,3)
\label{eq:encoding_2}
\end{IEEEeqnarray}
\item Decoding at node 3:
\begin{IEEEeqnarray}{rCl}
R_{2\rightarrow 13}^{(l)}&>& I\left(X_2;U_{2\rightarrow 13,l}\Big|X_3\mathcal{W}_{[2,l]}\mathcal{V}_{[13,l,1]}\mathcal{V}_{[23,l,2]}\right)+\delta_{dc}(3,l)\\
R_{2\rightarrow 13}^{(l)}+R_{1\rightarrow 23}^{(l)}&>& I\left(X_1X_2;U_{1\rightarrow 23,l}U_{2\rightarrow 13,l}\Big|X_3\mathcal{W}_{[1,l]}\mathcal{V}_{[13,l,1]}\mathcal{V}_{[23,l,2]}\right)+\delta''_{dc}(3,l) \label{eq:rates_decod_com_3} \\
\hat{R}_{2\rightarrow 3}^{(l)}&<& R_{2\rightarrow 3}^{(l)}+I\left(U_{2\rightarrow 3,l};X_3\mathcal{V}_{[13,l,3]}\Big|\mathcal{W}_{[3,l]}\mathcal{V}_{[23,l,2]}\right)-\delta_{dp}(3,l)\\
\hat{R}_{1\rightarrow 3}^{(l)}&<& R_{1\rightarrow 3}^{(l)}+I\left(U_{1\rightarrow 3,l};X_3U_{2\rightarrow 13,l}\mathcal{V}_{[23,l,3]}\Big|\mathcal{W}_{[2,l]}\mathcal{V}_{[13,l,1]}\right)-\delta'_{dp}(3,l)\\
\hat{R}_{1\rightarrow 3}^{(l)}+\hat{R}_{2\rightarrow 3}^{(l)}&<& R_{1\rightarrow 3}^{(l)}+R_{2\rightarrow 3}^{(l)}+I\left(U_{2\rightarrow 3,l};X_3\mathcal{V}_{[13,l,3]}\Big|\mathcal{W}_{[3,l]}\mathcal{V}_{[23,l,2]}\right)\nonumber\\
 \IEEEeqnarraymulticol{3}{c}{ +I\left(U_{1\rightarrow 3,l};X_3U_{2\rightarrow 13,l}\mathcal{V}_{[23,l,3]}\Big|\mathcal{W}_{[2,l]}\mathcal{V}_{[13,l,1]}\right)}\nonumber\\
 \IEEEeqnarraymulticol{3}{r}{ - I\left(U_{1\rightarrow 3,l};U_{2\rightarrow 3,l}\Big|\mathcal{W}_{[3,l]}\mathcal{V}_{[23,l,2]}\mathcal{V}_{[13,l,1]}X_3\right)-\delta''_{dp}(3,l)}
\label{eq:rates_decod_priv_3}
\end{IEEEeqnarray}
\item Encoding at node 3:
\begin{IEEEeqnarray}{rCl}
\hat{R}_{3\rightarrow 12}^{(l)}&>&I\left(X_3;U_{3\rightarrow 12,l}\Big|\mathcal{W}_{[3,l]}\right)+\delta_c(3,l,12)\\
\hat{R}_{3\rightarrow 1}^{(l)}&>&I\left(X_3;U_{3\rightarrow 1,l}\Big|\mathcal{W}_{[1,l+1]}\mathcal{V}_{[13,l,3]}\right)+\delta_c(3,l,1)\\
\hat{R}_{3\rightarrow 2}^{(l)}&>&I\left(X_3;U_{3\rightarrow 2,l}\Big|\mathcal{W}_{[1,l+1]}\mathcal{V}_{[23,l,3]}\right)+\delta_c(3,l,2)
\label{eq:encoding_3}
\end{IEEEeqnarray}
\item Decoding at node 1:
\begin{IEEEeqnarray}{rCl}
R_{3\rightarrow 12}^{(l)}&>& I\left(X_3;U_{3\rightarrow 12,l}\Big|X_1\mathcal{W}_{[3,l]}\mathcal{V}_{[12,l,2]}\mathcal{V}_{[13,l,3]}\right)+\delta_{dc}(1,l)\\
R_{3\rightarrow 12}^{(l)}+R_{2\rightarrow 13}^{(l)}&>& I\left(X_2X_3;U_{2\rightarrow 13,l}U_{3\rightarrow 12,l}\Big|X_1\mathcal{W}_{[2,l]}\mathcal{V}_{[12,l,2]}\mathcal{V}_{[13,l,3]}\right)+\delta''_{dc}(1,l)
\label{eq:rates_decod_com_1}
\end{IEEEeqnarray}
\begin{IEEEeqnarray}{rCl}
\hat{R}_{3\rightarrow 1}^{(l)}&<& R_{3\rightarrow 1}^{(l)}+I\left(U_{3\rightarrow 1,l};X_1\mathcal{V}_{[12,l+1,1]}\Big|\mathcal{W}_{[1,l+1]}\mathcal{V}_{[13,l,3]}\right)-\delta_{dp}(1,l)\\
\hat{R}_{2\rightarrow 1}^{(l)}&<& R_{2\rightarrow 1}^{(l)}+I\left(U_{2\rightarrow 1,l};X_1U_{3\rightarrow 12,l}\mathcal{V}_{[13,l+1,1]}\Big|\mathcal{W}_{[3,l]}\mathcal{V}_{[12,l,2]}\right)-\delta'_{dp}(1,l)\\
\hat{R}_{2\rightarrow 1}^{(l)}+\hat{R}_{3\rightarrow 1}^{(l)}&<& R_{2\rightarrow 1}^{(l)}+R_{3\rightarrow 1}^{(l)}+I\left(U_{3\rightarrow 1,l};X_1\mathcal{V}_{[12,l+1,1]}\Big|\mathcal{W}_{[1,l+1]}\mathcal{V}_{[13,l,3]}\right)\nonumber\\
 \IEEEeqnarraymulticol{3}{c}{ +I\left(U_{2\rightarrow 1,l};X_1U_{3\rightarrow 12,l}\mathcal{V}_{[13,l+1,1]}\Big|\mathcal{W}_{[3,l]}\mathcal{V}_{[12,l,2]}\right)}\nonumber\\
 \IEEEeqnarraymulticol{3}{r}{ - I\left(U_{2\rightarrow 1,l};U_{3\rightarrow 1,l}\Big|\mathcal{W}_{[1,l+1]}\mathcal{V}_{[12,l,2]}\mathcal{V}_{[13,l,3]}X_1\right)-\delta''_{dp}(1,l)\ .}
\label{eq:rates_decod_priv_1}
\end{IEEEeqnarray}
\end{itemize}
The final private rate equations in Theorem \ref{theo-main-theorem} follows from a rather simple Fourier-Motzkin elimination procedure \cite{ELGamal-Kim-book}.
 \bibliographystyle{IEEEtran.bst}
\bibliography{IEEEabrv,biblio}

\end{document}

%% file: packages.tex
\usepackage[latin1]{inputenc}
\usepackage[T1]{fontenc}
\usepackage[english]{babel}
\usepackage{amssymb,amsmath,amsfonts}
\usepackage{times}
\usepackage{graphicx}
\usepackage{color}
\usepackage{verbatim}
\usepackage{cite}

\usepackage{subfig}		

\usepackage{stmaryrd}

\usepackage{cancel}
\usepackage[normalem]{ulem}

\usepackage{verbatim}

\newtheorem{definition}	{Definition}
\newtheorem{theorem}	{Theorem}

\newtheorem{lemma}		{Lemma}
\newtheorem{corollary}	{Corollary}
\newtheorem{remark}		{Remark}

%% file: commandes.tex
\newcommand{\bE}{{\mathbb E}}

\newcommand{\mkv}{-\!\!\!\!\minuso\!\!\!\!-}

